%% file: main.tex
\documentclass[onecolumn]{aastex631}
\usepackage{natbib}
\usepackage{outlines}
\usepackage{verbatim}
\usepackage{tcolorbox}
\usepackage{placeins}
\usepackage{subfigure}
\usepackage{enumitem}
\usepackage{bm}

\newcommand{\kmss}{\mbox{km~s$^{-1}$}}
\newcommand{\kms}{\,\kmss}
\newcommand{\mss}{\mbox{m~s$^{-1}$}}
\newcommand{\ms}{\,\mbox{m~s$^{-1}$}}
\newcommand{\ci}{[\ion{C}{1}]}
\newcommand{\cii}{[\ion{C}{2}]}
\newcommand{\oiii}{[\ion{O}{3}]}

\newcommand{\ignore}[1]{{}}

\usepackage{nicematrix}
\usepackage{varwidth}
\BeforeBegin{NiceTabular}{\let\begin\BeginEnvironment\let\end\EndEnvironment}
\BeforeBegin{NiceArray}{\let\begin\BeginEnvironment}
\BeforeBegin{NiceMatrix}{\let\begin\BeginEnvironment}
\newcommand{\pastelorange}{F1CDB0}
\newcommand{\pastelred}{red!20}
\newcommand{\pastelpurple}{C7CEEA}
\newcommand{\pastelgreen}{E2F0CB}
\newcommand{\pastelgray}{E1DBD6}
\newcommand{\cellspacelimit}{1pt}
\newcommand{\headercolor}{blue!10}
\newcommand{\highlightcolor}{yellow!20}
\newcolumntype{Y}{>{\centering\arraybackslash}X}
\NiceMatrixOptions{cell-space-limits = \cellspacelimit}

\begin{document}

\title{ALMA Memo 621\\\vspace*{0.2in}The ALMA2030 Wideband Sensitivity Upgrade}

\author{John Carpenter}
\affiliation{Joint ALMA Observatory, Avenida Alonso de C\'ordova 3107, Vitacura, Santiago, Chile}

\author{Crystal Brogan}
\affiliation{National Radio Astronomy Observatory (NRAO), 520 Edgemont Road, Charlottesville, VA 22903, USA}

\author{Daisuke Iono}
\affiliation{National Astronomical Observatory of Japan (NAOJ), 2-21-1 Osawa, Mitaka, Tokyo 181-8588, Japan}

\author{Tony Mroczkowski}
\affiliation{European Southern Observatory (ESO), Karl-Schwarzschild-Str. 2, 85748, Garching bei Munchen, Germany}

\input{abstract}

\tableofcontents
\newpage

\input{introduction}
\input{ScienceCapabilities}
\input{origins_planets}
\input{origins_chemistry}
\input{origins_galaxies}
\input{summary}

\clearpage
\appendix
\input{AppendixTech}

\clearpage
\bibliography{main}

\end{document}

%% file: abstract.tex
\begin{abstract}

The Wideband Sensitivity Upgrade (WSU) is the top priority initiative for the ALMA2030 Development Roadmap. The WSU will initially double, and eventually quadruple, ALMA's system bandwidth and will deliver improved sensitivity by upgrading the receivers, digital electronics, and correlator. The WSU will afford significant improvements for every future ALMA observation, whether it is focused on continuum or spectral line science.  The continuum imaging speed will increase by at least a factor of 3 for the 2$\times$ bandwidth upgrade (and at least a factor of 6 for the 4$\times$ upgrade), plus any speed gains from improved receiver temperatures. The (single) spectral line imaging speed is expected to improve by a factor of 2--3 depending on the receiver band. The improvements provided by the WSU will be most dramatic for high spectral resolution observations, where the instantaneous bandwidth correlated at $\sim0.1\text{--}0.2$\kms\/ resolution will increase by 1--2 orders of magnitude in most receiver bands, with the largest gains at the lowest frequencies.
The improved sensitivity and spectral tuning grasp will open new avenues of exploration, increase sample sizes, and enable more efficient observations. The impact will span the vast array of astronomical topics that embodies ALMA’s motto {\em In Search of our Cosmic Origins}. 
The WSU will greatly expand the chemical inventory of protoplanetary disks that surround young stars, which will have profound implications for how and when planets form in disks and their composition. Observations of the interstellar medium in the Milky Way and nearby galaxies will simultaneously measure a variety of molecular species that will be used to build large samples of clouds, cores, and protostars in a variety of evolutionary states. The WSU will also enable efficient surveys of galaxies at high redshift to probe the origins of galaxies. 
The first elements of the WSU are now under development and will be available to the user community later this decade, including a wideband Band 2 receiver, a wideband upgrade to Band 6, new digitizers and digital transmission system, and soon a new correlator. 
Upgrades to other instruments and receiver bands are under study, including the newly developed ACA spectrometer and 2SB upgrades of Band 9 and 10 receivers for better rejection of atmospheric noise. 
The substantial gains in the observing efficiency enabled by the WSU will maximize the synergies between ALMA and facilities across the electromagnetic spectrum, and further enhance ALMA as the world leading facility for millimeter/submillimeter astronomy.

\clearpage
\end{abstract}

%% file: introduction.tex
\section{Introduction}
\label{sec:introduction}

The Atacama Large Millimeter/submillimeter Array (ALMA) is the most sensitive telescope ever built for high-resolution observations at millimeter and submillimeter wavelengths. 
The array consists of fifty 12-m diameter antennas that can be reconfigured to baselines as long as 16\,km (the ``12-m Array"), and the Atacama Compact Array (ACA), also known as the Morita Array. The Morita Array comprises twelve 7-m antennas (the ``7-m Array'') that sample short visibility spacings, and four 12-m antennas that provide total power capabilities (the ``Total Power'' or ``TP-Array'').
Each antenna is equipped with eight receiver bands that provide frequency coverage between 84 and 950\,GHz, with two additional bands under construction that will extend the coverage to as low as 35\,GHz. The ALMA site, located in the Atacama desert in northern Chile at an elevation of 5000\,m on the Chajnantor plateau, provides excellent observing conditions with low precipitable water vapor. The large number of antennas, the excellent receivers with low-noise performance, and the high-altitude site provide an extremely sensitive and flexible instrument for millimeter and submillimeter imaging.

To enhance ALMA as a world leading facility for millimeter/submillimeter astronomy, the ALMA partnership\footnote{ALMA is a partnership between the European Southern Observatory (ESO), the National Science Foundation (NSF) of the United States and the National Institutes of Natural Sciences (NINS) of Japan in collaboration with the Republic of Chile. ALMA is funded by ESO in representation of its member states, by NSF in collaboration with the National Research Council (NRC) of Canada and the National Science Council (NSC) of Taiwan, and by NINS in collaboration with the Academia Sinica (AS) in Taiwan, and the Korea Astronomy and Space Science Institute (KASI) of South Korea.} established the ALMA Development Program that promotes hardware, software, and infrastructure improvements for ALMA. Each of the ALMA partners administers the development program for their region, including holding workshops to discuss ideas for future developments, as well as soliciting and selecting proposals.
Descriptions of programs that have been funded through the ALMA Development Program are available through the NAOJ,\footnote{\url{https://researchers.alma-telescope.jp/e/report/development}} ESO,\footnote{\url{https://www.eso.org/sci/facilities/alma/development-studies.html}} and NRAO\footnote{\url{https://science.nrao.edu/facilities/alma/science_sustainability/NADevelopmentProgram}} websites.

To develop a vision for the future and maximize the impact of the ALMA Development Program, the ALMA Board tasked the ALMA Science Advisory Committee (ASAC) to recommend developments that ALMA should consider implementing by the year 2030. A working group prioritized those recommendations in a strategic ALMA2030 Development Roadmap \citep{Carpenter19}, which was endorsed by the ALMA Board in 2017. The scientific motivation of the Roadmap is centered around three broad themes: understanding the origin of galaxies from the first galaxies ($z>10$) to the peak of star formation at $z=1\text{--}3$, tracing the origin of chemical complexity through the star formation process, and probing the origins of planets. 
The top technical recommendation of the ALMA2030 Development Roadmap, which is needed to address all three science themes, is to increase the overall throughput of the ALMA system by both increasing the instantaneous frequency coverage and overall sensitivity. The Roadmap also recommended continued development of the ALMA Science Archive, which includes handling the data from the increased bandwidth. In addition, the Roadmap recommended further scientific study and technical evaluation of other long-term developments to examine if they are feasible for ALMA, including expanding the maximum baselines by a factor of 2--3, introducing focal plane arrays, and increasing the number of antennas.

The Wideband Sensitivity Upgrade (WSU) is a pan-ALMA initiative to implement the highest priority of the ALMA Development Roadmap: to at least double and eventually quadruple ALMA's system bandwidth\footnote{Throughout this document, for simplicity we will refer to ``2$\times$" and ``4$\times$" the present correlated bandwidth, rather than 16 and 32\,GHz per polarization as the future correlated bandwidth goals. However, strictly speaking the 2$\times$ and 4$\times$ factors only apply to the current digitized bandwidth ($4\times2$\,GHz, or 8\,GHz per polarization). The sensitivity roll-off induced by the current baseband anti-aliasing filters restricts the current usable correlated bandwidth to 3.75\,GHz per sideband per polarization for a 2SB receiver, or 7.5\,GHz in total per polarization.} and deliver improved sensitivity and scientific capabilities. This goal requires upgrading the receivers (front-end detectors) and most of the digital electronics, including the correlator, as well as the ALMA software that drives the system, and processes and stores the output. The WSU will afford significant improvements for every future ALMA observation for the full range of spatial scales provided by the 12-m, 7-m, and TP arrays. The impact will span the diverse astronomical topics that embodies ALMA’s motto {\em In Search of our Cosmic Origins}, whether it is focused on continuum or spectral line science (or both) or requires standard imaging/correlation or VLBI beamforming capabilities. The improved sensitivity and spectral tuning grasp (the ability to observe a wide range of spectral lines simultaneously) can be used to open new avenues of exploration, increase sample sizes, or simply to enable more efficient observations. In an era when the ALMA over-subscription rate has reached factors of $\sim6$ on the 12-m Array in Cycle 9, increased observing throughput is itself on the critical path towards enabling new discoveries.

This document highlights the science that will be enabled by the WSU. Section~\ref{sec:SC} summarizes the technical improvements and the overall scientific benefits of the WSU. The subsequent sections present the scientific impact of these upgrades on the ALMA2030 Roadmap's key science themes: The Origins of Planets (Section~\ref{sec:planets}), the Origins of Chemical Complexity (Section~\ref{sec:chemistry}), and the Origins of Galaxies (Section~\ref{sec:galaxies}).
Section~\ref{sec:summary} summarizes the WSU science case, and Appendix~\ref{sec:tech} provides detailed information on the technical aspects of the upgrade.

%% file: ScienceCapabilities.tex
\section{The Wideband Sensitivity Upgrade}
\label{sec:SC}

\begin{figure}[t]
\centering
\includegraphics[width=\textwidth]{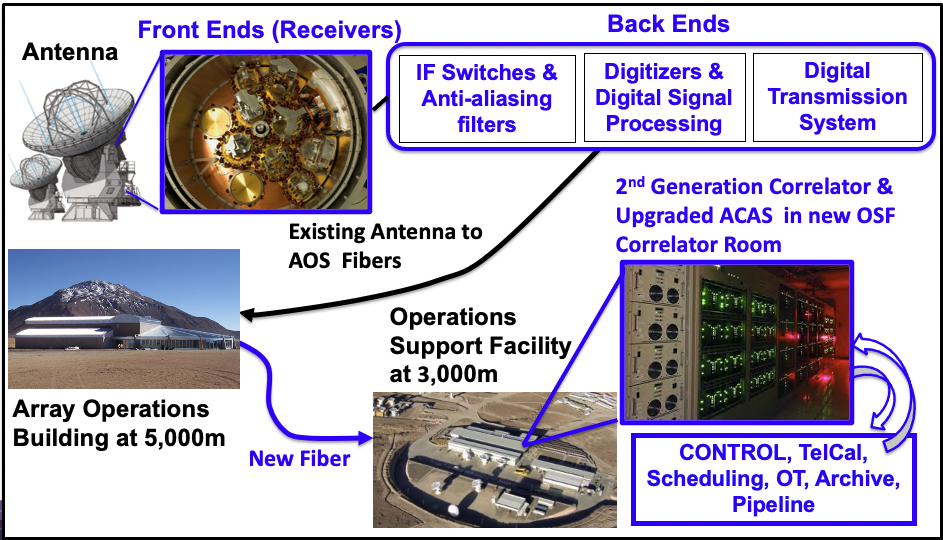}
\caption{Simplified overview of the ALMA signal chain post-WSU upgrade. Components that are either new or will be upgraded are shown in blue (Credit for inset graphics and images: ALMA (ESO/NAOJ/NRAO)).}
\label{fig:overview}
\end{figure}

The overall goal of the WSU is to upgrade the ALMA receivers to have 2--4$\times$ broader instantaneous bandwidth with lower noise and to also upgrade the digital signal chain, including the digitizers, data transmission system, and the correlator to process the full receiver bandwidth. Figure~\ref{fig:overview} shows a simplified view of the key parts of the ALMA signal chain, from the receivers in the antennas to the correlator room post-WSU upgrade, as well as some of the key software subsystems that must also be upgraded.  Components that will be upgraded as part of the WSU are indicated in blue. In addition to the digital components of the upgrade, ALMA also plans to upgrade gradually all of the front-end receivers.  A priority of the observatory is to complete the digital elements and install upgraded receivers while also minimizing any negative impact on ALMA science observing (i.e., technical downtime) during the course of the WSU deployment. ALMA aims to complete the digital elements and upgrade several of the receivers by the end of this decade. Technical details about the WSU are provided in Appendix~\ref{sec:tech}.

As summarized in Table~\ref{tbl:improvement} and described in the remainder of this section, the WSU will bring a number of dramatic improvements to the ALMA system, including far more flexibility for observing spectral lines, faster spectral scan observations (especially at high spectral resolution), more sensitive spectral line and continuum observations, ultra-high spectral resolution, and imaging and calibration improvements. The impact of these improvements on future ALMA science is described in Sections~\ref{sec:planets}--\ref{sec:galaxies}.  

\input{table_improvements}

\subsection{Increased Instantaneous Bandwidth}
\label{sec:available_bw}

\begin{figure}[h]
\setlength{\abovecaptionskip}{-10pt}
\centering
\includegraphics[width=0.95\textwidth]{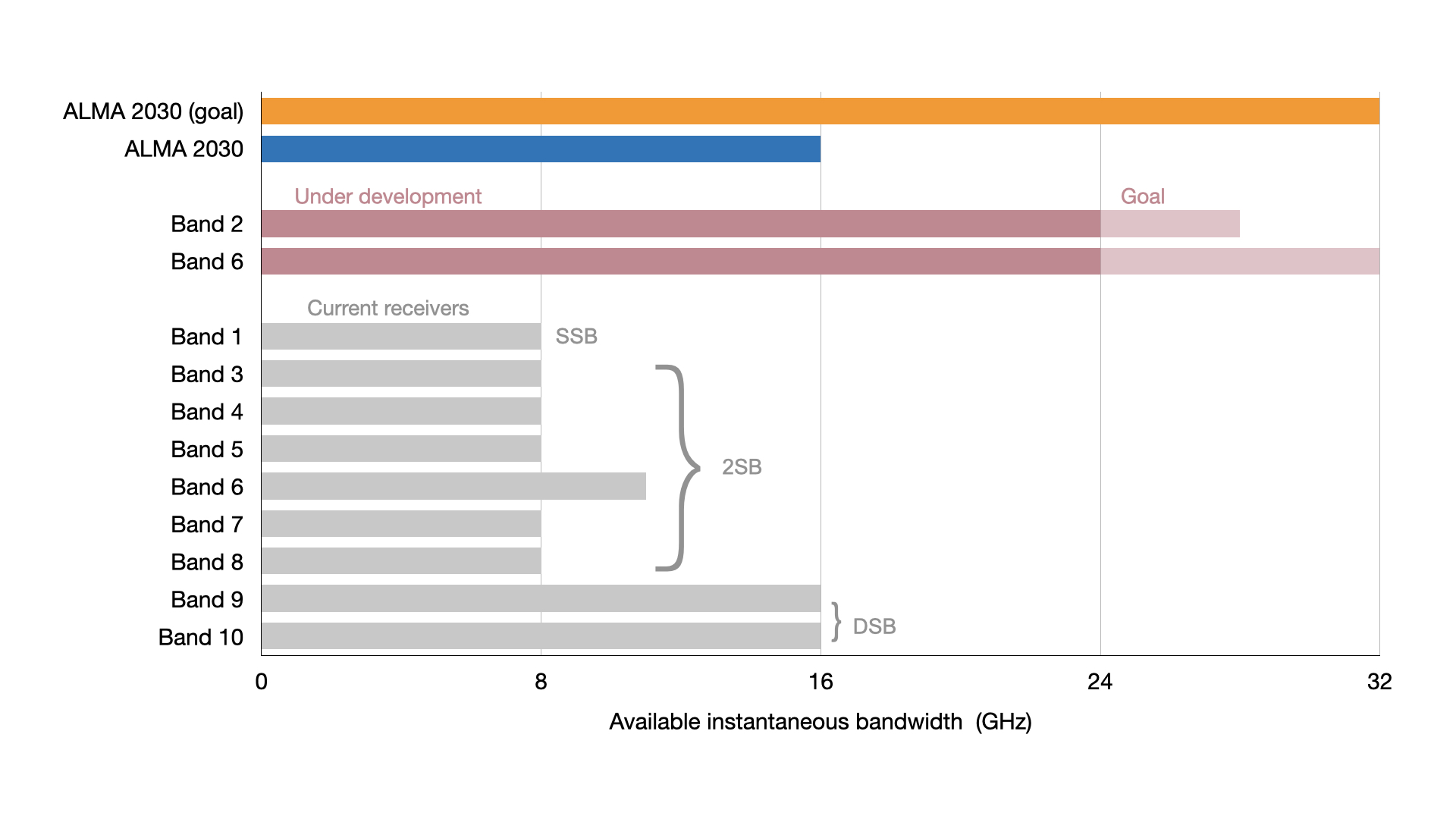}
\caption{Aggregate instantaneous bandwidth of the ALMA receivers (per polarization). At minimum the WSU receiver upgrades will increase the aggregate IF bandwidth (summed across sidebands for 2SB receivers) to at least 16\,GHz per polarization (blue bar), with a goal of 32\,GHz per polarization (orange bar). The gray bars show the aggregate bandwidth (per polarization) of the current ALMA receivers (summed across sidebands for 2SB receivers). The red bars show the two ALMA wideband receivers already under development (Bands 2 and 6), where the dark red is the minimum bandwidth (24\,GHz per polarization), and the light red shows the goal for these two projects. 
}
\label{fig:available_bw}
\end{figure}

The current ALMA receivers have a mixture of Single Sideband (SSB; Band 1), sideband separating (2SB; Bands 3--8), and Double Sideband (DSB; Bands 9 and 10) technologies. As shown in Figure~\ref{fig:available_bw}, the available aggregate instantaneous bandwidth per polarization\footnote{Table~\ref{tbl:FEstatus} in the Appendix lists the current and future Intermediate Frequency (IF) bandwidth (per polarization) of the ALMA receivers. The ``instantaneous'' bandwidth refers to the sum of the lower and and upper sidebands of the 2SB and DSB receivers.} of the ALMA receivers ranges between 8\,GHz per polarization for most receivers and up to 16\,GHz for Bands 9 and 10. The WSU will require future receiver upgrades to have at least 16\,GHz of instantaneous bandwidth per polarization with 2SB mixers, with a goal of 32\,GHz per polarization (in aggregate across sidebands). For most ALMA receivers, this will represent at least a factor of two increase in the bandwidth, and as much as a factor of four. The prototypes for the two wideband receivers under pre-production (Band 2 and Band 6v2) will exceed the minimum ALMA2030 requirements and produce at least 24\,GHz of bandwidth per polarization. The WSU upgrade will also require the receivers to have improved receiver noise temperatures over the current receivers (see Appendix~\ref{sec:receivers}). Throughout this paper, we assume that all receiver bands will be upgraded to at least 16\,GHz of instantaneous bandwidth per polarization (in aggregate across sidebands). We further assume that all upgraded receivers will be 2SB, apart from Band 1, which is likely to remain SSB even after upgrade.

\subsection{Unprecedented Spectral Resolution at Wide Bandwidth}
\label{sec:specres}

One of the principal requirements for making transformative strides in the Origins of Planets and Origins of Chemical Complexity science themes (Section~\ref{sec:planets} and Section~\ref{sec:chemistry}, respectively) 
is expanded spectral grasp --- the ability to simultaneously observe a wide range of diagnostic line transitions 
at the required spectral resolution with high spectral sensitivity. 
The WSU will make it possible to observe with high spectral resolution over the entire available bandwidth,
in contrast to the current situation in which the majority of Galactic spectral line projects must give up significant correlated bandwidth in order to achieve a suitable spectral resolution. Increased spectral grasp will not only decrease significantly  the observing time needed to obtain the required spectral transitions for a given science goal, spectral lines observed simultaneously bypass many of the effects that impede accurate spectral line analyses, such as relative flux calibration and variations in $uv$-coverage between the different tunings.

After the WSU, it will no longer be necessary to give up correlated bandwidth for spectral resolution, apart from specialized ``ultra-high'' spectral resolution use cases (see Section~\ref{sec:ultra}). The  WSU correlator can provide a native 13.5\,kHz spectral resolution (see Appendix~\ref{sec:correlator}) over 16\,GHz of 
receiver bandwidth at full polarization to achieve a velocity resolution of $0.12\, (35\,\mathrm{GHz}/\nu)$\kms, where $\nu$ is the observed frequency in GHz. A velocity resolution in the 0.1--0.2\kms\/ range cannot presently be achieved with the current ALMA Baseline Correlator (BLC) over the entire available IF bandwidth for {\em any} ALMA band. {\bf Achieving the ability to observe with 0.1--0.2\kms\/ spectral resolution over the full expanded correlated bandwidth (16 GHz per polarization at minimum, with goal of 32 GHz per polarization) at any ALMA frequency is the key science goal of the WSU.}

\begin{figure}[h]
\centering
\includegraphics[width=0.95\textwidth]{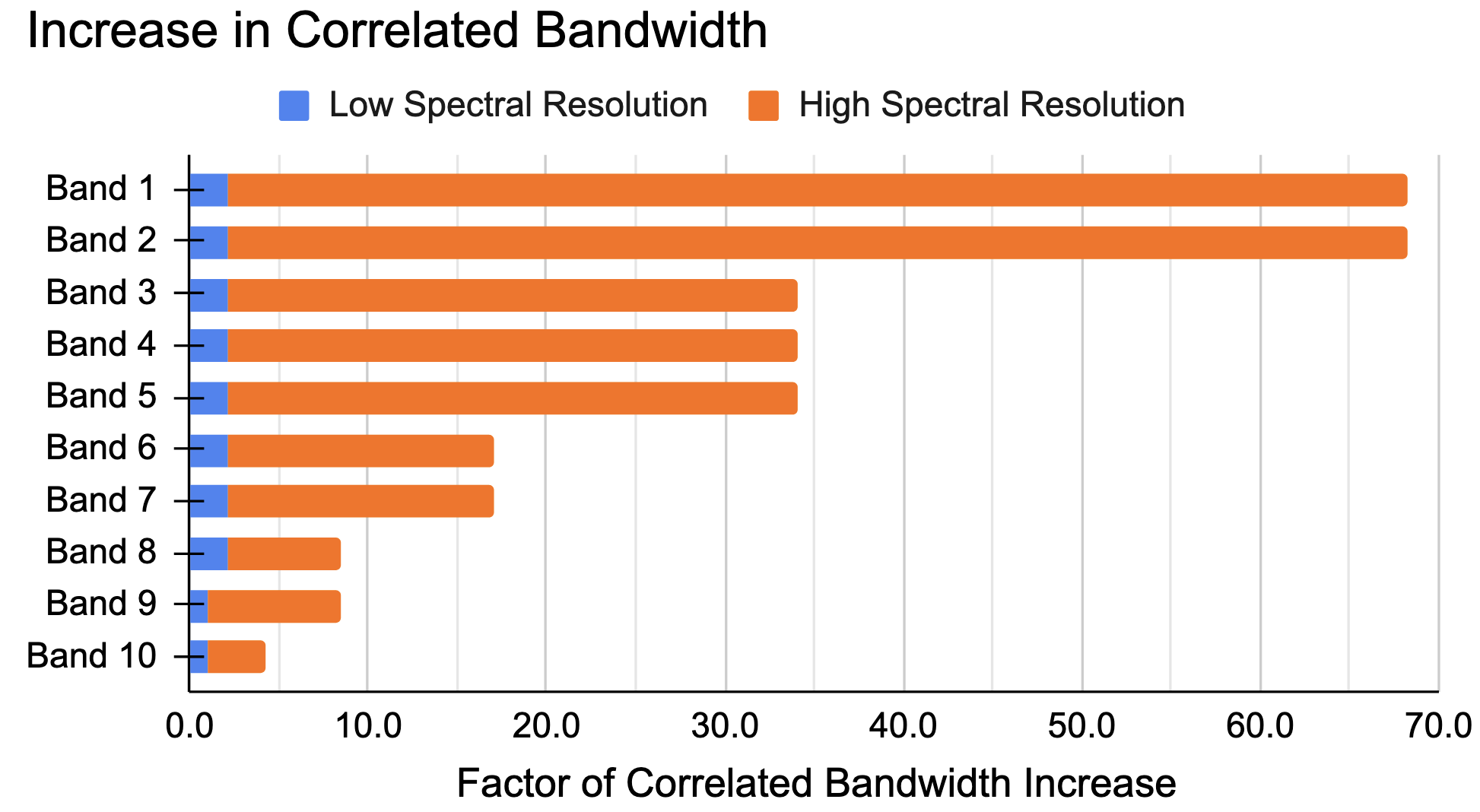}
\caption{Factor by which the correlated bandwidth will increase with the 2$\times$ bandwidth WSU relative to the current system for each band, in two spectral resolution regimes: ``Low Spectral Resolution'' (blue), which is defined by the best spectral resolution that the BLC can achieve at full correlated bandwidth (7.5\,GHz per polarization), and ``High Spectral Resolution'' (orange), which is defined by the WSU goal of reaching 0.1--0.2\kms\ in every ALMA band at the maximum correlated bandwidth (16\,GHz per polarization). The numerical values adopted for both the current system and WSU are provided in Table~\ref{tbl:bwincrease} and Table~\ref{tbl:SpecRes} for the two spectral resolution regimes. 
}
\label{fig:cbwgraph}
\end{figure}

Figure~\ref{fig:cbwgraph} shows the factor by which the correlated bandwidth will increase for each ALMA band for the initial 2$\times$ bandwidth upgrade for two velocity resolution regimes: 
\begin{itemize}
    \item ``Low Spectral Resolution'' is defined by the best spectral resolution that the current ALMA Baseline Correlator (BLC)  can achieve at full correlated bandwidth (7.5\,GHz dual polarization), and ranges from 8.4\kms\/ at Band 1 to 0.34\kms\/ at Band 10. The numerical values for all bands for this regime can be found in Table~\ref{tbl:SpecRes} in Appendix~\ref{sec:correlator}).
    \item ``High Spectral Resolution'' is defined by the WSU goal of achieving a spectral resolution between 0.1--0.2\kms\/ for every ALMA band across at least 16\,GHz bandwidth per polarization, with of goal of 32\,GHz per polarization. 
    The numerical values for all bands for this regime can be found in Table~\ref{tbl:bwincrease}. Note that at Band 1, a spectral resolution $<$0.2 \kms\/ is presently impossible.
\end{itemize}

{\bf In the Low Spectral Resolution regime, the WSU correlated bandwidth will initially increase by a factor of 2.1$\times$ for Bands 1--8  when these bands are all upgraded to instantaneous bandwidths $\geq$ 16\,GHz per polarization} (see Figure~\ref{fig:cbwgraph}). 
Although the current Band 9 and 10 DSB receivers already provide 16\,GHz of bandwidth per polarization, they will be significantly more sensitive with 2SB configurations after their upgrades (see Appendix~\ref{sec:rxother}). 
As shown in Figure~\ref{fig:cbwgraph} and Table~\ref{tbl:bwincrease}, the WSU will enable extraordinary increases in the amount of correlated bandwidth at high spectral resolution. {\bf Depending on the band, the bandwidth correlated at 0.1--0.2\kms\/ will increase by a factor between 4.3 (Band 10) and 68.3 (Bands 1 \& 2) for the minimum required 2$\times$ bandwidth upgrade!} An additional factor of up to two will be gained for both high and low spectral resolution observations when the 4$\times$ correlated bandwidth expansion is accomplished, depending on the instantaneous bandwidth achieved for each upgraded receiver band (see Appendix~\ref{sec:receivers} for details about the receiver upgrades).

\input{table_bwincrease}

To gauge the potential impact of this new capability on ALMA observations, we assessed the spectral properties of the accepted Cycle 8 proposals. Table~\ref{tbl:cycle8} presents a comparison between the total number of proposals per science category, as well as the number of projects that employed at least one narrow ($<1$\,GHz) spectral window.  {\bf In Cycle 8, 37\% of accepted projects had to give up some bandwidth to achieve their required spectral resolution, with proposal science categories 4 and 5 needing to make this trade-off for $\sim75$\% of projects (see Table~\ref{tbl:cycle8}).} Only 15\% of accepted proposals requested all spectral windows in TDM\footnote{The current ALMA correlator operates in two basic modes: Time Division Mode (TDM) that provides coarse spectral resolution, and Frequency Division
Mode (FDM) with fine spectral resolutions. The current system has 4 basebands,  each of which be used in either TDM or FDM.} (very coarse spectral resolution), and 48\% of the accepted projects request all of the spectral windows in FDM to have higher spectral resolution.

\input{table_cycle8spectral}

\subsection{Increased Observing Efficiency for Spectral Surveys}
\label{sec:specscan}

For some science cases, even with the expanded instantaneous and correlated bandwidth of the WSU, multiple tunings must be employed to cover the required frequency range, either to observe the full range of spectral line diagnostics or to search the plausible range of possible redshifts.
Moreover, some astronomical targets are so uniquely line-rich that full spectral scans are warranted to characterize their complex chemistry. Indeed, spectral scans are essential ingredients for all three of the ALMA Roadmap key science themes.
One of the most significant improvements from the WSU will be the increased observing efficiency of spectral scans as receiver bands are upgraded to wider bandwidths. 

\begin{figure}[h]
\centering
\includegraphics[width=0.9\textwidth]{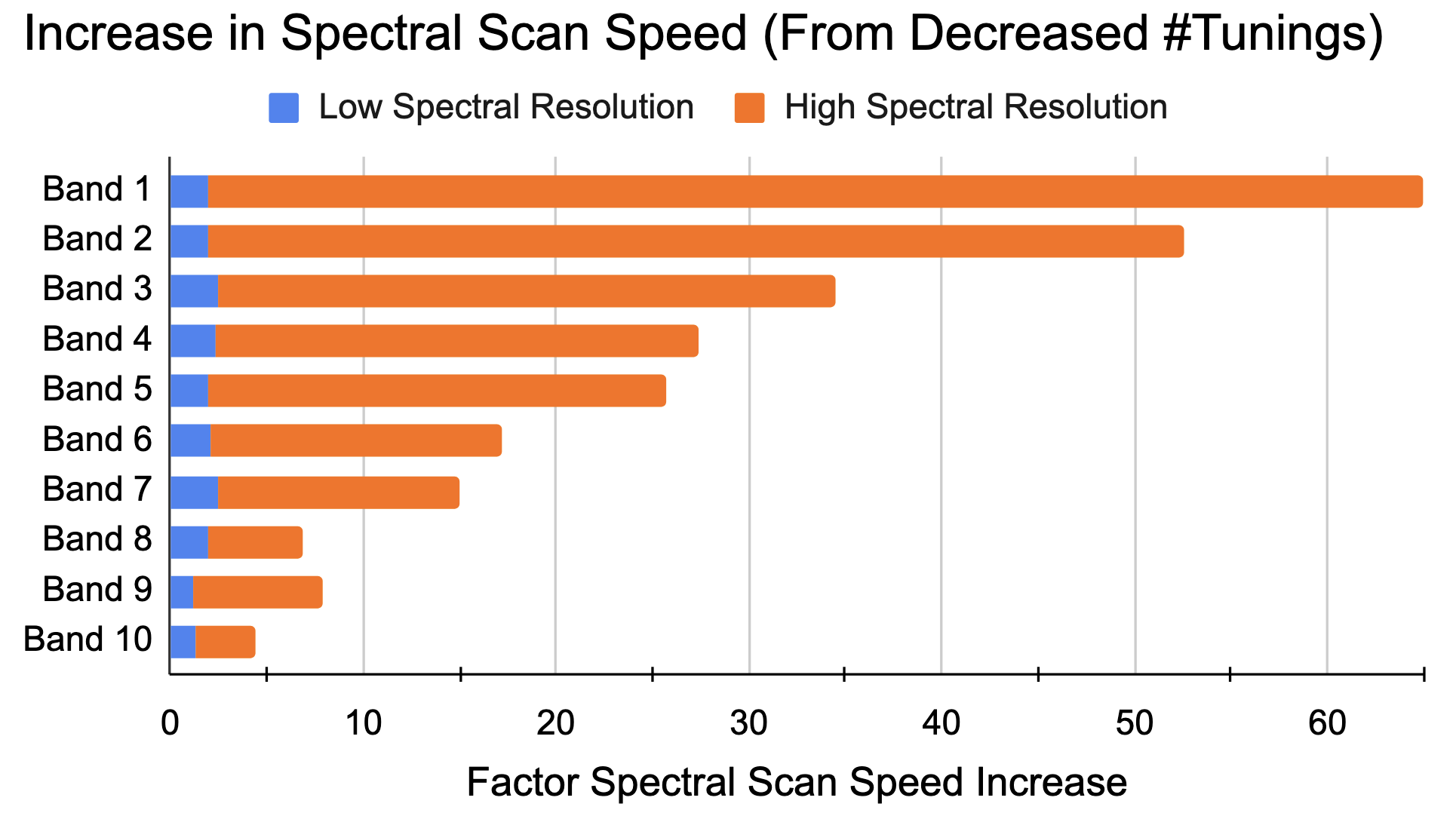}
\caption{Graphical demonstration of the spectral scan speed increase that will be afforded by the 2$\times$ WSU for the ``low'' (blue) and ``high'' (orange) spectral resolution regimes (see Section~\ref{sec:specres}). The exact values adopted are shown in Table~\ref{tbl:SSTune}.}
\label{fig:ScanSpeed}
\end{figure}

\begin{figure}[ht]
\centering
\includegraphics[width=0.74\textwidth]{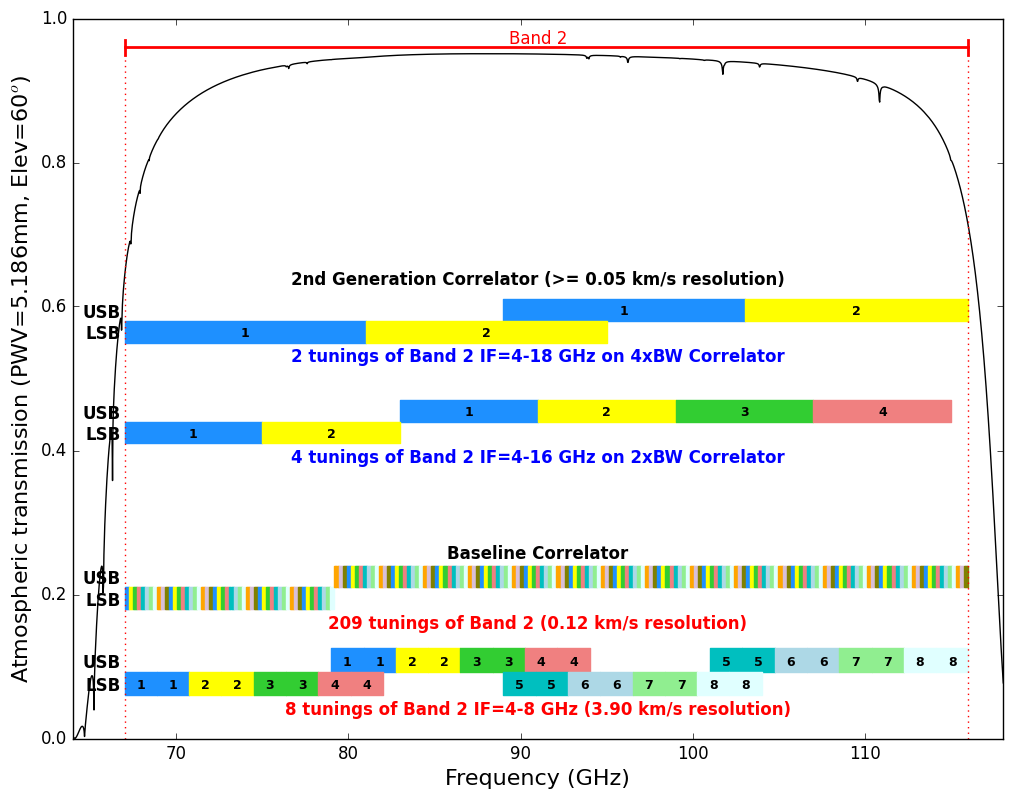}
\includegraphics[width=0.74\textwidth]{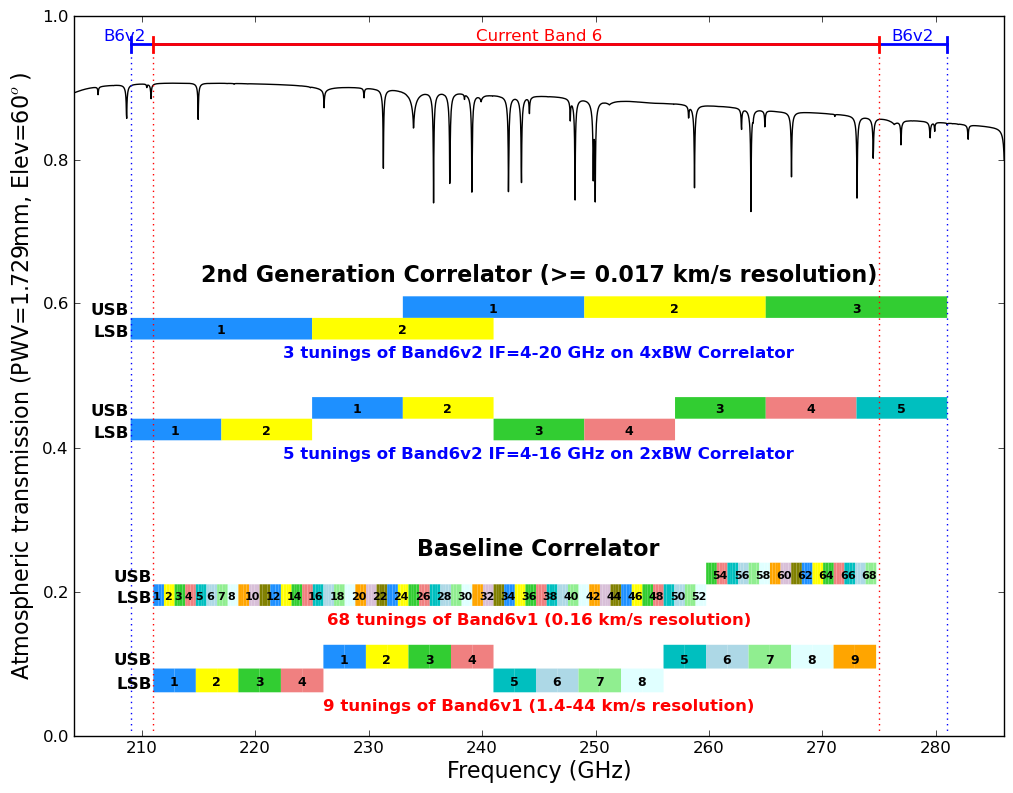}
\caption{Example spectral scan setups for (top) Band 2 and (bottom) Band 6. The lower two spectral setups in each plot show spectral scans for the current system for the indicated spectral resolution, while the upper two examples show the number of tunings that will be needed for the 2$\times$ and eventually 4$\times$  bandwidth upgrade, in the latter case assuming both bands under development meet their stretch goals for IF bandwidth (see Appendix~\ref{sec:receivers}). Credit: Todd Hunter (NRAO).
}
\label{fig:SpecScan}
\end{figure}

Figure~\ref{fig:ScanSpeed} shows the increase in the spectral scan speed that will be afforded by the $2\times$ bandwidth WSU as a result of the decreased number of tunings required to cover each receiver's full radio frequency (RF) range.\footnote{We note that for receiver bands still in the study phase for upgrade, we have retained the current RF ranges; see Appendix~\ref{sec:receivers}.} As in Section~\ref{sec:specres}, the increases are shown for the ``low'' and ``high'' spectral resolution regimes. The numerical values adopted to create this figure are provided in Table~\ref{tbl:SSTune}. For this demonstration, we assume that all of the receiver bands will be upgraded to an IF bandwidth of at least 16\,GHz per polarization (2SB). For the ``low resolution" velocity regime, the improvement in observing speed is typically a factor of 2, though larger values can occur due to the current inefficient match between the gap between sidebands (8\,GHz for most bands) and the maximum correlated bandwidth of 7.5\,GHz. {\bf The ``high resolution" case (0.1--0.2\kms\/) reveals the full power of the WSU, with time savings for spectral scans ranging from 4.3 (Band 10) to an impressive factor of 52.3 for Band 2.} Additional gains (up to a factor of 2 more than shown in Figure~\ref{fig:ScanSpeed}) will be afforded by the ultimate $4\times$ WSU upgrade, though the exact details will depend on the final specifications of each upgraded receiver band. Figure~\ref{fig:SpecScan} shows the detailed spectral tuning setups for the two wideband receivers already under development, Band 2 and Band 6v2 for both the $2\times$ and $4\times$ bandwidth WSU scenarios.

\input{table_spectral_scan_speed}

\FloatBarrier
\subsection{Unique Access to Ultra-high Spectral Resolution}
\label{sec:ultra}

The narrowest dual polarization spectral resolution of the BLC is $2\times15.3$\,kHz, due to the requisite online Hanning smoothing required to damp ringing, equivalent to a spectral resolution of 92\,\ms\/ at 100\,GHz. The resolution degrades to 184\ms\/ for full polarization observations (see Table~\ref{tbl:SpecRes}). Further, the best current BLC spectral resolution comes at the cost of a very narrow bandwidth -- only 0.234\,GHz in aggregate bandwidth if all four basebands use the highest spectral resolution mode. In contrast the AT.CSP can provide 13.5\,kHz resolution (40.5\ms\/ at 100\,GHz) for the full correlated bandwidth of 8\,GHz per sideband per polarization (eventual goal 16\,GHz per sideband per polarization), already surpassing the best spectral resolution that can be achieved with the BLC (see Appendix~\ref{sec:correlator} for additional details). In addition, correlated bandwidth can be traded for even higher spectral resolution independently for each of the 80 (eventually 160) 200\,MHz wide Frequency Slices that make up the AT.CSP correlated bandwidth through the use of the Zoom Mode. {\bf As an example, if all 80 of the AT.CSP Frequency Slices are configured for Zoom, the AT.CSP can process up to 2\,GHz of correlated bandwidth (full polarization) at a resolution of 15\ms\/ at 35\,GHz using the zoom capability of AT.CSP.} Alternatively, individual spectral lines can be targeted for ultra-high spectral resolution, while the majority of the available bandwidth can be spectrally averaged online to aid calibration and to lower the overall data rate and volume.

\begin{figure}[ht]
\centering
\begin{minipage}{0.5\linewidth}
\includegraphics[width=\textwidth]{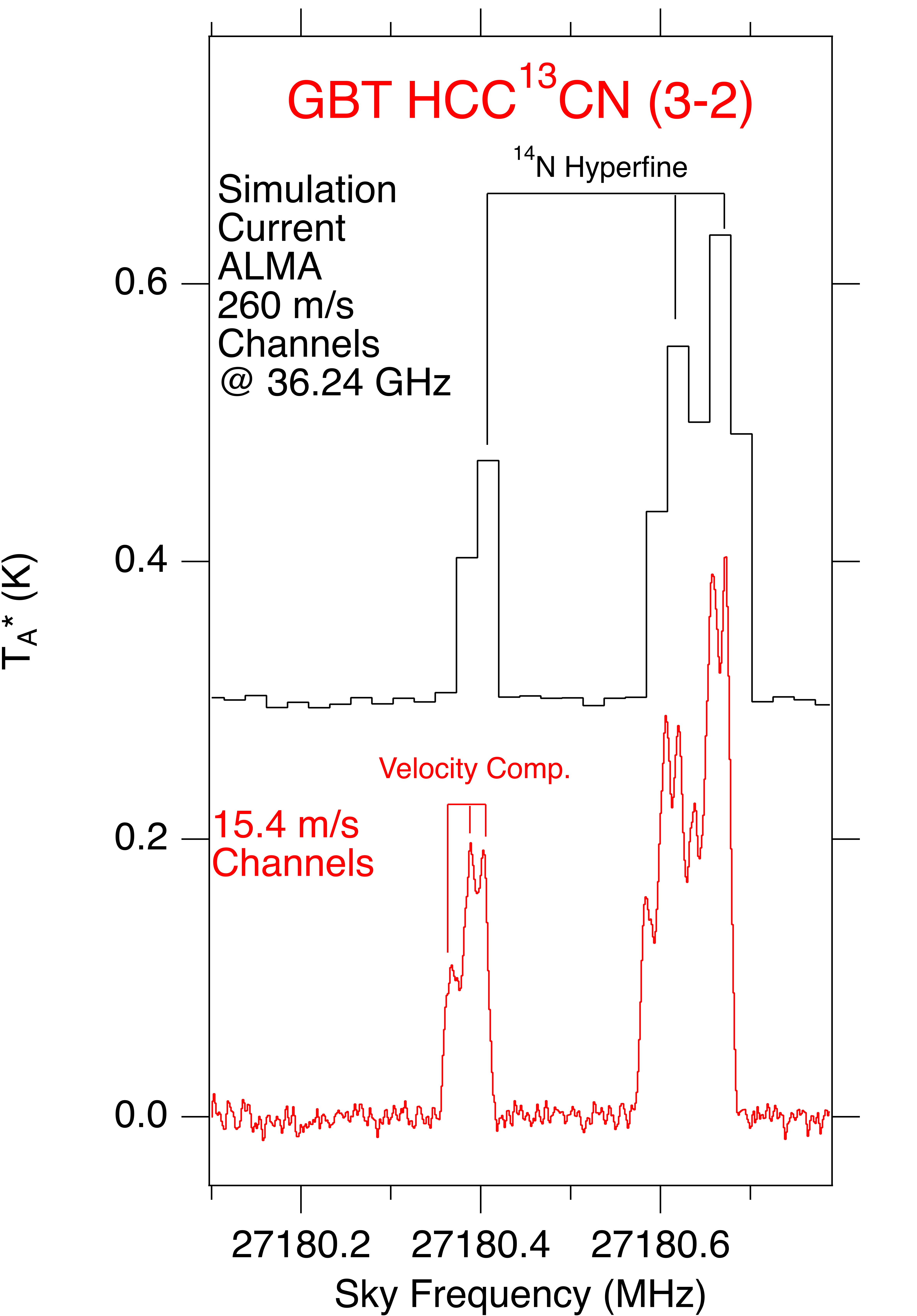}
\end{minipage}
\begin{minipage}{0.48\linewidth}
\caption{The red spectrum shows a recent GBT HCC$^{13}$CN $J=3-2$ spectrum from the GOTHAM survey \citep{McGuire2021} toward the cold dark molecular cloud TMC-1 with an “ultra-high” spectral resolution of 15.4\ms, revealing complex underlying kinematics. In contrast, the black spectrum shows a simulation of what the ALMA Band 1 spectrum of HCC$^{13}$CN $J=4-3$ would look like at the current best Band 1 spectral resolution at 36.24\,GHz of 260\ms. The hyperfine structure can still be resolved, but the complex $< 200$\ms\/ kinematics are completely lost. With the WSU, the study of such clouds can move from the mostly spectral (i.e., low angular resolution) domain with single dish telescopes to the spatial domain with ALMA. Credit: Brett McGuire (MIT/NRAO).}
\label{fig:ultra}
\end{minipage}
\end{figure}

As an example of the power of ultra-high spectral resolution observations, recent upgrades to the spectrometers on the Green Bank Telescope (GBT) and the Nobeyama 45-m  telescope afford ``ultra-high'' spectral resolution that have yielded insights into the heretofore ``hidden'' complex kinematics present in the very coldest and darkest regions of molecular clouds, which are candidate birth sites for long carbon chain molecules. Figure~\ref{fig:ultra} shows a spectrum from the GBT GOTHAM survey towards TMC-1 \citep{McGuire2021} that reveals complex kinematics that are only visible with better than 20\ms\/ spectral resolution. Only with the sensitivity of ALMA will it become possible to image spectrally these complex kinematics on smaller scales.

Examples of science use cases that require ``ultra-high'' spectral resolution ($< 100$\ms) imaging include:
\begin{itemize} 
\item Measuring subtle deviations from Keplerian rotation, including signatures of Keplerian sheer, vertical disk structure, and sub-Keplerian motions towards the mid-plane of protoplanetary disks \citep{Oberg2021,Teague2020};
\item Probing the kinematics of very dark and cold molecular clouds \citep[see Figure~\ref{fig:ultra};][] {McGuire2021}; 
\item Detecting the infall signatures (observed in absorption) toward the cold molecular envelopes of protostars \citep{Oya2018,Cabedo2021}; 
\item Characterizing the line-of-sight magnetic field strength through observations of the molecular Zeeman effect \citep{Vlemmings2019,Mazzei2020,Harrison2021};
\item Spectrally resolving the motions of atmospheric winds in Solar System objects \citep{Lellouch2019}. 
\end{itemize}

\noindent These use cases involve topics related to the ALMA2030 Roadmap key science goals: Origins of Planets (Section~\ref{sec:planets}) and the Origins of Chemical Complexity (Section~\ref{sec:chemistry}). Most of these use cases also share the property that they are best done at longer ALMA wavelengths (lower frequencies) to avoid the impacts of high continuum opacity and to access lower-lying rotational transitions which can best characterize cold gas. In the case of the Zeeman effect, the size of the magnetic field induced frequency splitting is proportional to 1/frequency, again preferencing the lower ALMA bands. However, as described above and demonstrated in Figure~\ref{fig:ultra}, Bands 1--3 are precisely where high spectral resolution are not currently possible with ALMA. 

\subsection{Improved Spectral Line and Continuum Sensitivity}
\label{sec:sensitivity}

A number of factors will enable the WSU to improve ALMA's overall spectral line and continuum sensitivity to the benefit of all projects. These improvements include:

\begin{itemize}
    \item Increase in the  digital efficiency of the ALMA system. Increasing the number of bits used in each stage of the digital process system will improve the sensitivity by a factor of 1.2 (see Appendix~\ref{sec:digeff}).
    \item Lower receiver noise temperatures. While the noise performance of the ALMA receivers is already excellent, advances in receiver technology will allow the noise temperature of the future Band 3-8 receivers to be further reduced by an estimated 20--30\% (see Appendix~\ref{sec:receivers}).
    \item Upgrading Bands 9 and 10 to sideband separation receivers. Bands 9 and 10 are currently double sideband (DSB) systems. Upgrading to sideband separating (2SB) systems will improve the spectral line sensitivity by 70--80\% by reducing the noise from the opposite sideband (see Appendix~\ref{sec:rxother}).
    \item Increased continuum bandwidth. The most important improvement to continuum science from the ALMA2030 upgrade will be the increased digitized and correlated bandwidth, with sensitivity theoretically improved initially by a factor of 1.46 for $2\times$ correlated bandwidth ([16\,GHz/7.5\,GHz]$^{0.5}$), and eventually 2.06$\times$ after the final 4$\times$ correlated bandwidth goal is reached. 
\end{itemize}

For spectral lines, the {\em minimum} gain in observing speed (defined as the time to reach a fixed sensitivity) is a factor of 1.44 as given by the improvement in the digital efficiency. For continuum observations, the $2\times$ bandwidth upgrade combined with the bandwidth increase will lead to at least a factor of 3 improvement in the continuum imaging speed for the $2\times$ bandwidth upgrade and a factor of 6 improvement for the $4\times$ bandwidth upgrade.

The net improvement in the system sensitivity will depend on the improvement in the receiver temperatures and the atmospheric conditions as well as the actual receiver bandwidth. We use the Band 6 as a representative case based on the Band 6 prototype (see Appendix~\ref{sec:Band6}). Including the improved receiver performance, the new Band 6 receiver will be 2.2$\times$ faster for spectral line observations compared to the most sensitive portion of the current Band 6 receiver, and 4.6$\times$ faster for lines that are near the edge of the current IF. For continuum observations, the upgraded Band 6 receiver will be 4.8$\times$ faster for the $2\times$ bandwidth upgrade and 9.6$\times$ faster for the $4\times$ bandwidth upgrade. Upgrades to the other current 2SB receivers (Bands 3--8) are expected to yield similar improvements.

Bands 9 and 10 will see significant gains in sensitivity when they are upgraded to 2SB systems. As described in Appendix~\ref{sec:rxother}, the system temperatures will improved by 50--70\% over the current DSB performance. Combined with the 20\% improvement in the digital efficiency, Band 9 and 10 spectral line observations will be 3.2--4.2$\times$ faster to reach the same sensitivity as the current receivers.

\subsection{Improved Image Fidelity and Calibration}
\label{sec:imaging}

\begin{figure}[tbh]
\centering
\includegraphics[width=0.49\textwidth]{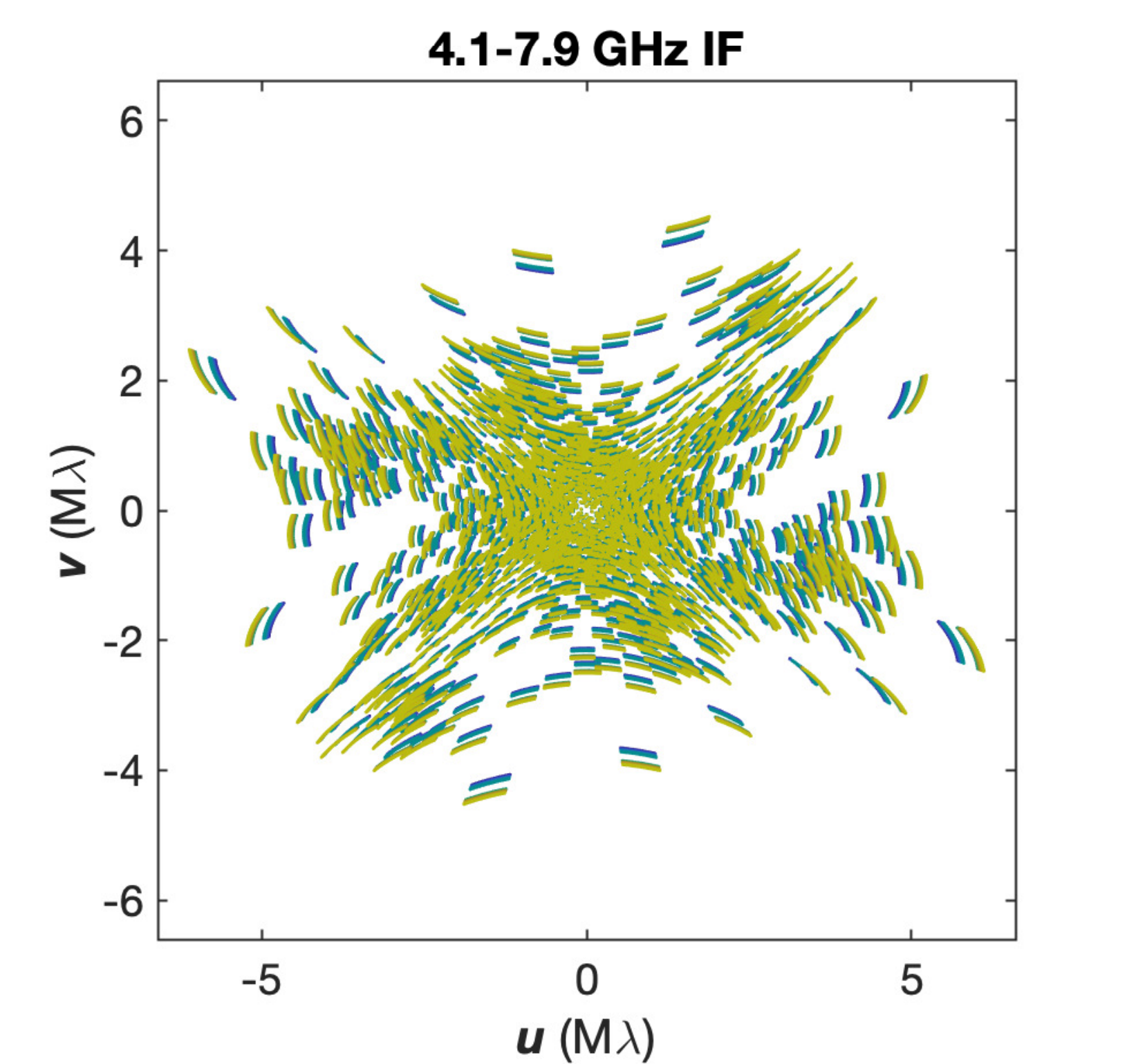}
\includegraphics[width=0.49\textwidth]{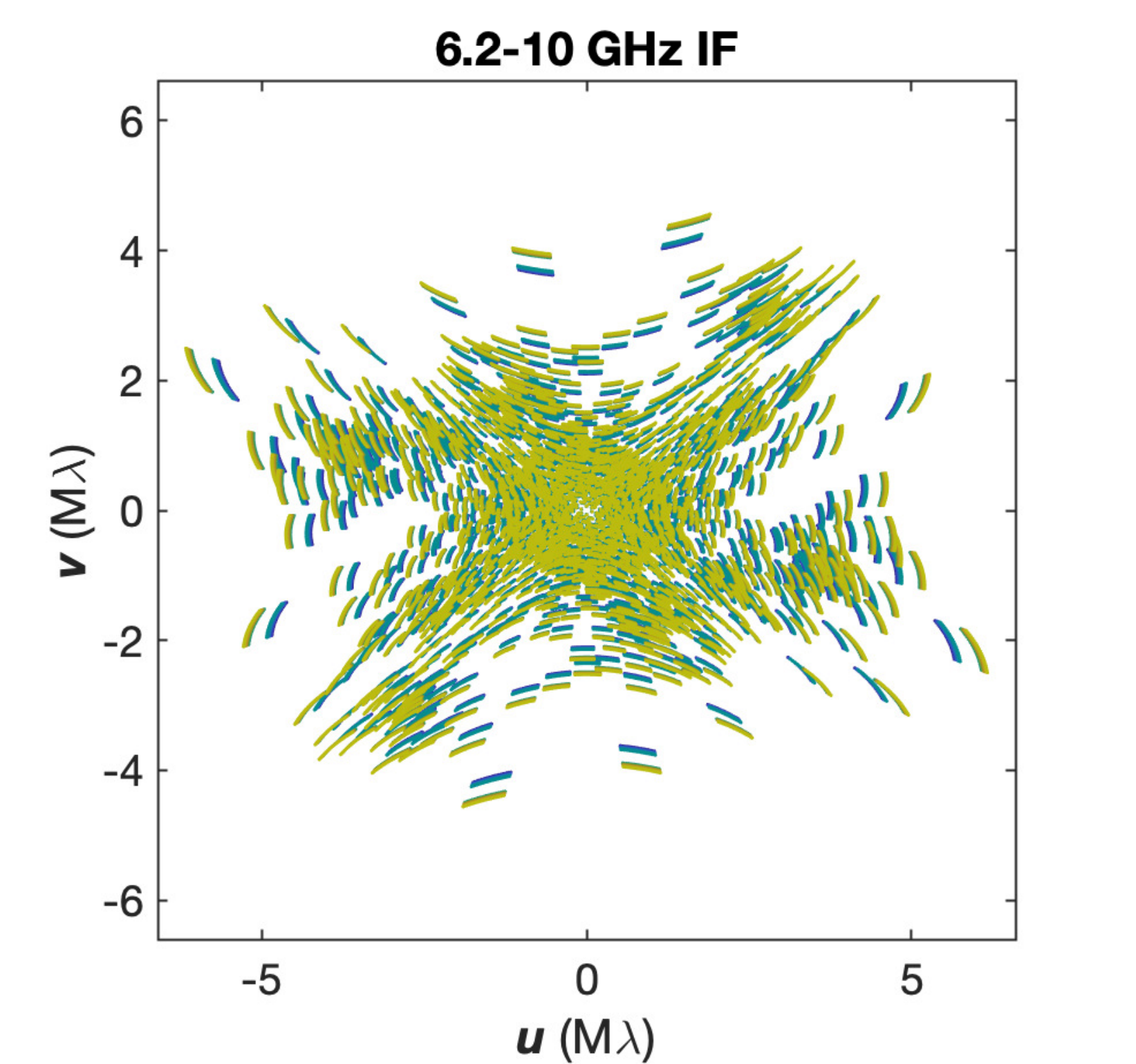}
\includegraphics[width=0.49\textwidth]{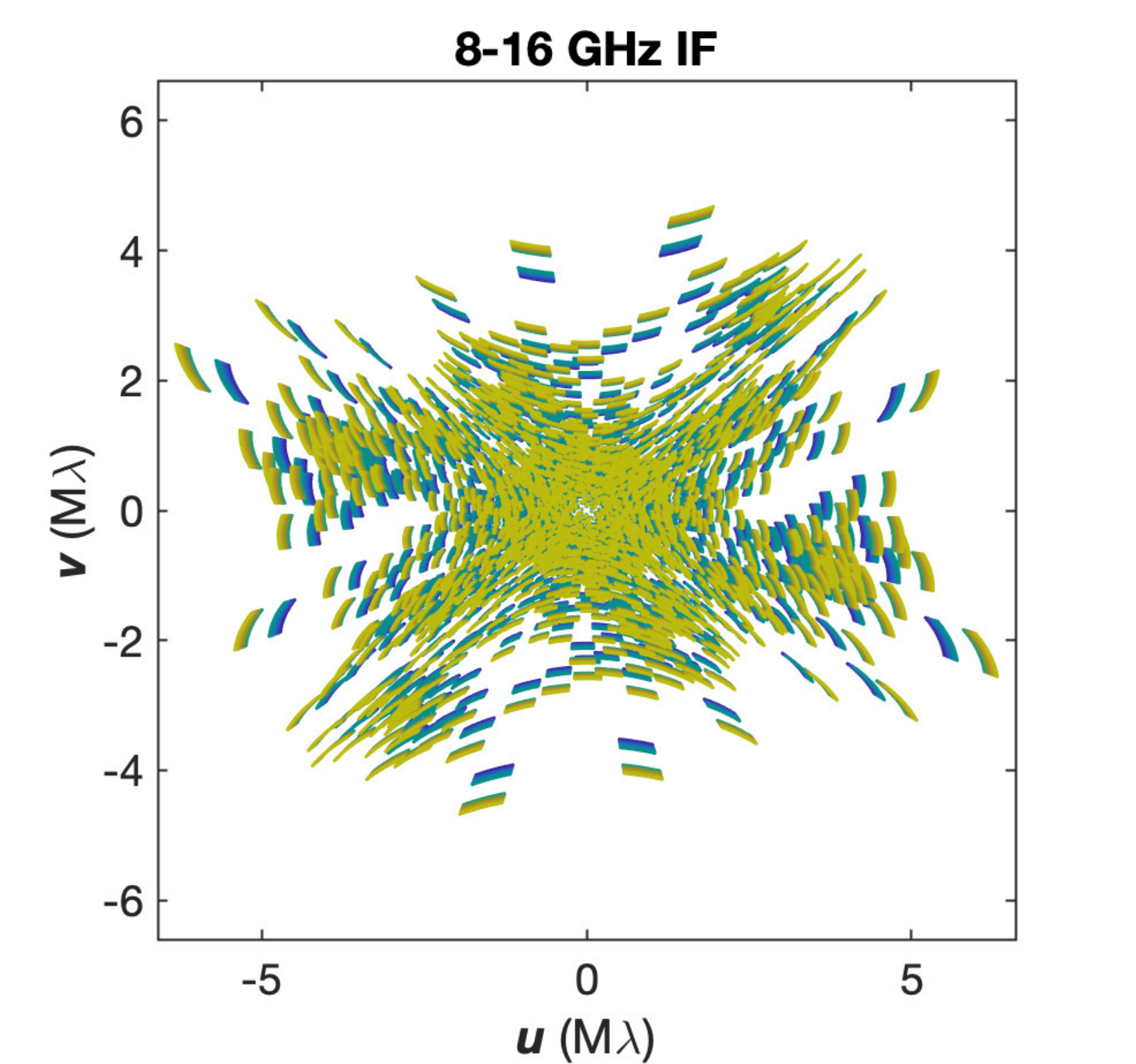}
\includegraphics[width=0.49\textwidth]{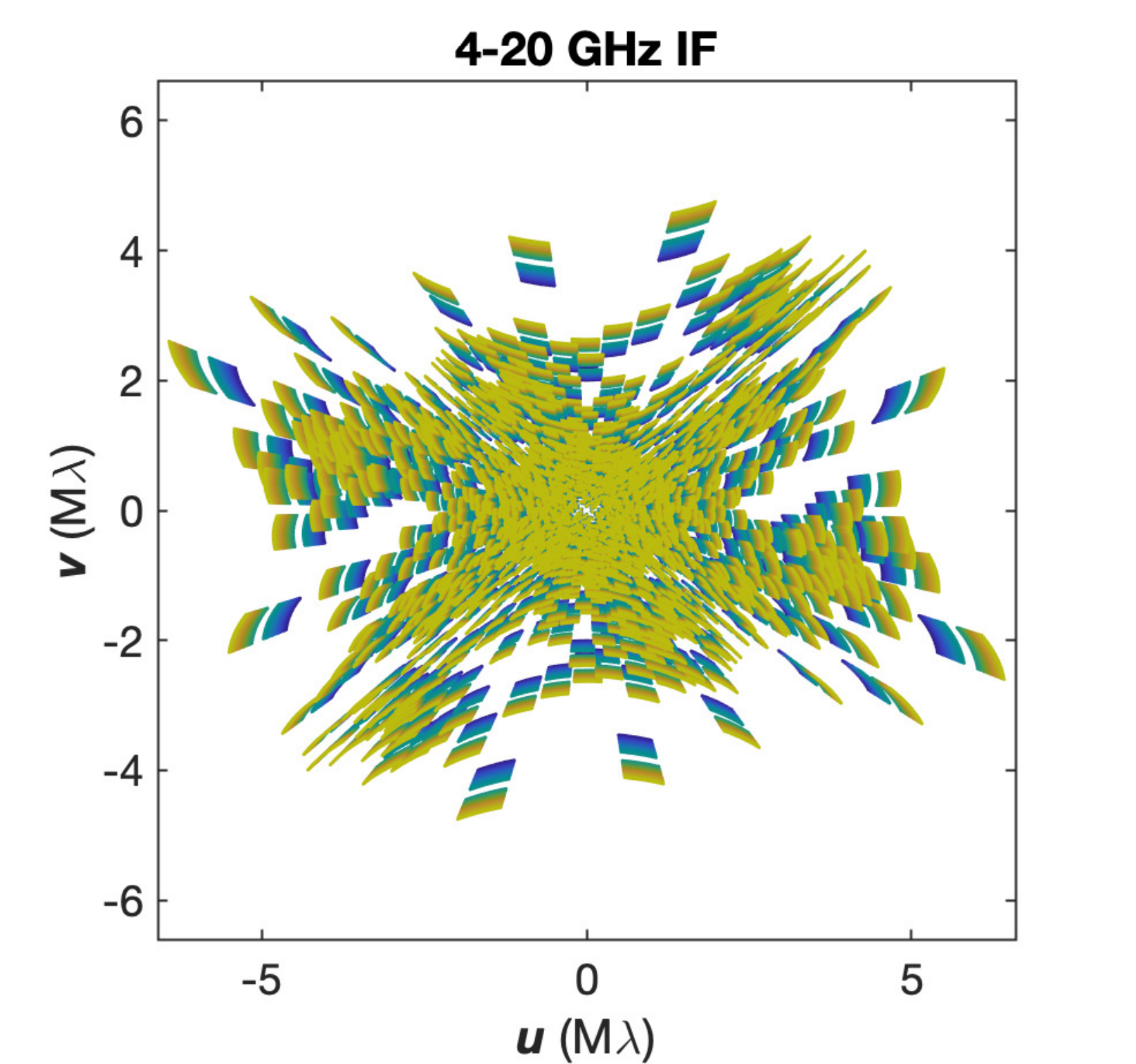}
\caption{$uv$-coverage plots that result from 45-minutes of simulated data in configuration C43-8 for the indicated IF bandwidths and an assumed frequency of 230\,GHz.  Note that while representative of Band 6 observations, the configuration and sky frequency are choices that scale. The bluer and greener points show the LSB and USB, respectively. The top-left panel shows an observational setup possible using the current IF bandwidth available to most current bands.  The top-right shows an observational setup possible using the current IF bandwidth of Band 6 coupled to the current backend system.  The bottom-left panel shows an observational setup possible with the $2\times$ bandwidth WSU, and the bottom-right panel shows an observational setup possible with the $4\times$ bandwidth WSU. The $uv$-coverage is more complete and the LSB/USB gap is fractionally less pronounced in the final ($4\times$) plot. The chosen source declination $\delta = +24^\circ$, corresponding to the Taurus region shown in Fig.~\ref{fig:fidelity}.}
\label{fig:uvcoverage}
\end{figure}

Beyond the raw improvement in continuum sensitivity due to wider bandwidth, continuum images will also have improved image fidelity due to the denser $uv$-coverage afforded by both the expanded frequency coverage (fractional bandwidth) and the wider correlated bandwidth. This is because modern continuum imaging is spectral in nature and leverages ``multi-frequency synthesis" algorithms to fill in the $uv$-plane. This effect is demonstrated in Figure~\ref{fig:uvcoverage}  for ALMA configuration C43-8 and 45 minutes of observing time. The top two panels show IF bands possible with the current system with a bandwidth of 3.75\,GHz per sideband, with the latter specific to the current Band 6 showing a sideband separation by 12\,GHz. The bottom panel in the figure shows the $uv$-coverage for the interim goal of 16\,GHz total bandwidth per polarization for the 2$\times$ WSU for an IF band of 8--16\,GHz. The final panel shows the dramatic improvement in $uv$-coverage when a total of 32\,GHz per polarization is achieved, using an IF band of 4--20\,GHz, which is the stretch goal of Band 6v2; see Table~\ref{tbl:FEstatus} in the Appendix).

\begin{figure}[h]
\centering
\includegraphics[width=0.9\textwidth]{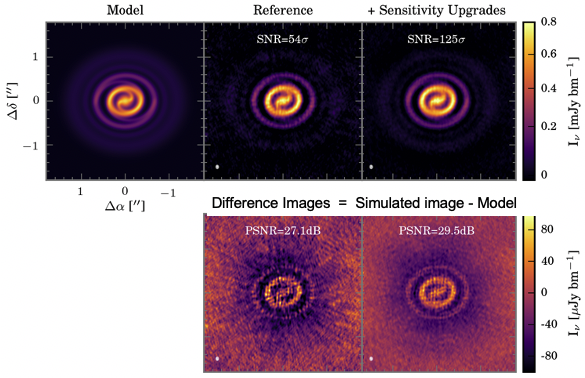}
\caption{Model and simulations of the 1.3\,mm continuum emission from a protoplanetary disk similar in morphology to IM Lup. The “Reference” simulated image panel shows the image quality that could be obtained today (assuming perfectly calibrated data) with the current Band 6 receiver and 3.75 GHz of bandwidth in each sideband, separated by 12\,GHz (i.e., IF=6.2--10 GHz, which is the “best” IF range of the current Band 6 receiver; see the corresponding $uv$-coverage in the top-left panel of Figure~\ref{fig:uvcoverage}). The final panel, “+Sensitivity Upgrades”, shows the image quality afforded by the WSU with more widely separated sidebands, 16\,GHz of bandwidth per polarization, the factor of 1.25$\times$ better sensitivity from improved Band 6v2 receiver temperatures, and the digital sensitivity improvement factor of 1.2 (see the corresponding $uv$-coverage in the bottom-right panel of Figure~\ref{fig:uvcoverage}). Credit: Ryan Loomis (NRAO).}
\label{fig:fidelity}
\end{figure}

To further demonstrate the improved image fidelity, Figure~\ref{fig:fidelity} shows simulations for a protoplanetary disk model at 1.3\,mm (Band 6) that incorporates a bright axis-symmetric ring-gap structure superposed with a fainter spiral structure. This model is analogous to the observed morphology recently discovered for the protoplanetary disk IM~Lup (Huang et al.\ 2018).
Like Figure~\ref{fig:uvcoverage}, these simulations used 45 minutes of on-source observing time in configuration C43-8, resulting in an angular resolution of $\sim50$\,mas. The top row of Figure~\ref{fig:fidelity} shows the model image, a ``Reference'' image simulated using current Band 6 ALMA capabilities,\footnote{We note the IF bandwidth for Band 6, as for most ALMA bands, is limited to a total of 7.5\,GHz by the current backend electronics. The distinction here is that Band 6 allows a choice of e.g.\ 6.2--10\,GHz for the IF band, which as seen in Figure~\ref{fig:uvcoverage}  can yield a modest improvement in $uv$-coverage.} and finally a simulation of the ``+Sensitivity Upgrades'' image that will be possible after the 2$\times$ bandwidth WSU, using Band 6v2 as an example.  These simulations include thermal random noise and otherwise assume perfect calibration. The bottom panels show a measure of image fidelity called the Peak Signal-to-Noise Ratio (PSNR{\footnote{\url{https://en.wikipedia.org/wiki/Peak\_signal-to-noise\_ratio}}}) for the residual image (i.e., image minus model). A dramatic improvement in the artifacts visible in the residual image is apparent. Indeed, the WSU 
improves the signal-to-noise by a factor of 2.3$\times$ (better than predicted by the sensitivity improvements alone 2.2$\times$) due to the image fidelity improvements.

The wider IF bandwidth from the WSU will also enable more accurate spectral index determination from single tuning observations.  Monte Carlo simulations of fitting the spectral index ($\alpha$) of a target with a signal-to-noise ratio of 50 for a 2\,GHz spectral window and $\alpha=0$, 1, 2, 3, 4 demonstrate that the uncertainty of $\alpha$ is reduced by a factor of 1.75 merely from the larger spread in frequency (i.e., from $\pm$10\,GHz to $\pm$16\,GHz, appropriate for Band 6 and the upcoming Band 2).  The improvement grows to factor of 2.15 when the gain in SNR from the 2$\times$ correlated bandwidth is also included. The results are similar in both bands, regardless of the value of $\alpha$.

In addition to the improvements described above for continuum science observations, the WSU will also expand the range of possible calibration options and facilitate new calibration strategies. The general increase in the aggregate continuum bandwidth of ALMA datasets will raise the signal-to-noise ratio of spectrally-averaged calibrator gain solutions for the same calibrator scan duration.  For example, in some cases, fainter quasars significantly closer to the science target will become available to use as phase calibrators, leading to more accurate phase transfer, particularly in the high frequency bands where the residual antenna position errors represent larger fractions of the observing wavelength.  For other cases where a nearby calibrator is already available, the higher signal-to-noise ratio means that the pipeline calibration will more often be able to calibrate narrow spectral windows individually, providing resilience against possible instrumental drifts between basebands and spectral windows. Alternatively, a shorter integration time on the phase calibrator could be chosen, enabling faster calibration cycle times and a subsequent reduction in the residual phase error, particularly on long baselines and at high frequencies. The larger bandwidth also means that continuum self-calibration will be possible on a greater fraction of science targets.  It is notable that even if the signal-to-noise ratio of the science target is so low that it allows only a single solution per execution block to be applied, this correction alone will remove the bulk of residual antenna position errors, which can be significant for longer baselines.  

%% file: table_improvements.tex
\begin{table}[h]
\centering
\caption{Improvements provided by the Wideband Sensitivity Upgrade}
\begin{NiceTabular}[width=\textwidth]{X[l]X[2,l]}[hvlines]
   \RowStyle[rowcolor=\headercolor]{\bfseries}
   Capability &
   Improvement\\
Instantaneous Bandwidth & 
\vspace*{0.1cm}\begin{minipage}{11.2cm}
\begin{itemize}[leftmargin=0.3cm]
\vspace*{-0.3cm}
\item Factor of 2 to 4 increase in the {\em available} instantaneous bandwidth (16 to 32\,GHz per polarization) compared to existing receivers.\vspace{0.2cm}
\end{itemize}
\end{minipage}\\
Correlated bandwidth & 
\vspace*{0.1cm}\begin{minipage}{11.2cm}
\begin{itemize}[leftmargin=0.3cm]
\vspace*{-0.3cm}
\item Factor of 4 to 68 increase in the {\em correlated} bandwidth at high spectral resolution (0.1--0.2\kms), with larger gains in the lower frequency bands.
\item Observers will no longer need to trade off high spectral resolution for bandwidth.\vspace{0.2cm}
\end{itemize}
\end{minipage}\\
Spectral scan speed   &
\vspace*{0.1cm}\begin{minipage}{11.2cm}
\begin{itemize}[leftmargin=0.3cm]
\vspace*{-0.3cm}
\item Increase of at least a factor of 2, and up to a factor of 4 (Band 10) to 64 (Band 1) for a spectral resolution of 0.1--0.2\kms.\vspace{0.2cm}
\end{itemize}
\end{minipage}\\
Spectral line imaging speed         & 
\vspace*{0.1cm}\begin{minipage}{11.2cm}
\begin{itemize}[leftmargin=0.3cm]
\vspace*{-0.3cm}
\item Increased spectral line speed from lower receiver noise temperatures ($\sim20$\%; up to $\sim50\%$ at edge of RF band in some receivers), improved digital efficiency ($\sim20$\%), and upgrade to 2SB mixers (Bands 9 and 10 only).
\item Net gain in spectral line imaging speed of $\sim2\text{--}3$.\vspace{0.2cm}
\end{itemize}
\end{minipage}\\
Continuum imaging speed         & 
\vspace*{0.1cm}\begin{minipage}{11.2cm}
\begin{itemize}[leftmargin=0.3cm]
\vspace*{-0.3cm}
\item Increase by at least a factor of 3 with 2$\times$ bandwidth increase and at least 6 for 4$\times$ bandwidth, including digital efficiency improvements.
\item Additional gains from improved receiver temperatures.\vspace{0.2cm}
\end{itemize}
\end{minipage}\\
Ultra-high spectral resolution  &
\vspace*{0.1cm}\begin{minipage}{11.2cm}
\begin{itemize}[leftmargin=0.3cm]
\vspace*{-0.3cm}
\item Provide for the first time unique access to ultra-high spectral resolution observations --- better than 0.01\kms\ at all ALMA frequencies.\vspace{0.2cm}
\end{itemize}
\end{minipage}
\label{tbl:improvement}
\end{NiceTabular}
\end{table}

%% file: table_bwincrease.tex
\begin{table}[ht]
\centering
\caption{Factor increase in the correlated bandwidth at high spectral resolution enabled by the WSU}
\begin{NiceTabular}{lYYYYYYYYYY}[code-before=\rowcolor{\highlightcolor}{7-7}][tabularnote=Notes: Comparison of the maximum correlated bandwidth that can be obtained at a spectral resolution between 0.1 and 0.2\kms\/ in dual polarization for the current ALMA correlator (BLC) and the WSU correlator (AT.CSP) for the $2\times$ bandwidth upgrade. 
The level of channel averaging in the WSU that will produce a similar velocity resolution as the BLC for these cases is also indicated. It is notable that a velocity resolution in the 0.1--0.2\kms\/ range cannot presently be achieved with the BLC at full correlated bandwidth for any ALMA band{,} and that a spectral resolution $<0.2$\kms\/ cannot be reached in Band 1 even for the narrowest correlated bandwidth of the BLC.]
\hline
\hline
   \RowStyle[rowcolor=\headercolor]{\bfseries}
   Band & 1 & 2 & 3 & 4 & 5 & 6 & 7 & 8 & 9 & 10\\
   \RowStyle[rowcolor=\headercolor]{\bfseries}
   Reference frequency (GHz) & 35 & 75 & 100 & 150 & 185 & 230 & 345 & 460 & 650 & 870\\
\hline
Velocity resolution (\kmss) & 0.26 & 0.12 & 0.18 & 0.12 & 0.10 & 0.16 & 0.11 & 0.16 & 0.11 & 0.17\\
BLC correlated bandwidth (GHz) & 0.234 & 0.234 & 0.469 & 0.469 & 0.469 & 0.938 & 0.938 & 1.875 & 1.875 & 3.750\\
AT.CSP correlated bandwidth (GHz) & 16 & 16 & 16 & 16 & 16 & 16 & 16 & 16 & 16 & 16\\
Channel averaging & 2 & 2 & 4 & 5 & 5 & 8 & 10 & 14 & 18 & 32\\
\bf Factor increase in correlated bandwidth & \bf 68.3 & \bf 68.3 & \bf 34.1 & \bf 34.1 & \bf 34.1 & \bf 17.1 & \bf 17.1 & \bf 8.5 & \bf 8.5 & \bf 4.3\\
\hline
\end{NiceTabular}
\label{tbl:bwincrease}
\end{table}


%% file: table_cycle8spectral.tex
\begin{table}[h]
\centering
\caption{Accepted Cycle 8 projects that use at least one narrow-band\tabularnote{A narrow-band spectral window is defined as a window with a bandwidth of $<$1\,GHz, which is less than the full available bandwidth of 1.875\,GHz.} spectral window}

\begin{NiceTabular}{lYYYY}
\hline
\hline
   \RowStyle[rowcolor=\headercolor]{\bfseries}
   Science category &
   Category number &
   Total number of projects &
   Number with narrow spectral windows &
   Percent with narrow spectral windows\\
\hline
Cosmology and the high redshift universe & 1 & 140 & 5 & 3.6\%\\ 
Galaxies and galactic nuclei  & 2 & 101 & 13 & 12.9\%\\
ISM, star formation, and astrochemistry  & 3 & 116 & 86 & 74.1\%\\
Circumstellar disks, exoplanets, and the solar system  & 4 & 77 & 58 &  75.3\%\\
Stellar evolution and the Sun   & 5 & 23 & 6 &  26.1\%\\
\hline
Total &  & 457 & 168 & 36.8\%\\
\hline
\end{NiceTabular}
\label{tbl:cycle8}
\end{table}

%% file: table_spectral_scan_speed.tex
\begin{table}[ht]
\centering
\footnotesize
\caption{Increase in the spectral scan speed with the $2\times$ bandwidth WSU}
\begin{NiceTabular}{ccc|YYYY|YYYYY}[code-before=\rectanglecolor{\highlightcolor}{2-12}{15-12} \rectanglecolor{\highlightcolor}{2-7}{15-7} \rectanglecolor{\headercolor}{1-1}{5-3} \rectanglecolor[HTML]{\pastelorange}{1-4}{1-7} \rectanglecolor[HTML]{\pastelpurple}{1-8}{1-12} \rectanglecolor[HTML]{\pastelgreen}{2-6}{5-6} \rectanglecolor[HTML]{\pastelgreen}{2-11}{5-11} \rectanglecolor[HTML]{\pastelgray}{2-4}{5-5}
\rectanglecolor[HTML]{\pastelgray}{2-8}{5-10}][tabularnote=\small Notes: Comparison of the number of spectral tunings currently required to span each band's RF range{,} for the ``low'' and ``high'' spectral resolution regimes. The exact spectral resolution is defined by what the BLC can achieve. The time savings only considers the efficiency of spectral scan tunings{,} and does not include the expected improvements in the digitizer and correlator efficiencies or receiver performance.]
\hline
\hline
  \RowStyle{\bfseries}
  \Block{1-3}{Band Properties} & & & \Block{1-4}{Low Spectral Resolution} & & & &  \Block{1-5}{High Spectral Resolution (0.1--0.2\kms)}\\ \cline{1-3} \cline{4-7}  \cline{8-12}
  \RowStyle{\bfseries}
   & & & \Block{1-2}{Current} & & WSU & \Block{3-1}{Time Savings Factor} & \Block{1-3}{Current}& &  & WSU & \Block{3-1}{Time Savings Factor}\\ \cline{4-5} \cline{8-10}
   \RowStyle{\bfseries}
   Band & Rep. & RF & Velocity & Number & Number &  & Velocity & Max & Number & Number & \\
   \RowStyle{\bfseries}
   & Freq. & BW & Res. & Tunings & Tunings & & Res. & CBW & Tunings & Tunings & \\ 
   \RowStyle{\bfseries}
   & (GHz) & (GHz) & (\kmss) & & & & (\kmss) & (GHz) & & & \\
\hline
1 & 35 & 15 & 8.36 & 2 & 1 & 2.0 & 0.26 & 0.234 & 64 & 1 & 64.0\\
2 & 75 & 49 & 3.90 & 8 & 4 & 2.0 & 0.12 & 0.234 & 209 & 4 & 52.3\\
3 & 100 & 32 & 2.93 & 5 & 2 & 2.5 & 0.19 & 0.469 & 68 & 2 & 34.0\\
4 & 150 & 38 & 1.95 & 7 & 3 & 2.3 & 0.12 & 0.468 & 81 & 3 & 27.0\\
5 & 185 & 48 & 1.58 & 8 & 4 & 2.0 & 0.10 & 0.468 & 103 & 4 & 25.8\\
6v2 & 230 & 64 & 1.27 & 9 & 4 & 2.1 & 0.16 & 0.938 & 68 & 4 & 17.0\\
7 & 345 & 98 & 0.85 & 17 & 7 & 2.4 & 0.11 & 0.938 & 105 & 7 & 15.0\\
8 & 460 & 115 & 0.64 & 18 & 9 & 2.0 & 0.16 & 1.875 & 61 & 9 & 6.8\\
9 & 650 & 118 & 0.45 & 10 & 8 & 1.3 & 0.11 & 1.875 & 63 & 8 & 7.9\\
10 & 870 & 163 & 0.34 & 13 & 10 & 1.3 & 0.17 & 3.75 & 43 & 10 & 4.3\\
\hline
\end{NiceTabular}
\label{tbl:SSTune}
\end{table}

%% file: origins_planets.tex
\FloatBarrier
\section{Origins of Planets}
\label{sec:planets}

\begin{figure}[h]
\centering
\includegraphics[width=\textwidth]{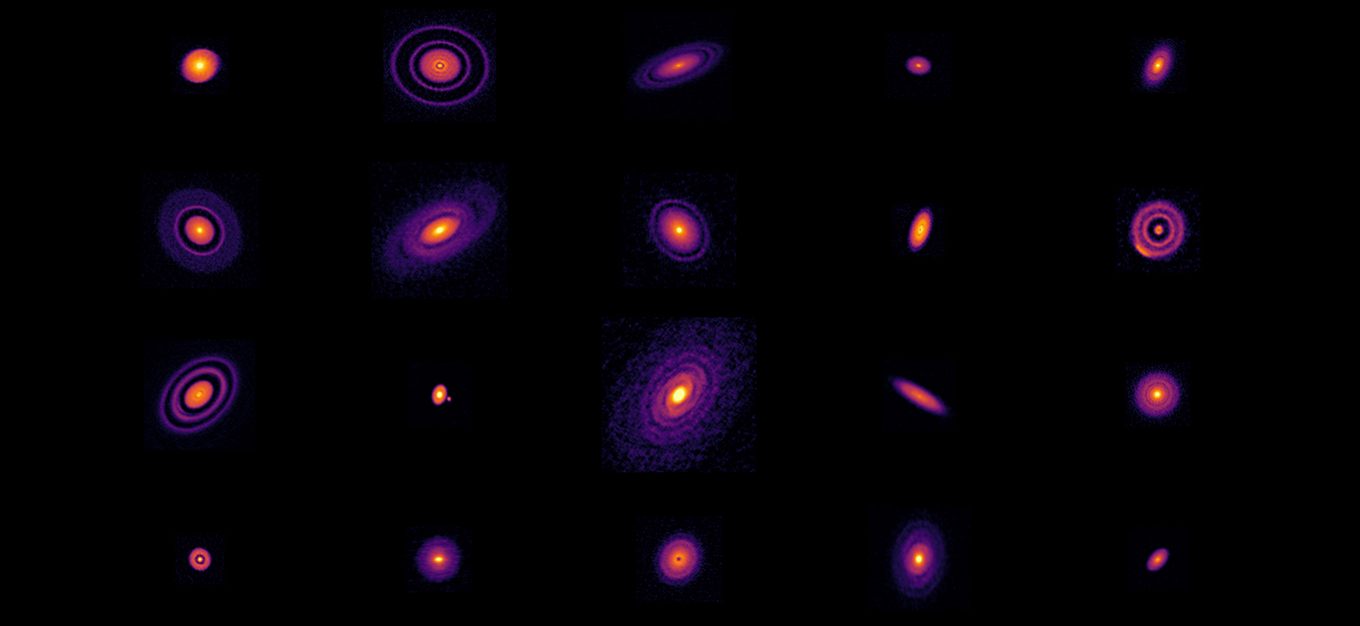}
\caption{ALMA 1.3\,mm dust continuum images of protoplanetary disks from the Disk Substructures at High Angular Resolution Project \citep[DSHARP;][]{Andrews18}. Credit: ALMA (ESO/NAOJ/NRAO), S. Andrews et al.; N. Lira. These high resolution images reveal the prevalence of rings, gaps, and spirals in protoplanetary disks. 
}
\label{fig:dsharp}
\end{figure}

The rotating disks of gas and dust that surround most young stars ($\sim$1\,Myr) are the birth sites of planets. The properties of these disks --- mass, size, and composition --- are  among the primary factors that will determine the properties of newly formed planetary systems. ALMA has already revolutionized our understanding of planet formation. The now iconic dust continuum image of HL Tau \citep{HLTau15} revealed a stunning network of gaps and rings that suggested planet formation is well underway at stellar ages of $\sim$1\,Myr. Further studies have shown that these features are ubiquitous in protoplanetary disks (Figure~\ref{fig:dsharp}). While the exact origin of these dust substructures is still debated, it is clear they must be a critical aspect of the planet formation process.

\begin{figure}
\centering
\includegraphics[width=\textwidth]{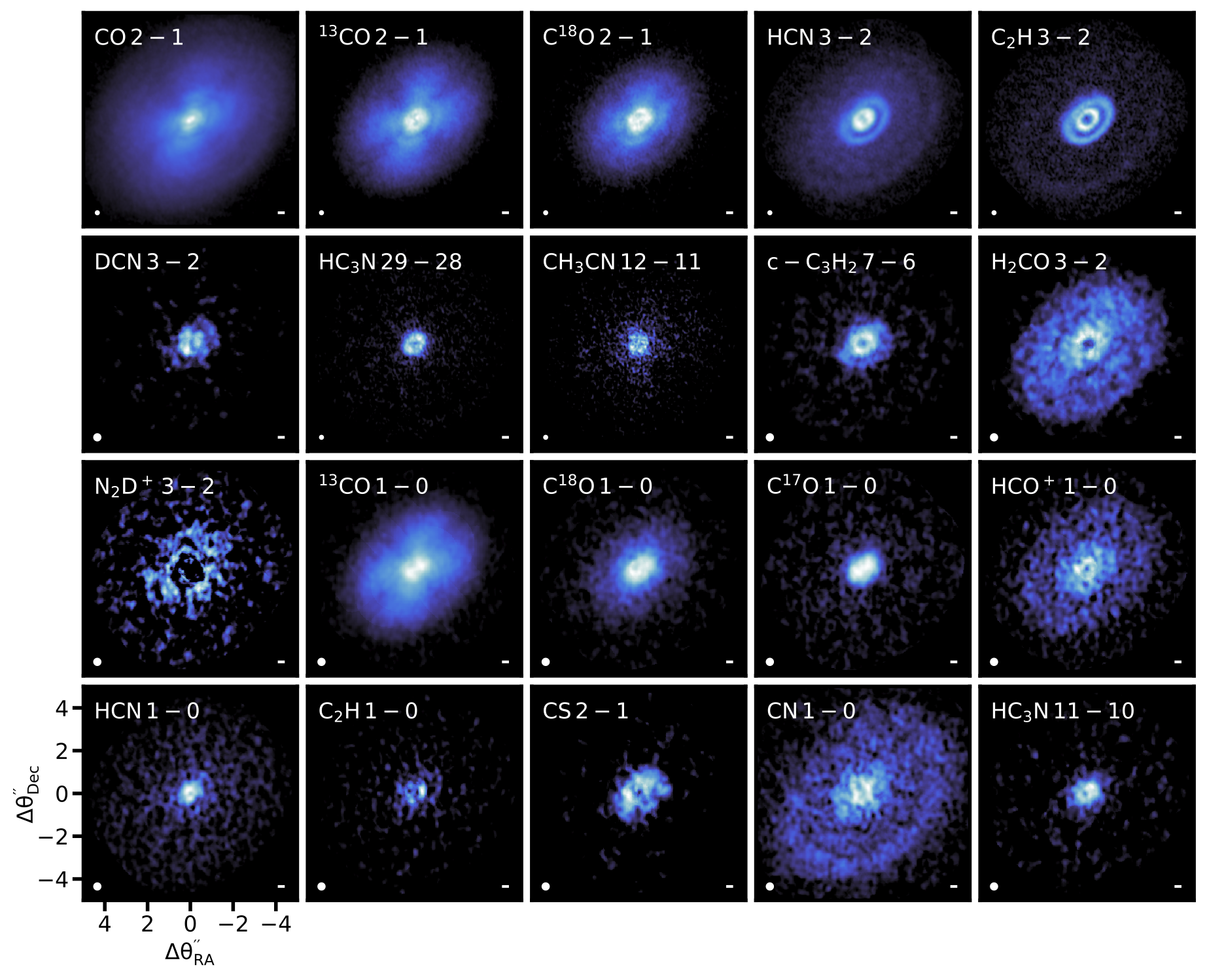}
\caption{Image of the disk surrounding HD~163296 in 20 different molecular transitions obtained by the MAPS Large Program \citep{Oberg21b}. Reproduced by permission of the AAS.
}
\label{fig:maps_hd163296}
\end{figure}

A complete understanding of the planet formation process requires understanding the gas as well as the dust. The gas contains the majority of the disk mass ($\sim99$\%, primarily in molecular hydrogen), sets the composition of planetary atmospheres, influences the dynamics of planets, and impacts the planet formation timescales. The gas also provides unique information on the kinematics within the disk, including large scale flows, localized velocity distortions from embedded planets, turbulence, and disk winds. Since molecular hydrogen cannot be detected in the cold regions of disks, observations of rare molecules are used to trace the kinematics, chemistry, and physical conditions. As shown in Figure~\ref{fig:maps_hd163296}, the appearance of a disk depends dramatically on the observed molecule and transition. The differences reflect variations in abundances, optical depth, and excitation, which can be used to decipher the physical, chemical, and dynamical properties of the disk.

The WSU will vastly increase ALMA's capabilities for spectral line and dust continuum observations of circumstellar disks. The Molecules with ALMA at Planet-forming Scales (MAPS) Large Program illustrates the enormous potential of the WSU. MAPS imaged five protoplanetary disks in a variety of molecular transitions in Band 3 and 6, and is the most comprehensive line survey of disks completed to date \citep{Oberg21b}. For the Band 6 observations, 13 transitions were observed using two receiver tunings at a velocity resolution of $\sim0.1\text{--}0.2$\kms. At 0.1\kms\ resolution, the maximum correlated bandwidth in Band 6 with the current correlator is 0.47\,GHz per polarization, which can be split between 4 to 8 spectral windows. The minimum requirements for the WSU correlator will process 80 independent spectral windows for a total of 16\,GHz per polarization, with a goal to process 32\,GHz per polarization. Thus the number of spectral lines that can be observed at high-spectral resolution will increase by more than an order of magnitude. In addition, since users will no longer need to sacrifice spectral resolution for bandwidth, the WSU will simultaneously provide 16--32\,GHz per polarization of continuum bandwidth. Combined with the digital improvements, continuum observations will be 3--6$\times$ faster to reach the same sensitivity as the current system, plus any additional gains from improved receiver sensitivity. The result will be a continuum and spectroscopic machine that will enable comprehensive studies of the chemical composition, structure, and kinematics of disks. 

\begin{figure}
\centering
\includegraphics[width=\textwidth]{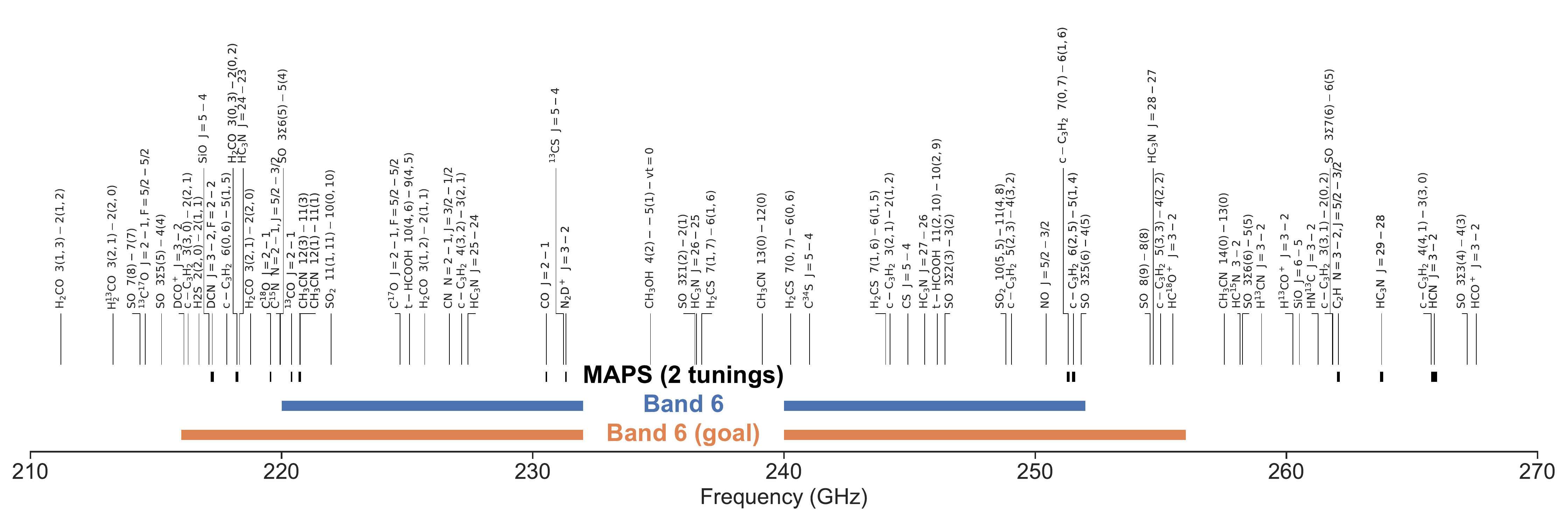}
\caption{Example observation in Band 6 for the current correlator and the Wideband Sensitivity Upgrade. Also indicated are a select list of available molecular transitions. The black bars indicate the spectral coverage for the two high-resolution (0.1--0.2\kms) spectral settings used by the MAPS Large Program \citep{Oberg21b}. The IF coverage for the Band 6 upgrade is shown in blue (12\,GHz per sideband as the minimum requirements) and orange (16\,GHz per sideband). With the WSU, at least 16\,GHz per polarization with 80 spectral windows can be correlated at high spectral resolution with a goal of 32\,GHz. The WSU provides up to a factor of 64 increase in the amount of bandwidth that can be correlated at high-spectral resolution in Band 6, vastly increasing the spectral throughput.
}
\label{fig:maps_corr}
\end{figure}

\begin{figure}
\centering
\includegraphics[width=\textwidth]{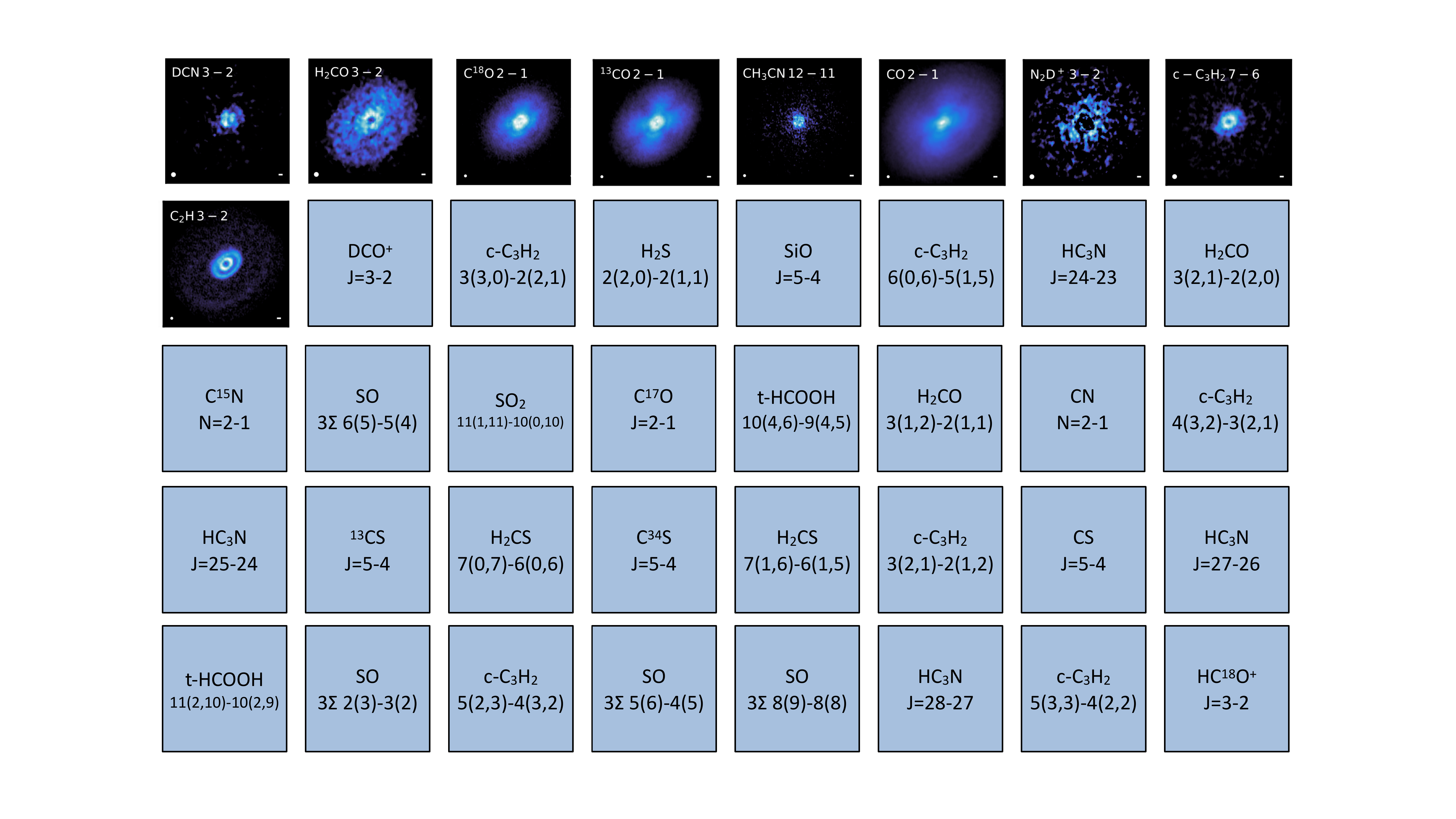}
\caption{Illustration of the throughput power of the Wideband Sensitivity Upgrade that can be obtained with a receiver and correlator that can process 16\,GHz of bandwidth per polarization (see Figure~\ref{fig:maps_corr}). Nine images of HD~163296 obtained with the MAPS Large Program \citep{Oberg21b} are shown. The remaining squares show the additional lines that can be observed {\it simultaneously} with the upgrade. Moreover, an additional 120 spectral windows would be available to target other lines.
}
\label{fig:maps_lines}
\end{figure}

\subsection{Physical and Chemical Structure of Disks}

The properties of disks have a direct impact on the planet formation process. The most obvious impact is that the mass and size of a disk will determine where in the disk planets form and their masses \citep[e.g.,][]{Benz14}. Moreover, gradients in the disk temperature and density create a rich, time-dependent radial and vertical chemical structure that will influence the composition of planets \citep{Oberg21a}. Chemical processes may also have significant impact on planet formation itself. As molecules freeze out on dust grains at the snowlines of various molecular species, the surface density of solids can increase significantly, which can facilitate the growth of planetesimals and planets \citep{Armitage16}.

\begin{figure}
\centering
\includegraphics[width=\textwidth, trim=0in 3in 0in 3in, clip]{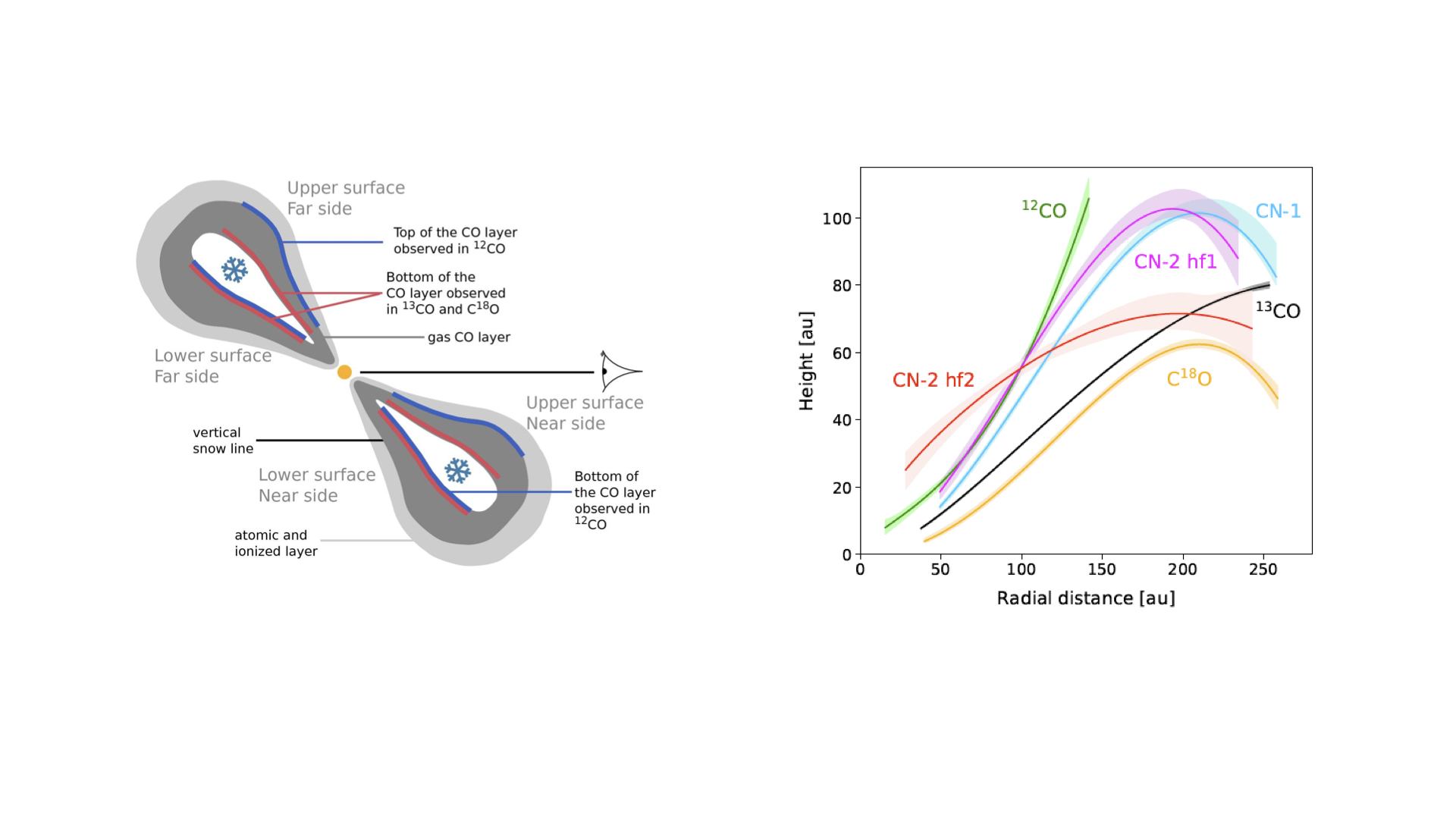}
\caption{ALMA observations to measure the vertical temperature and chemical structure of disks.
Left panel: Schematic of how  molecular lines can trace various layers within a disk \citep{Pinte18a}.
Right panel: Location of the emission surfaces in $^{12}$CO, $^{13}$CO, C$^{18}$O, and CN hyperfine components in the disk Elias 2-27 \citep{Paneque22}. With the WSU, a single correlator setting can be used to probe many lines simultaneously to obtain a comprehensive census of the vertical and chemical structure of disks.
}
\label{fig:pinte_surface}
\end{figure}

ALMA has led to a resurgence in observing molecular lines to study the three dimensional structure of disks. The sensitive, high spatial and spectral resolution observations allow ALMA to detect directly the emission surface of various molecular species. Figure~\ref{fig:pinte_surface} illustrates the technique. As illustrated in the left panel of this figure, optically thick lines such as CO can be used to probe a surface layer within the disk. Moreover, because high angular resolution observations with ALMA can resolve the scale height of the disk, the surface emission layer from both the back and front side of the disk can be detected spatially and kinematically. The emission surfaces of the disk can then be mapped as a function of radius within the disk for various molecules as illustrated in the right hand side of the figure. The emission surfaces can be used to trace the disk temperature at various layers for optically thick lines, or the chemical abundances of species for optically thin lines.

\begin{figure}
\centering
\includegraphics[width=\textwidth]{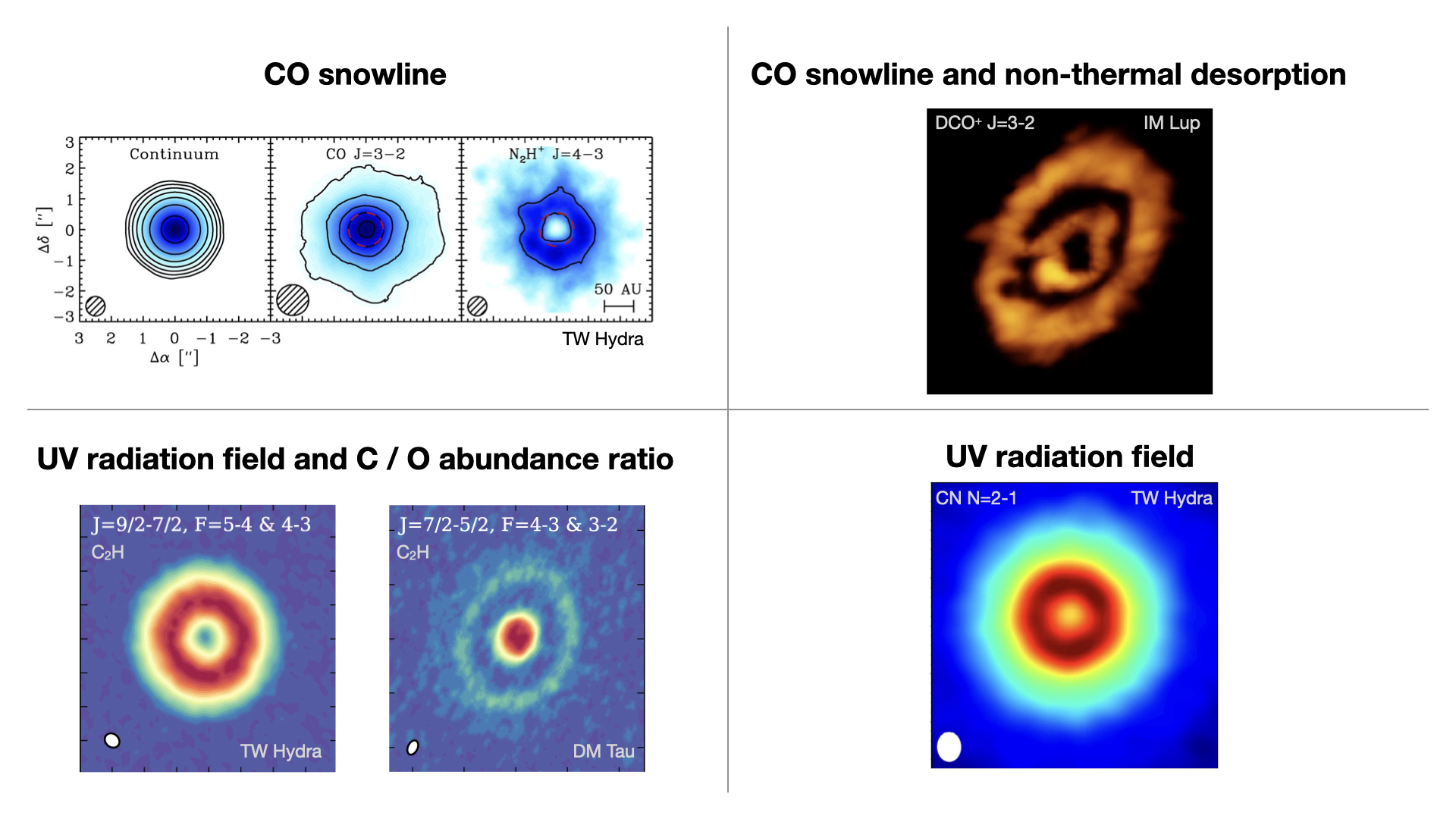}
\caption{Illustration on how ALMA spectral lines images are used as diagnostics in protoplanetary disks. 
Upper left: Continuum, $\mathrm{^{12}CO}$, and $\mathrm{N_2H^+}$ images of TW Hydra; adapted from \citet{Qi13}.
Upper right: Double-ring of $\mathrm{DCO^+}$ emission in IM~Lup \citep{Oberg15}. Credit: K. 
\:Oberg, CfA; ALMA (NRAO/ESO/NAOJ); B. Saxton (NRAO/AUI/NSF).
Lower left: Observations of rings of the hydrocarbon $\mathrm{C_2H}$ in TW~Hydra and DM~Tau \citep[reproduced by permission of the AAS]{Bergin16}. 
Lower right: Ring of CN emission in TW~Hydra \citep{Cazzoletti18}.
}
\label{fig:molecules_diagnostics}
\end{figure}

ALMA molecular line observations can also trace chemical and physical processes within disks (see Figure~\ref{fig:molecules_diagnostics}). For example, \citet{Qi13} found that the $\mathrm{N_2H^+}$ emission in TW Hydra is distributed in a ring, where the inner radius of the ring is coincident with the expected freeze-out location of CO. Such dichotomy in the spatial distributions is consistent with theoretical models where  $\mathrm{N_2H^+}$ is destroyed in the gas phase by CO. Observations of DCO$^+$ in IM~Lup revealed a double-ring structure, where the inner ring can be explained by the freeze-out of CO and the outer ring by the non-thermal desorption of CO in the outer regions of the disk \citep{Oberg15}. In TW~Hydra and DM~Tau, the hydrocarbon $\mathrm{C_2H}$ shows a ring structure, which can be explained by a combination of a strong UV radiation field and a high carbon-to-oxygen abundance ratio \citep{Bergin16}. A ring of CN emission was also found in TW~Hydra, and was found to be a sensitive tracer of the UV radiation field \citep{Cazzoletti18}.

Increasing the number of density-tracing spectral lines will improve disk mass estimates and scrutinize long-held assumptions such as the gas to dust ratio = 100 and the molecular abundance ratios with empirical data. The most common tracer of the disk mass is the dust continuum, which relies on uncertain assumptions for the dust mass opacity and the gas-to-dust ratio. Measurements of the disk mass through spectral lines such as $^{12}$CO alone give inaccurate results due to optical depth and chemical evolution in disks \citep{McClure16}. Simultaneous observations of multiple CO isotopologues, small organic molecules, deuterated species, and other molecules will be used to understand the complex time-dependent chemistry in disks, measure radial variations in key abundance ratios (e.g., C/O, D/H), and constrain the gas masses in disks \citep{Oberg21a}.

The WSU will greatly advance observational studies of the physical and chemical structure of disks. A single correlator setting will probe many tens of molecular transitions simultaneously (see Figures~\ref{fig:maps_corr} and \ref{fig:maps_lines}). In addition to the gains in spectral throughput, the improved system from the WSU will enable ALMA to reach the same sensitivity in less than half the time in the key Band~6 receiver upgrade. 
The observations will yield extensive information on gas masses in disks, organic chemistry which is key to pre-biotic chemistry and the origin of life, and spatial variations abundances of gas-phase carbon, oxygen, and nitrogen, and how it impacts the composition of planetary atmospheres. The WSU will enable vast surveys of disks in a variety of evolutionary stages and ages. These observations will be invaluable inputs to physical and chemical disk models to decipher the physical conditions present for planet formation.   

\vspace{0.8in}
\subsection{Kinematic Signatures of Planet Formation}

\begin{figure}
\centering
\includegraphics[width=\textwidth]{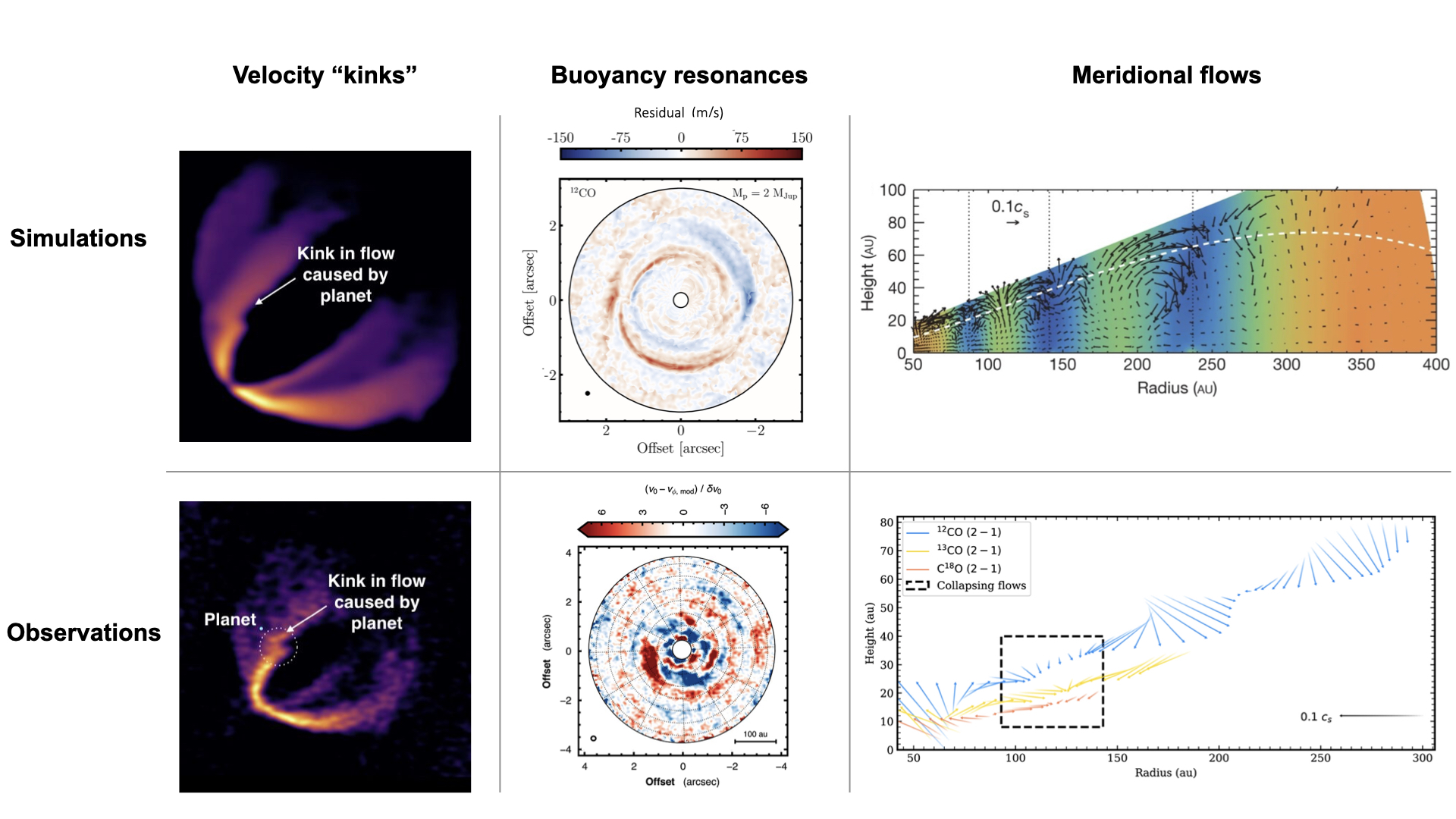}
\caption{Numerical simulations of the velocity distortions caused by embedded planets in protoplanetary disks, and the ALMA observations that suggest such processes are present in actual disks.
The numerical simulations in the top show, from left to right, 
(a) localized ``kink'' in the velocity field caused by an embedded planet \citep[reproduced by permission of the AAS]{Pinte18b};
(b) velocity residuals produced by the buoyancy resonance \citep[reproduced by permission of the AAS]{Bae21}; and 
(c) meridional circulation flows as the spiral wakes raises gas from the mid-plane, which then flows back onto the embedded planet \citep{Teague19a}. The observations in the bottom row show, from left to right,
(a) CO J=2--1 channel map for the disk surrounding HD~163296 \citep[reproduced by permission of the AAS]{Pinte18b}, where the dashed circle encompasses the kink in the velocity field that is indicative of an embedded planet;
(b) velocity residual map of TW~Hydra that reveal spiral patterns in the kinematics \citep[reproduced by permission of the AAS]{Teague22};
(c) velocity field of $^{12}$CO, $^{13}$CO, and C$^{18}$O J=2--1 toward the disk around HD~169142 that suggests a collapsing flow within the gas between a radius of 100 and 150\,au \citep[reproduced by permission of the AAS]{Yu21}.
}
\label{fig:flows_num_obs}
\end{figure}

The kinematics of gas in protoplanetary disks is dominated by Keplerian rotation, which is valuable on its own as a dynamical tracer of the stellar mass (see \citealt{Sheehan19} and references therein). Additional forces can cause deviations from pure Keplerian rotation that are diagnostic of the planet formation process (see the top row in  Figure~\ref{fig:flows_num_obs}). Embedded planets generate spiral wakes that create a localized distortion (``kink") in the velocity field \citep{Perez15,Perez18,Pinte18b}. These spiral wakes can also raise material in the midplane to the upper layers in the disk, which then flows back onto the planet to establish a meridional circulation flow \citep{Szulagui14,Morbidelli14,Fung16}. Resonances between orbiting planets and vertical motions of the gas (``buoyancy resonance") can produce tight spirals in the velocity field. Gravitational instabilities in massive disks distort the Keplerian velocity \citep{Hall20,Longarini21}. Other deviations from Keplerian motion may be present, including disk winds and self-gravity when the disk mass is an appreciable fraction of the stellar mass.

ALMA has made remarkable observations that provide tantalizing evidence of embedded planets in disks. By obtaining sensitive, high spectral resolution observations, the global Keplerian velocity can be modeled and subtracted from the observations to yield the residual velocity distortions. Examples of the residual velocity field maps obtained from ALMA observations are shown in the bottom row of Figure~\ref{fig:flows_num_obs}. Perhaps the most tantalizing evidence for planets is the detection of a localized ``kink'' in the velocity field of HD~163296 \citep{Pinte18b}. Similar kinks have now been detected in several other disks \citep{Pinte19,Casassus19,Pinte20,Perez20}, suggesting that massive planet formation is advanced even at stellar ages of a few million years. Large scale velocity flows in disks have also been detected. \citet{Teague19a} detected  {\it vertical} velocity flow in the upper layers of the disk around HD~163296 that are consistent with meridional flows \citep[see also][]{Casassus21}. By mapping these flows in different molecules that probe different surface layers of the disk, it is further possible to map the vertical velocity field at different heights within the disk to sample the 3-D velocity structure \citep{Yu21}. Observation of other disks have revealed large-scale arc and spiral type features in the kinematics \citep{Teague22}.

The key to these ALMA observations are high spectral resolution and high sensitivity. The kinematic perturbations created by planets and disk evolution can be 0.1\kms\ or less, such that observations with a velocity resolution comparable or better are needed. 
The WSU will advance these studies in two respects. First, the sensitivity of individual lines will be improved such that the  observing speed will be 2.25--4.7 faster with the Band 6 upgrade (see Appendix~\ref{sec:Band6}) after considering both the improved digital efficiency and the improved receiver temperatures over the full IF bandwidth. Second, the order of magnitude increase in the correlated bandwidth at high spectral resolution will observe multiple spectral lines with different optical depths and trace the kinematics of the gas at different heights in the disk. Such observations can  be diagnostic of the origin of the flows depending on how the kinematics varies with height in the disk \citep{Bae21,Teague22}.

\subsection{Disk Demographics}
The spectacular, high resolution dust continuum images of the rings and gaps in protoplanetary disks (see Figure~\ref{fig:dsharp}) are predominantly available for the largest, brightest disks. Demographic surveys have shown that most disks are fainter and more compact. In the Ophiuchus molecular cloud, for example, high resolution images are available for only the 5\% of the brightest disks \citep{Cieza21}. Whether or not the gaps and rings are prevalent in the more typical disk is unclear since high resolution continuum images of these fainter systems are challenging. The brightness profile of these fainter disks are similar to the cores of the brighter disks but without the extended disk emission \citep{Long19}, which suggested that either these compact disks were born small, or they lacked the mechanisms to prevent the inward radial drift of dust that produces the large dust rings. Advance imaging techniques have shown that the bright cores in these extended disks also exhibit gaps and rings \citep{Jennings22}, so it is plausible that such substructures are present in the more typical disks as well. The WSU will allow continuum images to be obtained 3--6$\times$ faster than currently possible. Appreciable samples of high resolution images of the fainter disks will become practical to determine if the more typical protoplanetary disks also contain rich substructures.

\subsection{Circumplanetary Disks}
During the formation of giant planets, a circumplanetary disk is expected to form, which will regulate the accretion of material onto the planet and will host the formation of moons. \citet{Isella19} and \citet{Benisty21} presented the first compelling detection of a circumplanetary disk with the submillimeter continuum detection coincident with the planet PDS~70c \citep{Haffert19}. The observations revealed a compact disk less than 1.2\,au in radius, and enabled estimates of the disk mass. 

\begin{figure}
\centering
\includegraphics[width=\textwidth, trim=8cm 13cm 8cm 11cm, clip]{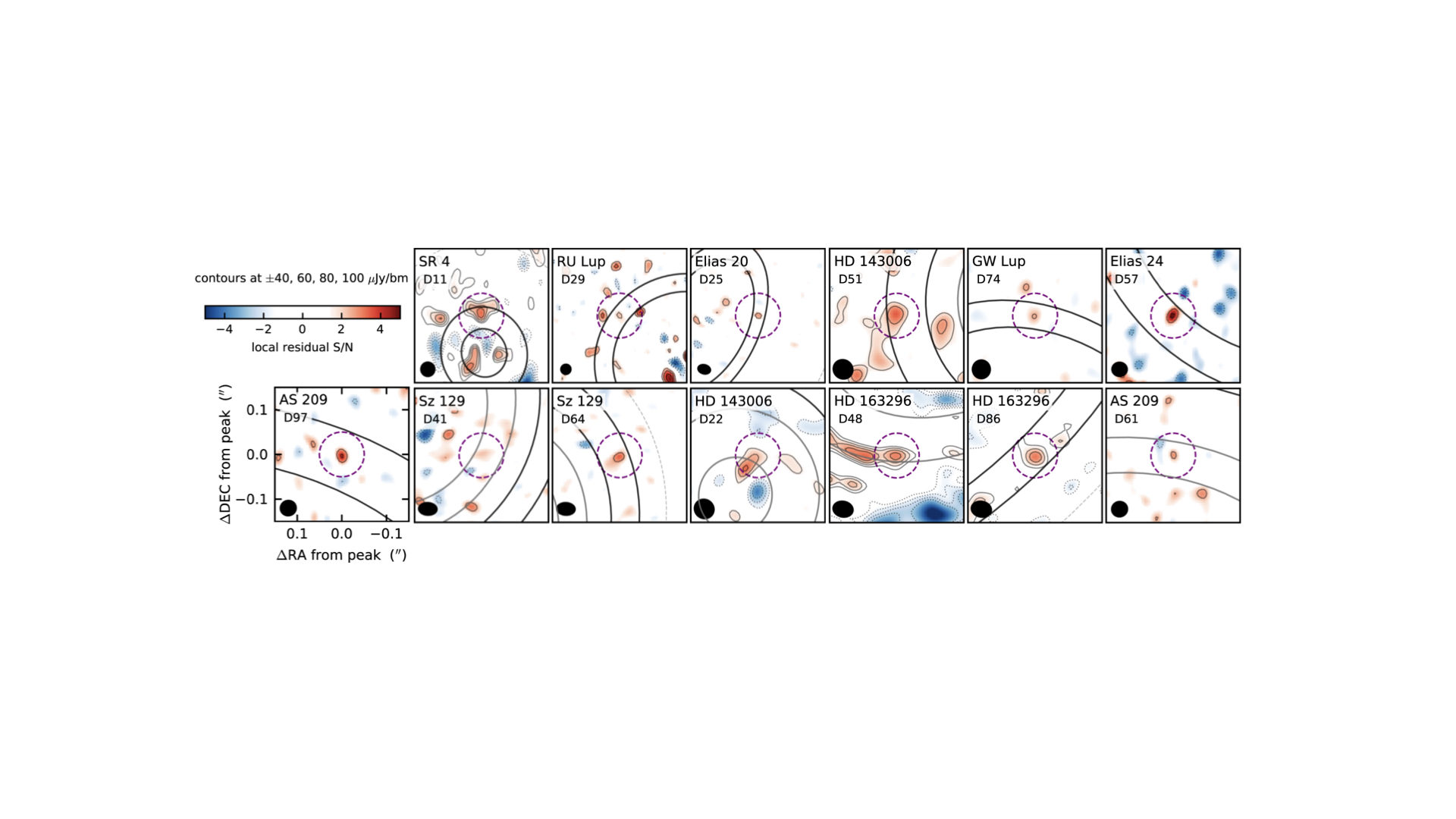}
\caption{Signal-to-noise images of the 1.3\,mm dust continuum emission after subtracting a model of the disk emission for a subset of the DSHARP disks. The images are zoomed in on the peaks in the residuals (highlighted by the purple dashed circles) of candidate circumplanetary disks. The black and gray arcs outline gaps in the continuum images. ALMA continuum images obtained after the Wideband Sensitivity Upgrade will have at least twice the continuum sensitivity and improved $uv$-coverage to enable better subtraction of the disk emission and identification of circumplanetary disks. Figure from \citet[reproduced by permission of the AAS]{Andrews21}. 
}
\label{fig:dsharp_cpd}
\end{figure}

\citet{Andrews21} searched for circumplanetary disks within the sample of 20 images obtained at high-resolution resolution by DSHARP \citep{Andrews18}. To search for circumplanetary disks, a model of the extended disk emission was subtracted to identify any compact residuals located within the gaps in the continuum emission, which are the suspected location of embedded planets. While several intriguing residual peaks are seen (see Figure~\ref{fig:dsharp_cpd}), the peaks are not compelling detections given the relatively low signal-to-noise ratio (3--5). The WSU provides two significant advantages in future searches for circumplanetary disks. First, the expanded bandwidth will improve the $uv$-coverage through multi-frequency imaging, which will enable a better subtraction of the larger-scale disk emission and more accurate identification of residuals. Second, the combination of expanded bandwidth, lower receiver noise temperatures, and higher efficiencies of the digital system will improve the continuum sensitivity by a factor of at least two for the same integration time, allowing for more robust identification of circumplanetary disks.

%% file: origins_chemistry.tex
\section{Origins of Chemical Complexity}
\label{sec:chemistry}

Given its already unprecedented spectral line sensitivity, ALMA has been leading a revolution in the study of (sub)millimeter spectral lines since its inception --- from the discovery of new and potentially biologically relevant astronomical molecules, to the use of multi-line diagnostics to reveal the kinematics and physical conditions across a vast range of astronomical environments. Although we highlight only a few of the relevant science cases here, the study of chemical complexity across all astrophysical environments is unified by common needs: to  survey spectral line emission and absorption across a broad frequency range, with adequate spectral resolution for the environment, with high sensitivity and high observing efficiency. Indeed, improved observing efficiency is key to making the next big leap forward in our understanding of chemical complexity, so that we can extend beyond the study of just a few especially bright (and/or nearby) exemplary objects, to statistically significant samples spanning relevant ranges of parameter space in age, evolutionary stage, chemical processing, dynamical influences, and external or inherited environmental factors. 

The advantages that the WSU brings to the study of circumstellar disks (see Section~\ref{sec:planets}) readily apply to observations of the chemical complexity across a range of environments. The 2--4$\times$ increase in the IF bandwidth and the large increase in correlated bandwidth allows for vastly increased spectral grasp that permits simultaneous observations of a multitude of spectral lines with improved sensitivity. The enhancements are particularly impressive for projects requiring high spectral resolution  ($\sim  0.1$\kms\/) in Bands 1 to 7, where the correlated bandwidth will increase by at least an order of magnitude (see Figure~\ref{fig:cbwgraph} and Table~\ref{tbl:bwincrease}). The net impact will be vastly improved observing efficiency in all phases in the life cycle of the interstellar medium, from the properties of molecular clouds, to the formation of dense cores and protostars, and to the regeneration of gas and dust in the envelopes of evolved stars.

\subsection{Star Formation}

A long-standing goal of star formation studies, and one of the key objectives for ALMA, is to understand how stars form in molecular clouds.  The basic narrative for star formation has long been established; see the reviews in Protostars and Planets~VI \citep{PPVI} and the recent Protostars and Planets~VII proceedings.\footnote{\url{http://ppvii.org}} Galaxies are populated with massive molecular clouds ($\sim 10^4$--$10^6\,\rm{M}_\odot$) that are cold, magnetized, and highly (super-sonic) turbulent \citep{Chevance22}. Molecular clouds exhibit a rich filamentary structure, which in turn contain compact dense regions (``cores") that gravitationally collapse to form protostars \citep{Hacar22,Pineda22}. Protostars are surrounded by an envelope of material that accretes onto a circumprotostellar rotating disk, which  serves as a conduit to accrete material onto the protostar and to form planetary systems \citep{Dunham14}. Protostar formation is also accompanied by high-velocity jets and outflows that remove angular momentum from the protostellar system \citep{Bally16}. In addition to gravity, magnetic fields are also thought to significantly influence the process of star formation both during the collapse phase and in the removal of angular momentum at later stages \citep{Pattle22, Tsukamoto22}. The end product of this process, when averaged over an ensemble of cores and clouds, is a characteristic spectrum of stellar masses that appears common in the local universe over a broad range of environments \citep{Hopkins18}.  

Despite the success of this general picture of the star formation process, it remains elusive to understand quantitatively what causes some regions within clouds to form stars and star clusters and not others, why the overall efficiency of star formation in galactic molecular clouds is low, what determines the final mass of the star, and why the star formation process produces a consistent stellar mass function across a range of environments. Theoretically, this remains a challenge because of the complex and diverse forces that are at play. The gravitational forces that lead to star formation are opposed by gas pressure, turbulence, magnetic fields, and feedback (e.g, stellar radiation, jets, outflows, supernova). Modelling the complex interactions between these forces remains a formidable challenge. Molecular lines that trace different physical conditions, together with continuum and polarization information, are among the key observables for understanding these processes.

\begin{figure}
\includegraphics[width=\textwidth,trim=0in 0.4in 0in 0.15in, clip]{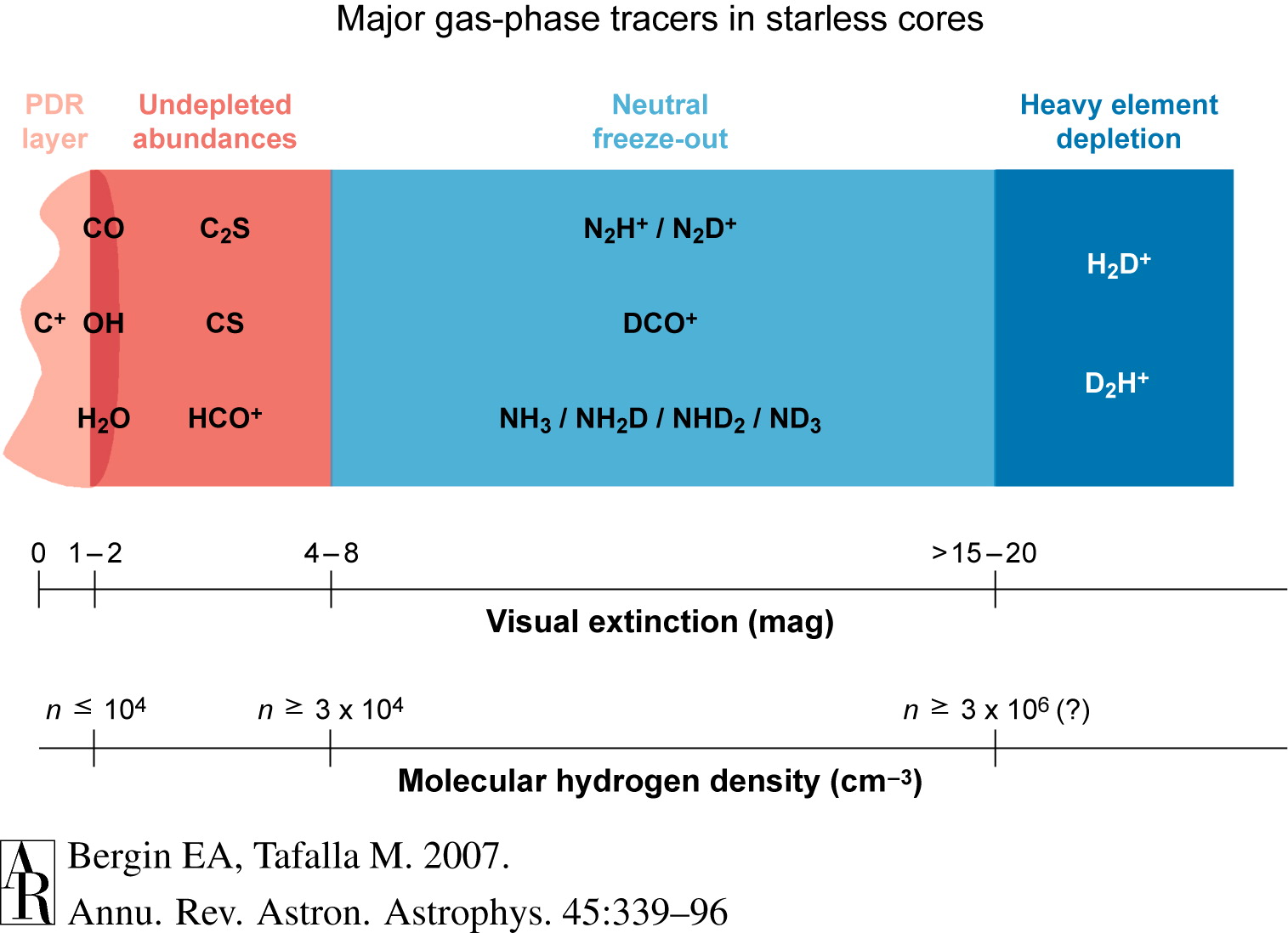}
\caption{Schematic of the major molecular probes of the molecular clouds as a function of depth and density. As the gas density increases and the temperature decreases, carbon-based molecules freeze out onto the grains, which leads to the enhancement of nitrogen and deuterated species. Figure from \citet{Bergin07}.}. 
\label{fig:starlesscores}
\end{figure}

\begin{figure}
\begin{center}
\includegraphics[width=0.8\textwidth]{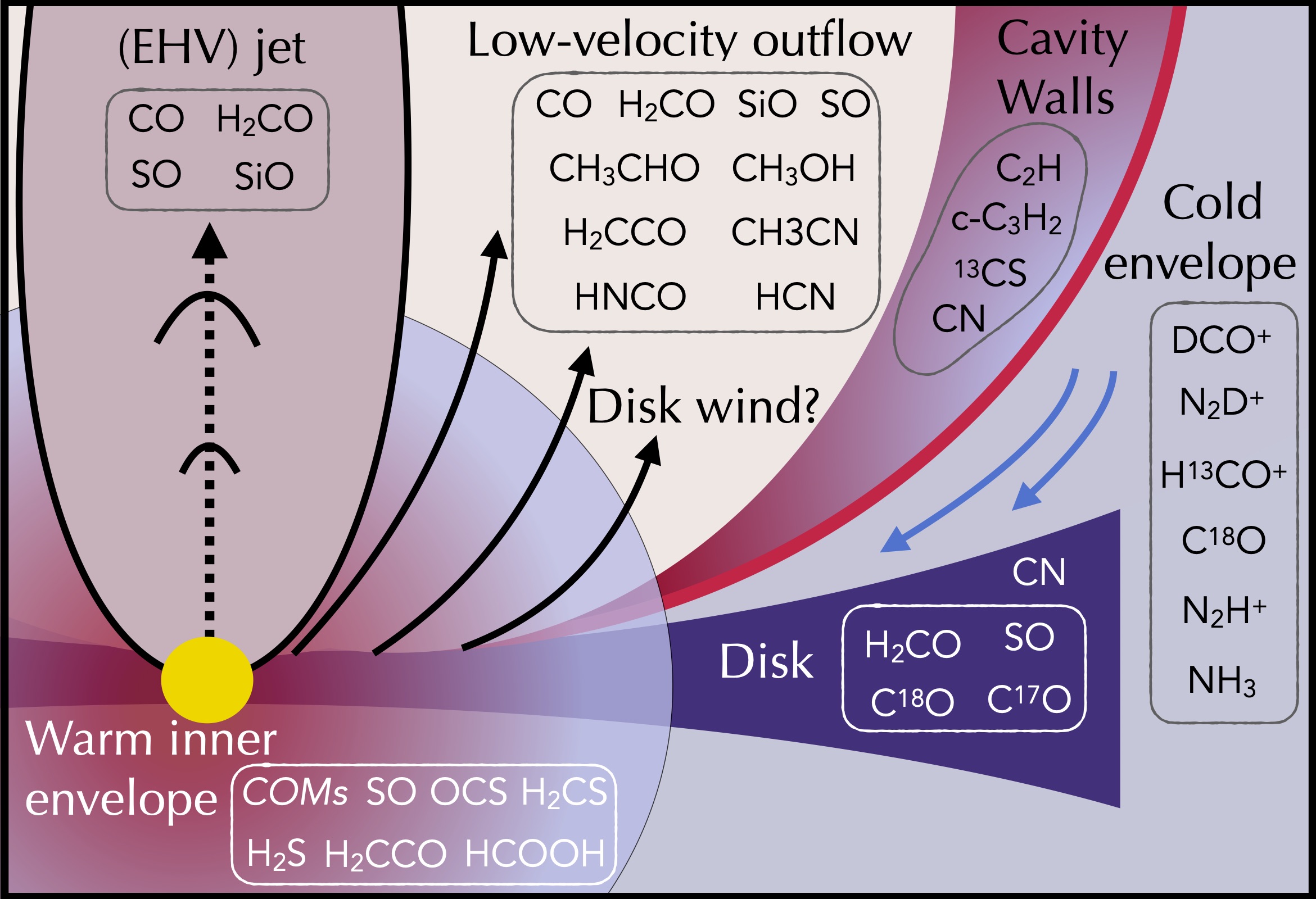}

\vspace{0.5cm}

\includegraphics[width=\textwidth]{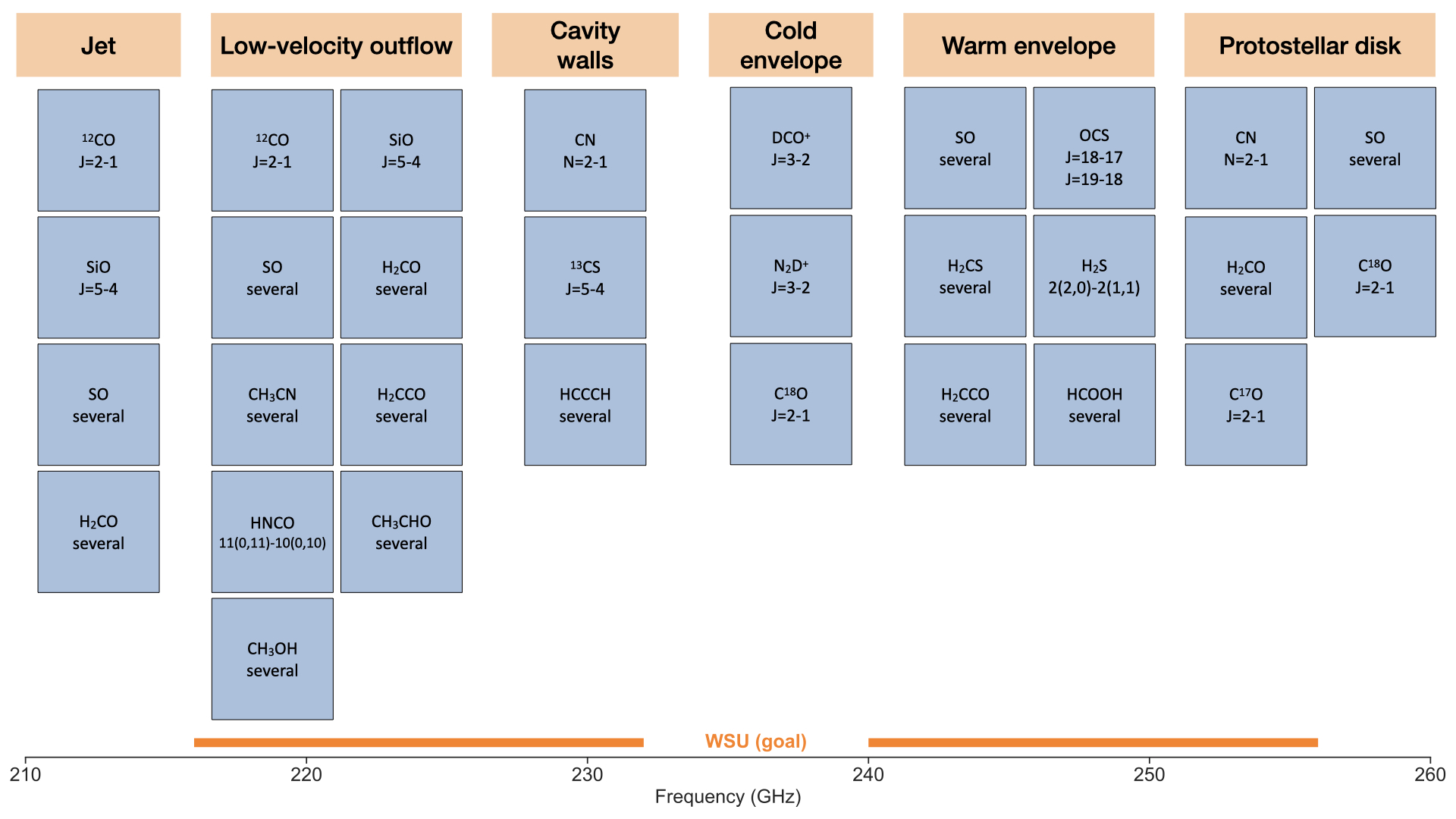}
\caption{Top: Schematic indicating a selected number of molecular species that can probe different parts of a protostellar environment \citep{Tychoniec21}. 
Bottom: An example Band 6 tuning illustrating the different lines in the above schematic that can be observed simultaneously with the WSU for a LO frequency = 236\,GHz and a 4--20\,GHz IF bandwidth. For this tuning, 86\% of the molecules in \citet{Tychoniec21} can be observed. Many additional lines, not shown here, can also be observed since the full IF bandwidth shown by the orange bar will be correlated.
\label{fig:tychoniec}}
\end{center}
\end{figure}

A major observational challenge to advance our understanding of the star formation process is the enormous range of scales involved, from molecular clouds that span up to $\sim100$\,pc in diameter, to dense cores ($\sim0.1$\,pc), to circumstellar disks ($\sim0.0001$\,pc). Thus observations over large angular scales are needed to map clouds and filaments, while high resolution is needed to map the structures of the circumprotostellar environment. An additional challenge is that the orders of magnitude in spatial scales are accompanied by orders of magnitudes in densities and variations in temperature, which drives a rich chemistry. Molecules that are abundant in the larger-scale cloud (e.g, CO, CS, HCO$^+$) freeze out onto grains in the dense, cold pre-stellar cores, which drives chemical reactions that enhance the abundance of other molecules  (e.g., $\mathrm{N_2H^+, N_2D^+}$, DCO$^+$; see Figure~\ref{fig:starlesscores}). As a protostar forms, the gas temperature once again increases, which is accompanied by shocks from jets and outflows. These processes release complex molecules from the grains and once again drive a new chemistry, where different molecules can be used to trace different components (e.g., envelope, jet, outflow, disk) of the protostellar system (see  Figure~\ref{fig:tychoniec}).

\begin{figure}
\includegraphics[width=\textwidth]{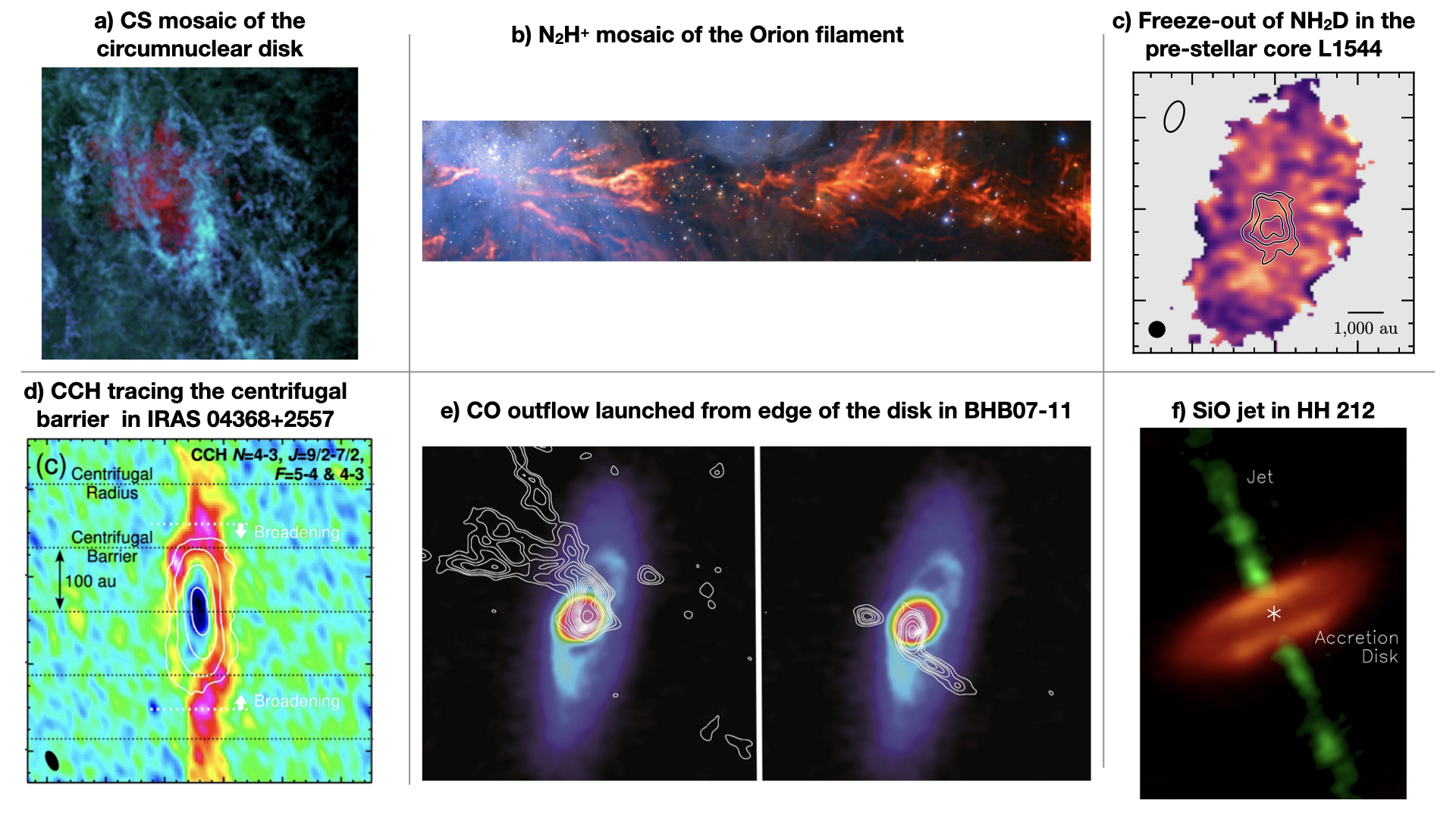}
\caption{A sampling of ALMA observations that illustrates the diverse array of molecules used to probe the interstellar medium and the star formation process. 
(a) CS mosaic \citep{Hsieh21} of the circumnuclear disk at the galactic center (blue and purple) compared with ionized gas from the Hubble Space Telescope (red). Credit: Dong, H. et al 2011 - ESA/Hubble | Hsieh, P.-Y. et al. - N. Lira - ALMA (EOS/NAOJ/NRAO).
(b) ALMA $\mathrm{N_2H^+}$ $J=1-0$ mosaic of the Orion Integral Shaped Filament combined with IRAM~30\,m single-dish observations \citep[Credit: ESO/H. Drass/ALMA (ESO/NAOJ/NRAO)/A. Hacar]{Hacar18}.
(c) Freeze-out of para-$\mathrm{NH_2D} (1_{11}$--$1_{01})$ emission (color) with respect to the 1.3\,mm dust continuum (contours) in the prestellar core L1544 \citep{Caselli22}.
(d) Observations of the CCH $N=4-3$, tracing the centrifigual barrier in the protostar IRAS~04368+2557 \citep[Figure 1c in][]{Sakai17}.
(e) CO $J=2-1$  (contours) of the outflow that is launched near the edge of the continuum disk (color) in the protostar BHB07-11 \citep[Credit: MPE]{Alves17}.
(f) SiO $J=8-7$ image (green) of the jet and the continuum emission from the circumstellar disk (red) around the protostar HH~212 \citep[Credit: ALMA (ESO/NAOJ/NRAO)/Lee et al.]{Lee17}.
}
\label{fig:images_sf}
\end{figure}

Because of the large range of physical scales that encompass the star formation process and the complex physical and chemical processes involved, ALMA users have conducted a vast array of observations including detailed observations of individual cores and protostars \citep[e.g.,][] {Jensen2019,Ligterink2022,Ginsburg2019}, comprehensive surveys of ensembles of sources \citep[e.g., the Large Programs FAUST and IMF:][]{FAUST,IMF}, and extensive mosaics of individual clouds and filaments in a variety of molecular species \citep[e.g.,][]{Barnes2021,Lu2021,Yang2021}. These observations have leveraged ALMA's unprecedented access to the full range of relevant spatial scales using the TP, 7-m, and 12-m arrays to sample large, intermediate, and fine angular scales, respectively. 
Figure~\ref{fig:images_sf} provides just a sampling of images obtained with ALMA that probe the various stages in the star formation process, from molecular clouds, to filaments, prestellar cores, and protostars. These images aptly illustrate the utility of various molecules to trace different components of the interstellar medium and the star formation process.  

\begin{figure}
\includegraphics[width=\textwidth]{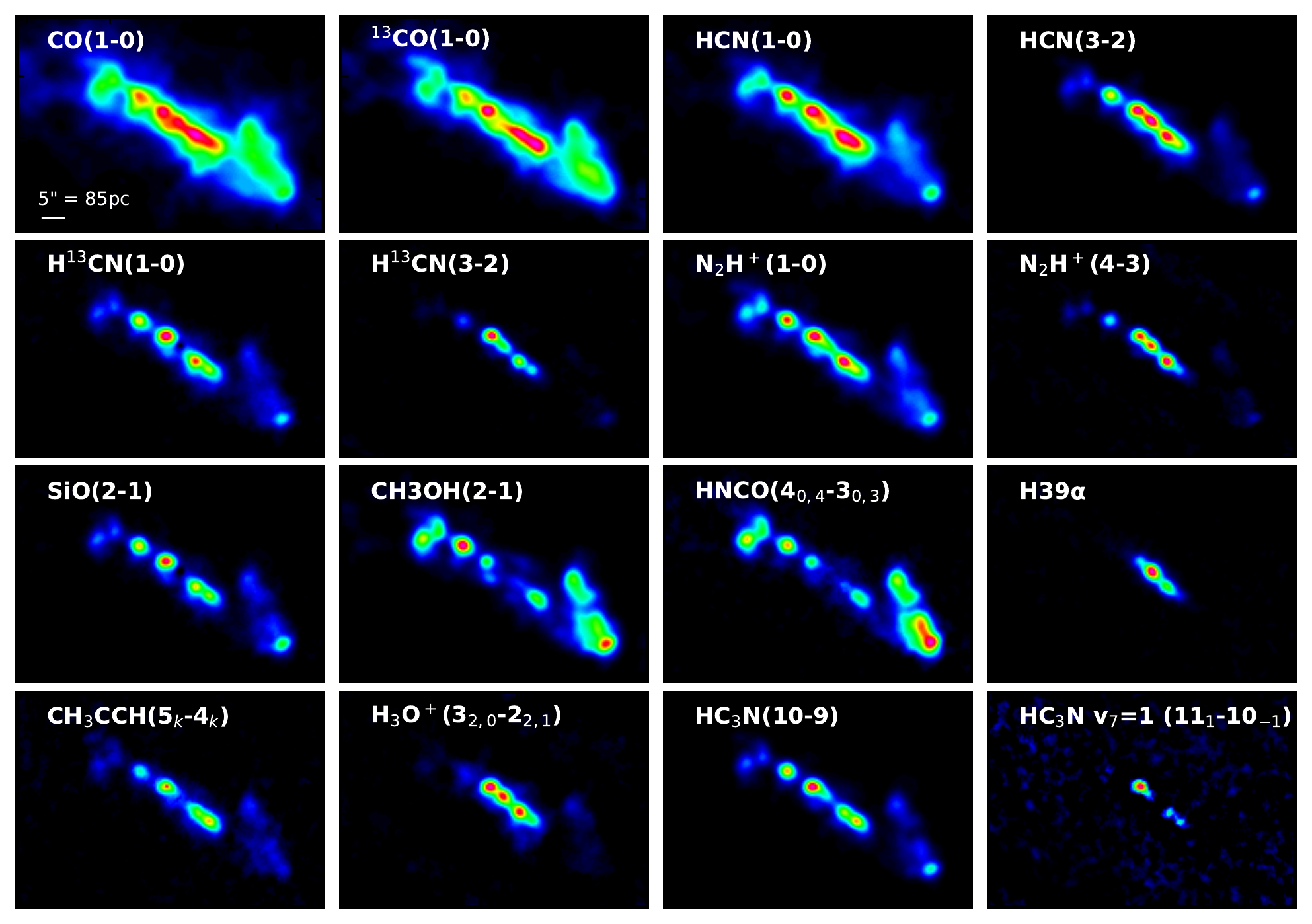}
\caption{Molecular line images of the nuclear region of the galaxy NGC~253 from the ALCHEMI Large Program \citep{Martin2021ALCHEMI}.}
\label{fig:alchemi_images}
\end{figure}

Analogous surveys of the molecular content in nearby galaxies have also been undertaken. For example, the PHANGS Large Program demonstrated that the properties of molecular clouds on $\sim$100\,pc scales correlate with their location in the host galaxy, with higher surface densities and turbulent pressure toward the center of galaxies and in spiral arms compared to interarm regions \citep[][]{Sun20,Leroy2021}. Detailed studies of the M33 and the Antennae galaxies (to name a few), have revealed that the large-scale filamentary structure of GMCs in the Milky Way is ubiquitous in both ``normal'' and starburst galaxies \citep[][]{M33,Antennae}, while studies of numerous dwarf galaxies have begun to explore the consequences of low-metallicity star formation on chemical complexity \citep[][]{SMC,Turner2019,LMC}. Detailed studies of the nuclear regions of galaxies have also been obtained by ALMA to provide a global perspective on cloud properties \citep{Costagliola2015,Meier2015,Harada2018,Harada19,Martin2021ALCHEMI,Sakamoto2021I,Sakamoto2021II}. As illustrated by the spectacular case of NGC~253 (see Figure~\ref{fig:alchemi_images}), the particular chemical routes to the formation and destruction of molecular species results in obvious morphological differences between molecular tracers. This differentiation can be used to disentangle the different physical and chemical mechanisms in play within the central region of galaxies \citep[e.g., Figure~3 from][]{Meier2015} and to trace the chemical evolution of molecular clouds due to collision-induced star formation in the circumnuclear disk of galaxies \citep[e.g., Figure~5 from][]{Harada2018}.

\begin{figure}
\includegraphics[width=\textwidth]{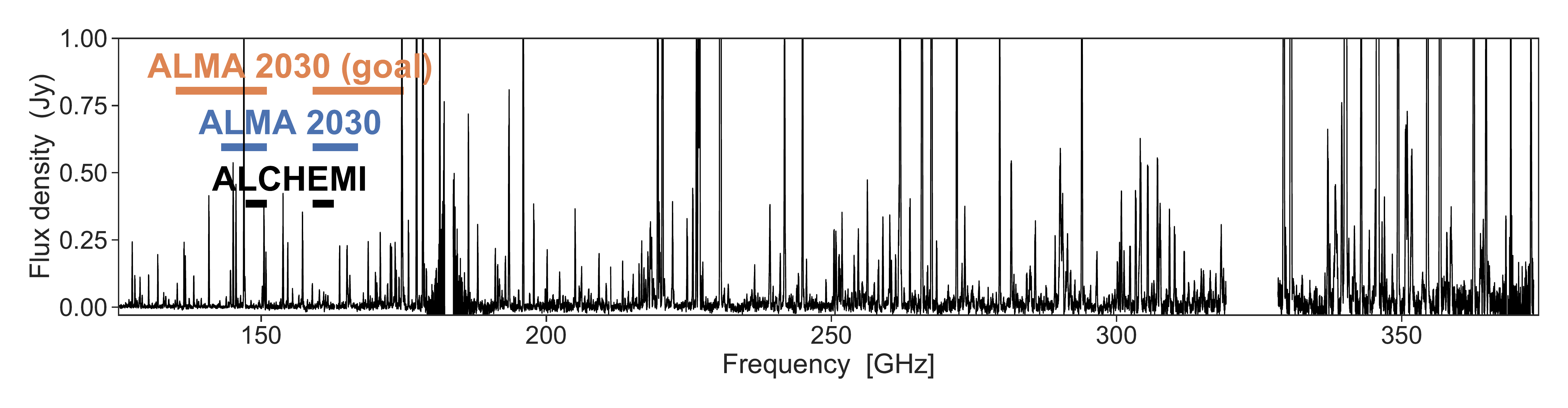}

\includegraphics[width=\textwidth]{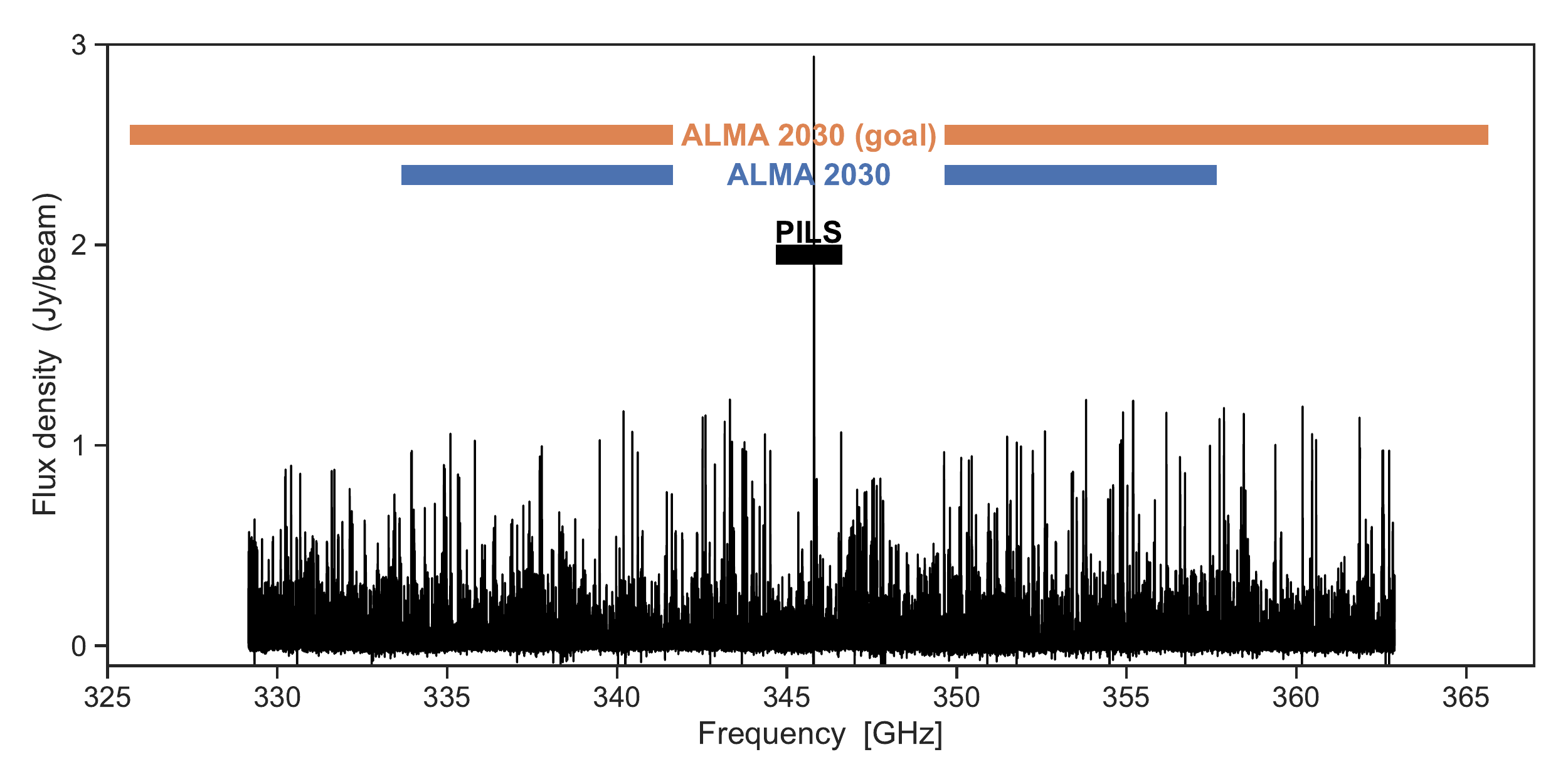}

\caption{Illustration of the improvement that the WSU will provide for spectral scan observations. The top panel shows the spectral scans across multiple bands obtained for NGC~253 \citep{Martin2021ALCHEMI}. The bottom panel shows the spectral scan for IRAS~16293-2422B from PILS \citep{Joergensen2016}. In each plot, the instantaneous spectral coverage of the actual observation is shown in black, while that of the ALMA2030 upgrade is shown in blue for the minimum requirements and in orange for the goal. The ALMA2030 upgrade will improve the instantaneous coverage by a factor of 2--4 for coarse spectral resolution scans such as NGC~253. For surveys such as PILS, which required 18 separate tunings at high spectral resolution, similar spectral coverage can be achieved in only two tunings.
}
\label{fig:spectralscans}
\end{figure}

The WSU will expand the already impressive ALMA results and improve dramatically the efficiency of observations of the interstellar medium. Observations of clouds and protostars in the Galaxy generally require high velocity resolution, such that using the current system, the correlated bandwidth needs to be compromised to cover the lines of interest. In addition, many studies require both continuum and line observations, which further limits the number of available spectral line windows. Such restrictions will no longer apply with the WSU, which will allow for a wide selection of lines to be observed simultaneously while obtaining the continuum at maximum sensitivity. The bottom panel in Figure~\ref{fig:tychoniec} illustrates just one possible correlator setup with the WSU, in which 86\% of the lines highlighted by \citet{Tychoniec21} can be observed in a single tuning to observe {\it all} protostellar components. The WSU will also enable far more efficient spectral scans that are often used to obtain chemical inventories of sources. Figure~\ref{fig:spectralscans} shows  two example spectral scans that have been obtained with ALMA: the nuclear region of NGC~253 \citep{Martin2021ALCHEMI} and the protostellar binary IRAS~16293-2422B \citep{Joergensen2016}. As shown in the figure, the ALMA2030 upgrade provides vastly increased bandwidth to improve the efficiency of spectral scans. The PILS survey of IRAS~16293-2422B can be achieved with 9 times fewer tunings even with the minimum goals of the WSU upgrade, while the  spectral survey of NGC~253 will be 2--4$\times$ faster, plus any additional gains from the improved sensitivity. 

The impact of the WSU on studies of the interstellar medium and protostars will be substantial. Observations of individual sources will have access to far more diagnostic lines to investigate the evolutionary states of clouds, cores, and protostars (Section~\ref{sec:specres}). Mosaics of clouds and filaments will provide a comprehensive census of cores and protostars over all evolutionary states. Spectral scans will be vastly more efficient, allowing comprehensive chemical inventory of cores and protostars (Section~\ref{sec:specscan}). The improved observing efficiency will ultimately enable comprehensive surveys of far more sources than currently possible, which is key to understanding the diversity of outcomes from the star formation process. 

\FloatBarrier

\input{evolved_stars}

\FloatBarrier

\subsection{Discovery of New Astronomical Molecules}

To date, ALMA has been used to discover seven new molecules in the ISM, as well as a host of isotopologues (including deuterated molecules) and vibrationally excited transitions of known species \citep{McGuireCensus}. The origin of life on our planet remains one of the great unsolved mysteries of human existence, and the discovery of biologically significant molecules in the ISM (or even their precursors) offers exciting prospects for understanding how the complex chemistry that led to life on our planet came to be. With at least 241 individual molecular species already discovered so far in space \citep{McGuireCensus}, the majority using radio telescopes, new and complex molecules are only to be found in the most chemically rich environments. Indeed, such environments create a surfeit of riches, such that one must first identify, model, and exclude the weeds (i.e., common abundant molecules) before new discoveries can be made. 
Thus, achieving more complete censuses of diverse chemically rich environments, along with the derivation of the underlying physical conditions and relative abundances, are the key to discovering new molecules, driving forward chemical modelling research, and potentially revealing the complex chemical pathways that led to the Universe we see today, and ultimately life.

\begin{figure}[h]
\centering
\includegraphics[clip,trim=0.0cm 0.0cm 0.0cm 0.1cm,width=\textwidth]{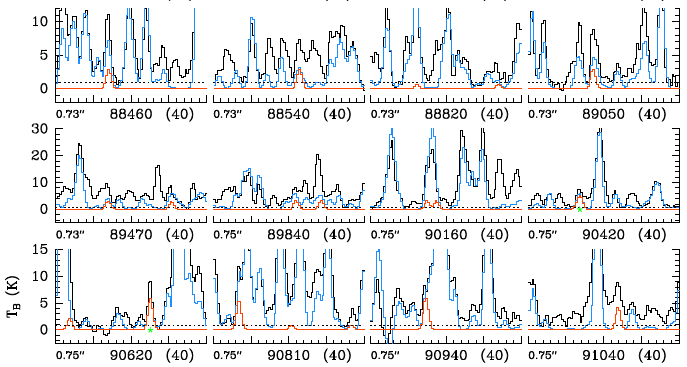}
\caption{Small portion of the total Band 3 ALMA frequency range (84.1 to 114.4\,GHz) required to identify convincingly the new astronomical molecule urea towards SgrB2(N1S). These data are from the ReMoCA survey \citep{Belloche2019}. The predicted emission from the full molecular line model is overlaid in blue, while the predicted urea emission (under the assumptions of the model) is shown in red. After the WSU, this full Band 3 spectral scan could be acquired in about a quarter of the time.
}
\label{fig:remoca}
\end{figure}

One critical missing ingredient to unleash the full potential of ALMA to discover new molecules is the ability to span {\em efficiently} a broad range of frequencies, with high spectral resolution, in order to sample multiple transitions of any potential new molecule. (Sensitivity is rarely an issue in these environments.) Figure~\ref{fig:remoca} from the Re-exploring Molecular Complexity with ALMA Survey \citep[ReMoCA,][]{Belloche2019} shows a small portion of the total Band 3 spectrum (84.1 to 114.4\,GHz) used to discover convincingly  the biologically significant molecule urea (NH$_2$C(O)NH$_2$)\footnote{On Earth, urea (also known as carbamide) is the result of the metabolic breakdown of proteins in the liver of all mammals and even some fish.} toward the massive star forming region SgrB2(N1S).  The extraordinary complexity of the rich line emission, and the need to cover many potential transitions over a broad frequency range in order to separate convincingly the new molecule from the line forest are readily apparent. 

The next challenge, for this particular example and other new molecule discoveries, will be to explore how urea behaves in other line-rich environments in the galaxy (SgrB2 lies near the Galactic center, at a distance $\sim 8$\,kpc) to answer questions like: is urea common in the hot cores around massive protostars, and if so, is the chemical model, molecular abundances, and physical conditions derived from these data widely applicable? The ReMoCA survey required five tunings and could only record a single polarization of data to achieve a relatively poor spectral resolution of 1.3--1.7\kms, which would be insufficient toward many regions with narrower line widths. After the 2$\times$ bandwidth WSU upgrade, the original ReMoCA frequency range could be covered with only 2 tunings (using the Band 2 receiver), with $\sim 0.12$\kms\/ spectral resolution and full polarization. After accounting for the additional sensitivity improvements (2 orthogonal polarizations, digital efficiency improvement, and assuming similar receiver noise performance), four additional line-rich sources could be explored for comparison with SgrB2(N1S) in the same amount of time as the original observations.

\begin{figure}[ht]
\centering
\includegraphics[width=\textwidth]{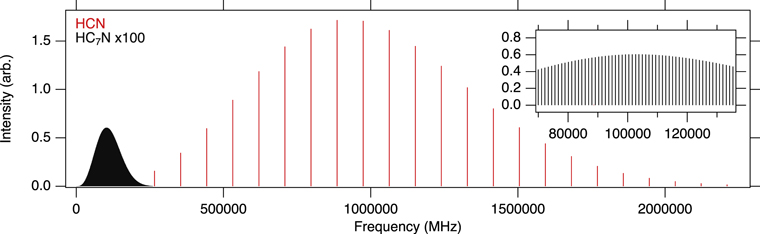}
\caption{Rotational spectra of HCN (red) and HC$_7$N (black) at $T = 150$\,K from \citet{McGuireCensus}, demonstrating that heavier molecules have their rotational transitions at lower radio frequencies. In this example, HC$_7$N transitions are concentrated in the future ALMA Band 2 RF range. The inset shows a magnified portion of the HC$_7$N to show detail as to the density of lines. In both the main plot and the inset, the intensities of the HC$_7$N lines are multiplied by a factor of 100.
}
\label{fig:hcn}
\end{figure}

A new avenue ripe for the discovery of new molecules after the WSU will be the two new lowest frequency bands of ALMA: Bands 1 and 2. For molecular transitions arising from rotational motion, larger molecules have smaller rotational constants (larger moment of inertia), which in turn results in closely spaced rotational energy levels with transitions starting at lower radio frequencies compared to smaller/lighter molecules \citep{McGuireCensus}. Figure~\ref{fig:hcn} adapted from \citet{McGuireCensus} illustrates this phenomenon by comparing HCN with HC$_7$N. 

Many undiscovered molecules that are of potential prebiotic significance can be best searched for specifically in Bands 1 and 2 \citep[e.g.,][]{Mcguireprebiotic}.  Demonstrating this point, in the last three years alone, more than 22 new molecules have been discovered using the Yebes 40-m and the Green Bank Telescope (GBT) single dish telescopes to perform spectral scans toward the nearby (122\,pc) starless core TMC-1 at frequencies less than 50\,GHz \citep[e.g., the large programs QUIJOTE and GOTHAM; see][and references therein]{McGuire2021Science,Cernicharo2022,McGuireCensus}. Many of the new molecules are long-chain hydrocarbons and sulfur-bearing species, that prior to their discovery were not predicted by existing chemical models (models that are now being retooled to account for this new complexity), reflecting the close symbiotic relationship between observational, lab, and modeling-based astrochemistry. Indeed, TMC-1 has revealed itself an extremely prodigious producer of large pre-biotic molecules, but two significant questions remain: (1) What is the spatio-kinematic distribution and differentiation of the rich molecular emission from TMC-1?; (2) How prevalent is such molecular richness in other starless core environments? 

\begin{figure}[h]
\centering
\includegraphics[width=\textwidth]{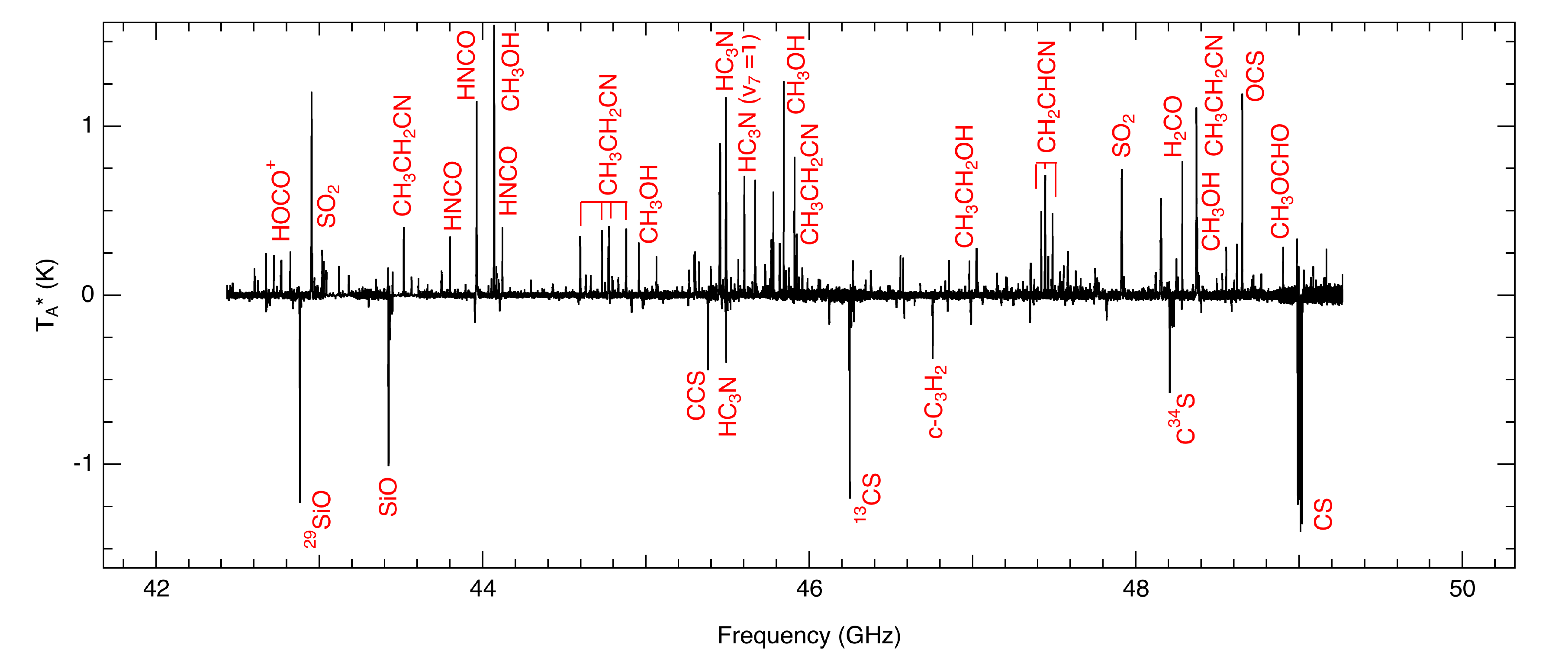}
\caption{GBT 42.4 to 49.2\,GHz spectrum toward the massive star forming region SgrB2 from the PRIMOS survey with $\sim 0.16$\kms\/ spectral resolution (data taken in 1\,GHz wide increments, see \citealt{Neill2012}), demonstrating the extraordinary chemical richness available in the ALMA Band 1 tuning range. Credit: Brett McGuire (MIT/NRAO).
}
\label{fig:Band1}
\end{figure}

With new low frequency millimeter-wavelength receivers and observing efficiency gains from the ability to observe with high spectral resolution at wide bandwidth, ALMA will be well placed to begin answering these questions. The spectrum in Figure~\ref{fig:Band1} from the GBT demonstrates the remarkable chemical richness that ALMA Band 1 spectral line observations may afford, including a wide range of complex organic molecules (COMs) and even Zeeman magnetic field tracers like CCS. Remarkably, after the ALMA WSU, imaging data (rather than a single spectrum as with the GBT) over the same spectral range (42.4 to 49.2\,GHz) could be acquired with $\sim 0.1$\kms\/ spectral resolution (full polarization) in a single tuning, 
compared to the 28 tunings that would be required to cover the same frequency range at $\sim 0.2$\kms\/ (dual polarization) with the current ALMA system. Interestingly, it would also take $\sim 28$ tunings to cover this frequency range with the JVLA at 0.2\kms\/ resolution (dual polarization).

\FloatBarrier

%% file: evolved_stars.tex
\subsection{Evolved Stars}

Low and intermediate-mass stars (0.8--8\,M$_{\odot}$) evolve off the Main Sequence to become Red Giants and later ascend the Asymptotic Giant Branch (AGB). Here, they eject most of their mass, forming circumstellar envelopes (CSEs) composed of molecular gas and dust. Material in the CSEs is accelerated outwards in the form of stellar winds and enriches the interstellar medium (ISM) with heavy atoms from the third stellar dredge up, and molecules formed in the CSEs themselves \citep{Herwig2005,Hoefner2018,Decin2021}. The chemistry of the CSEs is quite complex, depending on the interaction with dust grains and shock propagation \citep{Millar2016}. It is further complicated by the development of high-velocity collimated winds during the post-AGB phase and later on by the ionizing UV radiation from the hot central star, during the Planetary Nebula (PN) phase \citep{Balick2002}. ALMA study of AGB stars and PN has revolutionized understanding of multiple aspects of these stellar evolutionary phases \citep[for initial expectations, see][]{Olofsson2008}. The more massive red supergiant stars \citep[RSGs; e.g.,][]{Smith2014} have also been studied using ALMA, for example Betelgeuse \citep{Kervella2018}, VY CMa \citep{DeBeck2015} and Antares \citep{OGorman2020}. 

The WSU will be important for studies of evolved stars because (i) the improved continuum sensitivity will enable deeper and more efficient study of stellar surfaces and circumstellar dust; (ii) the increased line sensitivity will make it feasible to study the relatively low abundance molecular species/isotopologues and their circumstellar distributions, in addition to the study of gas in more distant stars \cite[e.g.,][]{Groenewegen2016}; and (iii) larger instantaneous bandwidth and faster spectral scans (performed in fewer tunings) will support investigation of the complex chemistry of these objects. The stellar evolution from the AGB through the post-AGB phase and finally the PN is characterized by the formation of astonishing circumstellar nebulae, whose shapes are still challenging the current evolutionary models. 
The improved capabilities will support the study of circumstellar structures; e.g., disks, magnetic fields and the effects of companions.   This will help in obtaining a more detailed understanding of the shaping process that takes place to form PN.

\subsubsection{Circumstellar Chemistry and Dust}

\begin{figure}
    \centering
    \includegraphics[width=0.75\textwidth]{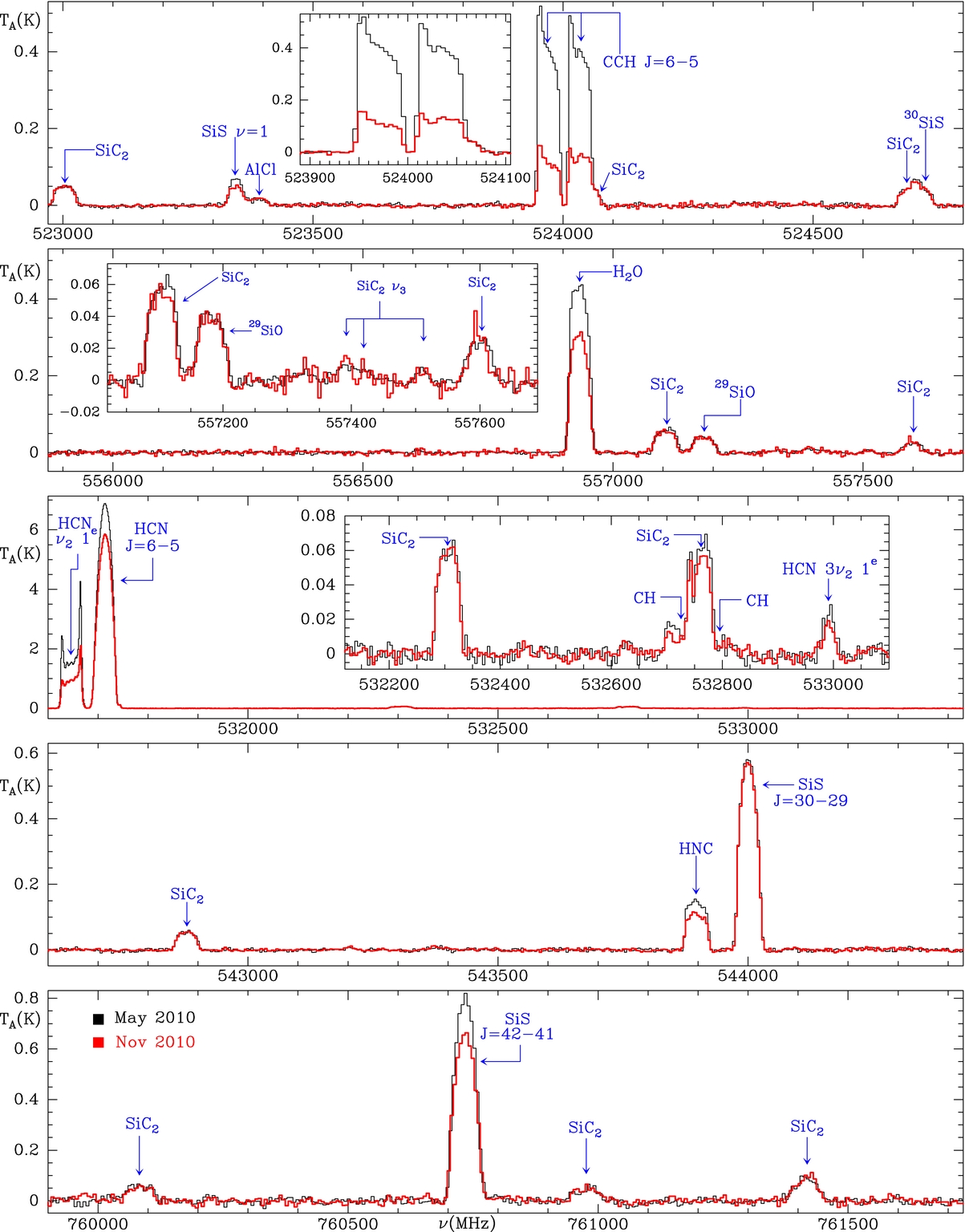}
    \caption{Variation over six months  in several lines in IRC+10216 \citep[reproduced by permission of the AAS]{Cernicharo2014}. The WSU will provide an efficient means to trace the temporal and spatial variations in the circumstellar environments around evolved stars from many molecules simultaneously.}
    \label{fig:IRC-time}
\end{figure}

Time variation in the intensity of thermal molecular lines, believed to originate in the inner CSE, has been detected towards IRC+10216 using the IRAM 30-m and Herschel, as shown in Figure~\ref{fig:IRC-time}  \citep{Cernicharo2014}, and monitoring campaigns have been carried out with ALMA \citep{He2019}. The variation of some of the lines appears to be either in-phase or in anti-correlation with the near-infrared light.
The large instantaneous bandwidth of the WSU will enable this variation to be studied more accurately due to truly simultaneous information over wider frequency ranges (as opposed to these being potentially acquired at quite different times in the stellar pulsation cycles). For the study of circumstellar chemistry, and abundance determinations in general, the WSU will have a high impact in making spectral scans possible in fewer tunings. Combined with spatial resolution sufficient to resolve the CSE, this will enable the efficient mapping out of dust parent molecules to contribute to understanding of dust formation processes in oxygen-rich and carbon-rich stars \citep{Gobrecht2016}. Gas phase dust precursors include aluminium and titanium oxides \citep[e.g.,][]{Kaminski2019b,Gobrecht2022}.

\begin{figure}
  \centering
  \begin{subfigure}[]{}
    \centering
    \includegraphics[clip=True,viewport=0 197 522 605,height=2.85in]{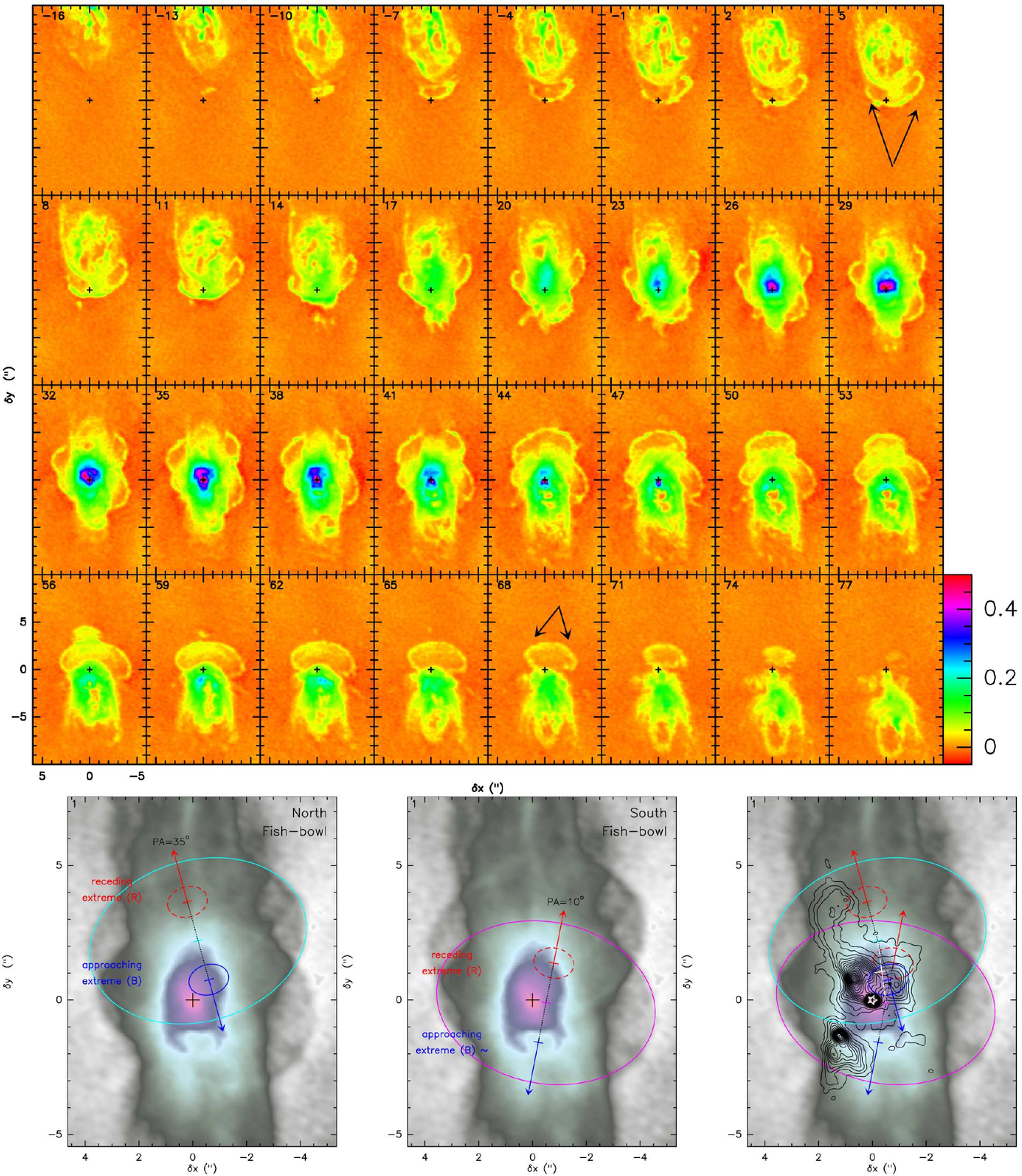}
  \end{subfigure}
  \qquad

  \begin{subfigure}[]{}
    \centering  
    \includegraphics[clip=True,viewport=3 227 485 675,width=0.45\textwidth]{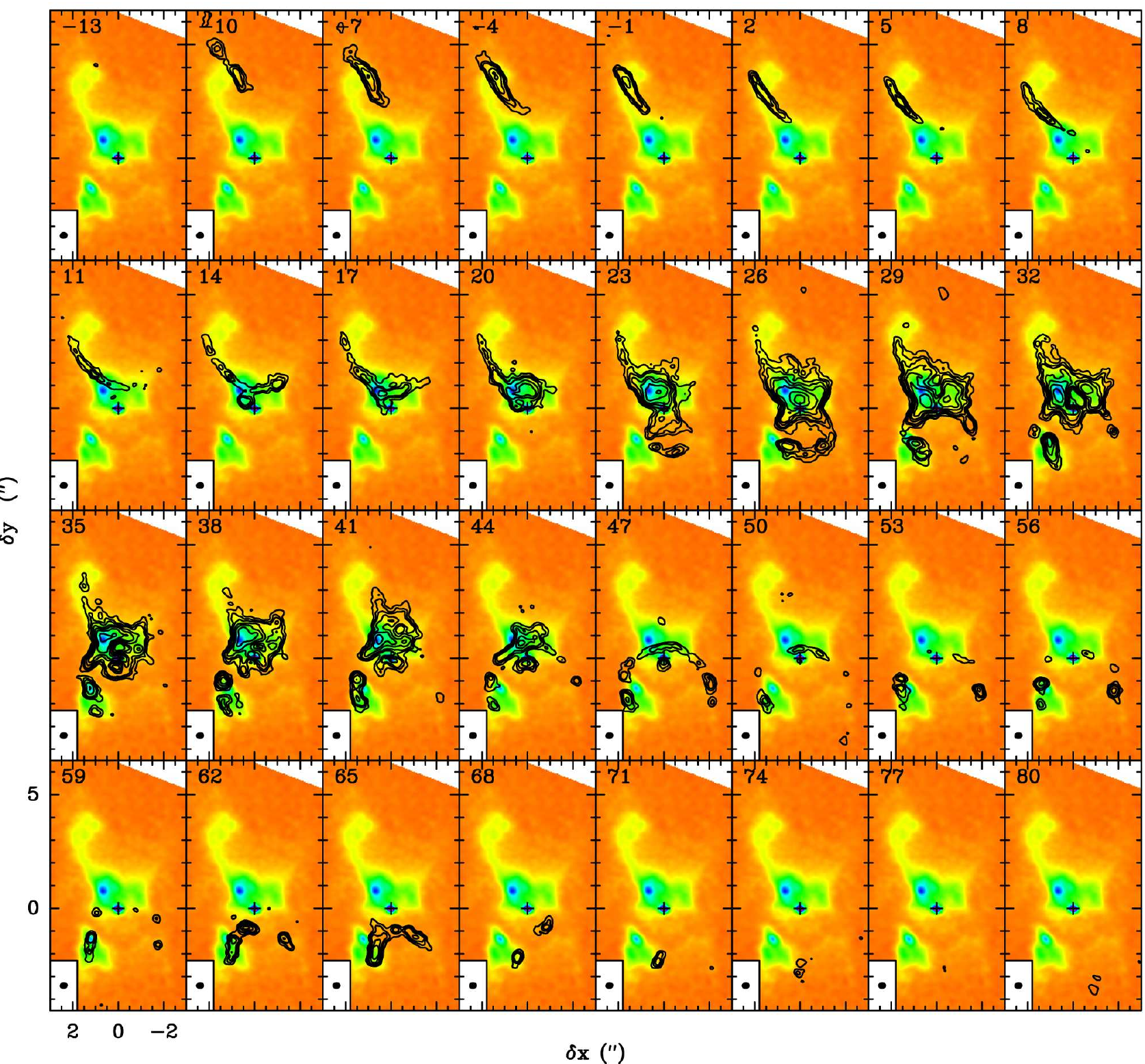}
  \end{subfigure}
  \hfill
  \begin{subfigure}[]{}
    \centering  
    \includegraphics[clip=True,viewport=3 220 445 645,width=0.42\textwidth]{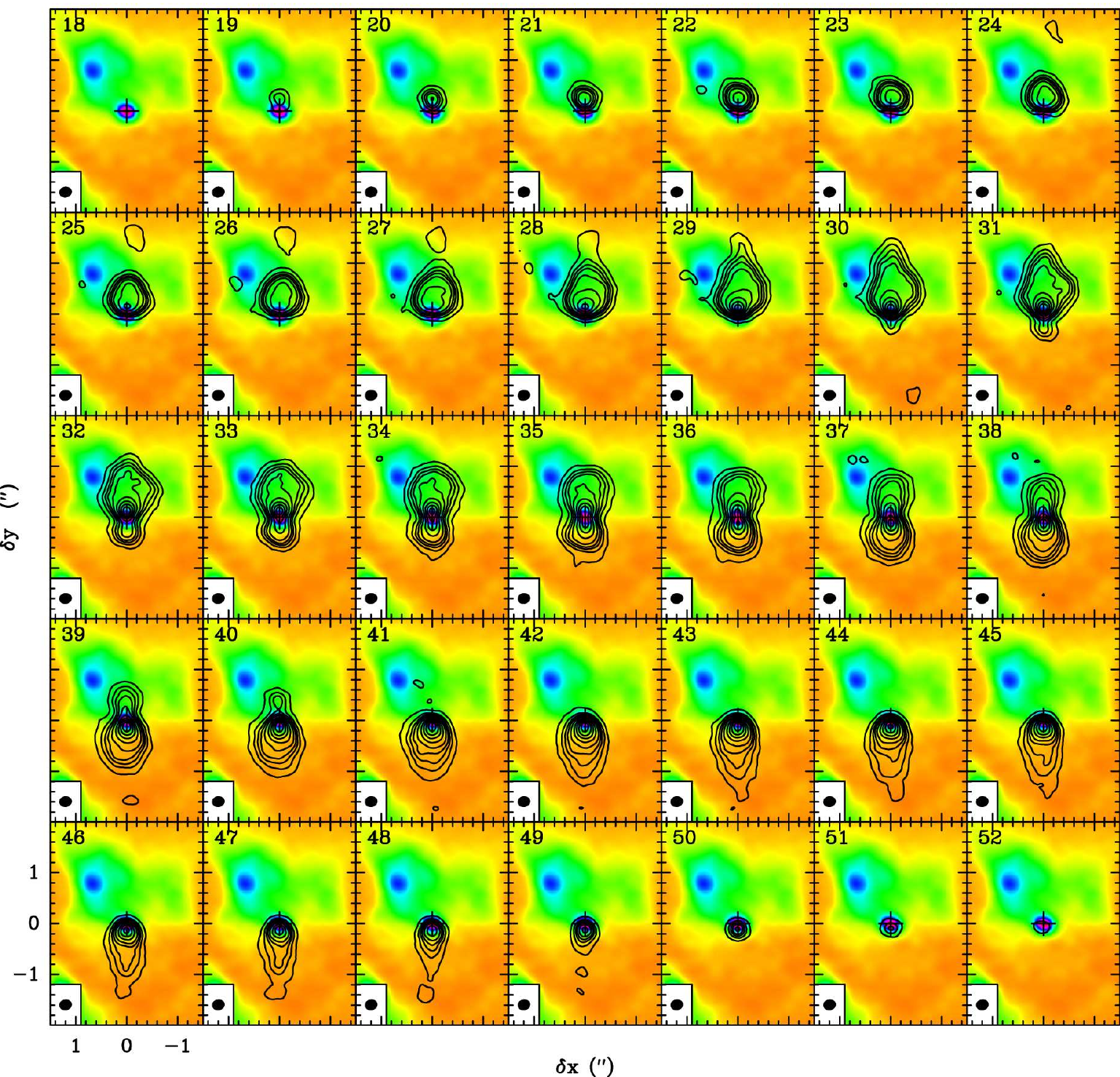}
  \end{subfigure}
  
  \begin{subfigure}[]{}
    \centering  
    \includegraphics[clip=True,viewport=12 325 492 474,height=2in]{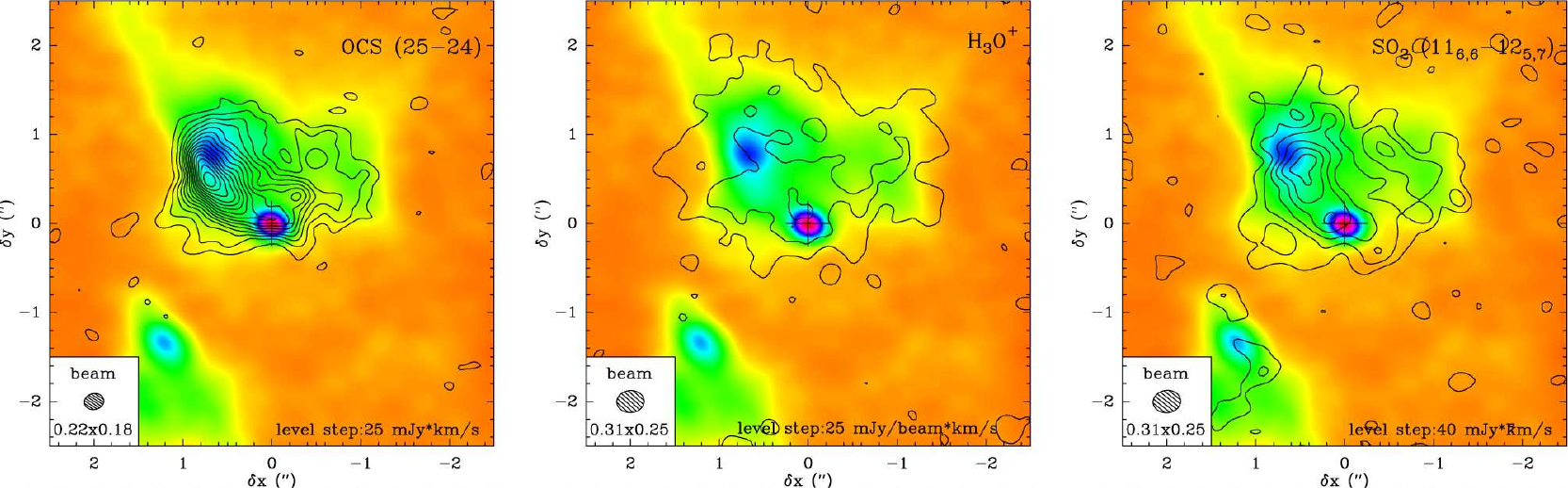}
  \end{subfigure}
  \caption{Channel maps of (a) CO, (b) CS, and (c) SiO, and (d) integrated maps of OCS, H$_3$O$^+$, and SO$_2$ toward OH~231.8+4.2. This figure illustrates the necessity to observe evolved stars in multiple line tracers to understand the chemical and dynamical complexity in the circumstellar environment. Figure from \citet{Sanchez2018}.}
  \label{fig:rotten}
\end{figure}

In addition, the increased line sensitivity of the WSU will make possible the study of relatively low abundance species  \citep[e.g., halides;][]{Danilovich2021} and of isotopologues.  An example of the chemical and dynamical complexity of the environments in post-AGB stars is given by the observation of several species done with ALMA in OH~231.8+4.2 \citep{Sanchez2018}, where it is evident how the conditions for the line excitation and the abundance distribution make different molecules trace different regions (Figure~\ref{fig:rotten}).

With respect to circumstellar dust, massive dust clumps have been detected towards the RSG star VY CMa using ALMA \citep{Kaminski2019a}. The increased continuum sensitivity provided by the WSU will enable mapping of dust distributions towards fainter objects and, in conjunction with gas phase precursor information, provides the prospect of improved knowledge of the dust formation process \citep{Kaminski2019b}.

\subsubsection{Evidence of Companions}

\begin{figure*}[tbh]
   \centering
    \includegraphics[width=12cm]{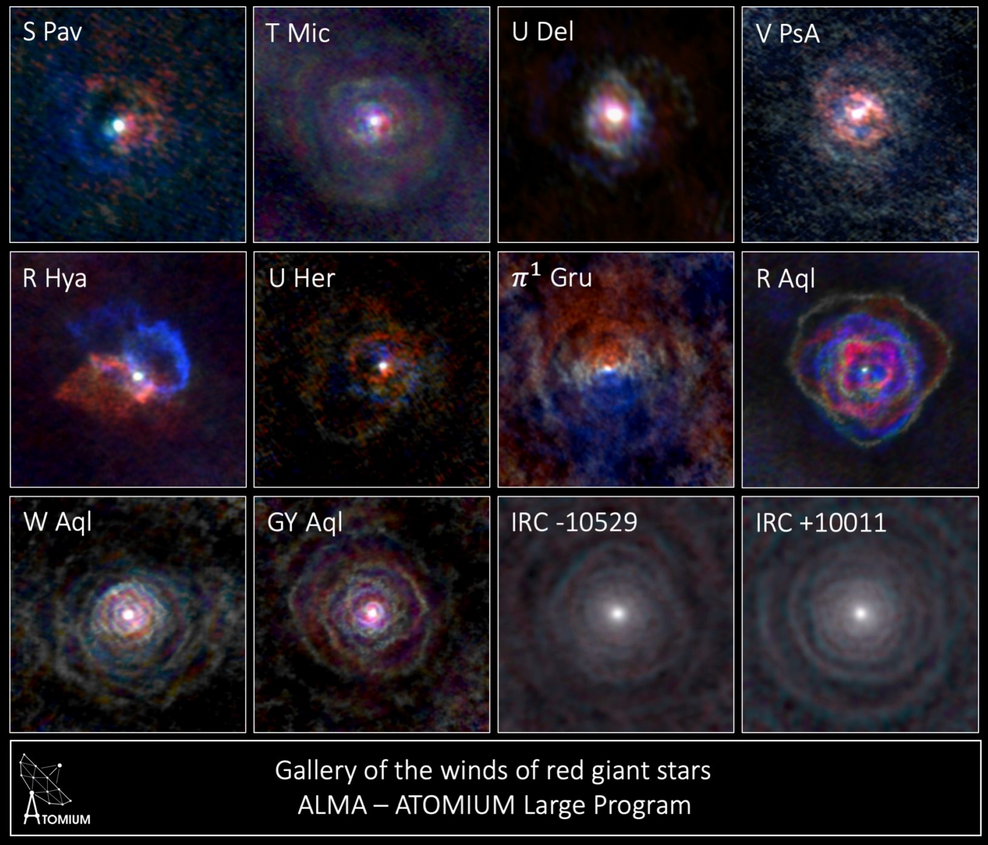}
      \caption{ALMA CO $J=2-1$ observations of winds from AGB stars. Red-shifted emission with respect to the star is shown in red, blue-shifted emission in blue, and emission at the rest velocity in white \citep[Credit: L. Decin – ESO – ALMA; also see][]{Decin2020}. 
              }
         \label{fig:spirals}
   \end{figure*}

Spirals and arcs of CO emission detected using ALMA (see Figure~\ref{fig:spirals}) have been used to infer the presence of companions towards multiple targets \citep[e.g.,][]{Maercker2012,Kim2017,Decin2020,Randall2020}. Towards AGB star L$_2$ Pup, ALMA was also used to make a direct detection of a candidate planet \citep{Kervella2016}. The WSU increased line sensitivity will aid in the detection of molecular signatures of companions and can be used to gain a better understanding of unusual spatial features seen in low abundance molecules, such as the SO emission detected towards GX Mon by \citet{Randall2020}. The WSU increased continuum sensitivity could support the direct detection of further candidate planets in these systems.  \citet{Ramstedt2014,Ramstedt2017,Ramstedt2018} also used ALMA to study known binary systems with different characteristics. 

For AGB stars with initial mass $\ge$ 1.5\,M$_{\odot}$ the expected multiplicity of (sub-)stellar companions is $\ge$ 1 \citep{Decin2020}. The observational signatures of the companions include disks, bipolar outflows and spirals. The interaction with a binary companion may be a dominant factor in shaping the wind of these stars, which then evolve to complex PN morphologies. Over the next decade, ALMA observations of larger samples of AGB stars can help to confirm the incidence and effects of companions. 
Such observations will therefore contribute to the understanding of the long-debated origins of complexity in PN morphologies.

\subsubsection{Post-AGB and RGB Objects}

ALMA has been used to study the structures and chemistry of post-AGB objects such as the Red Rectangle \citep{Bujarrabal2016}, in which the presence of a rotating equatorial disk and an outflow has been confirmed. 
Keplerian disks have been detected so far only in four post-AGB stars \citep{Bujarrabal2018} and display low velocities ($\Delta v\approx5$\kms). The enhancement in sensitivity at high spectral resolution provided by the WSU will be fundamental to hunt for more of such structures, linked to the bipolar shapes commonly seen after the AGB.

\begin{figure*}[h]
\vspace{-3cm}
   \centering
    \includegraphics[width=8cm,clip]{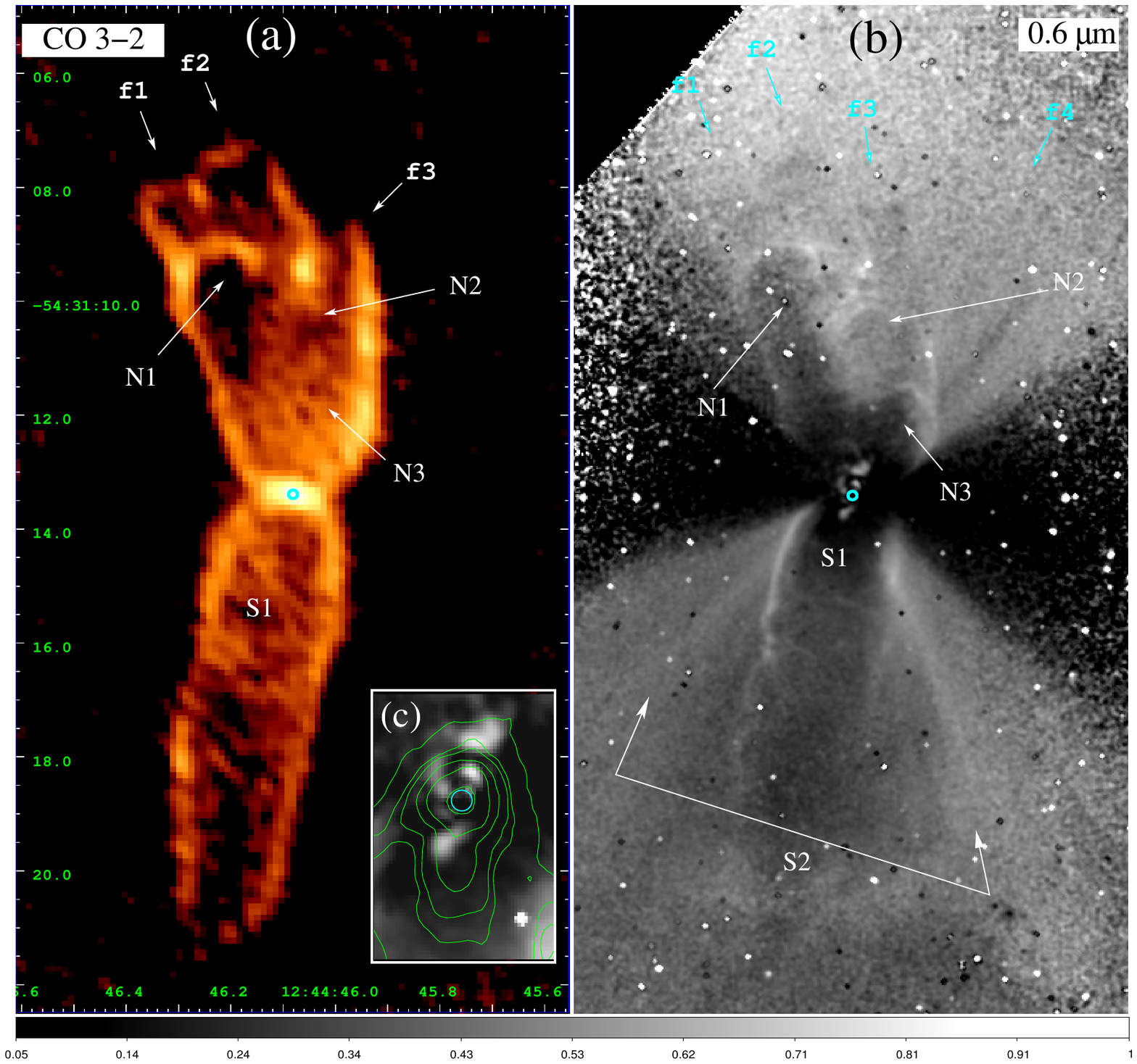}
   \includegraphics[width=8cm,clip]{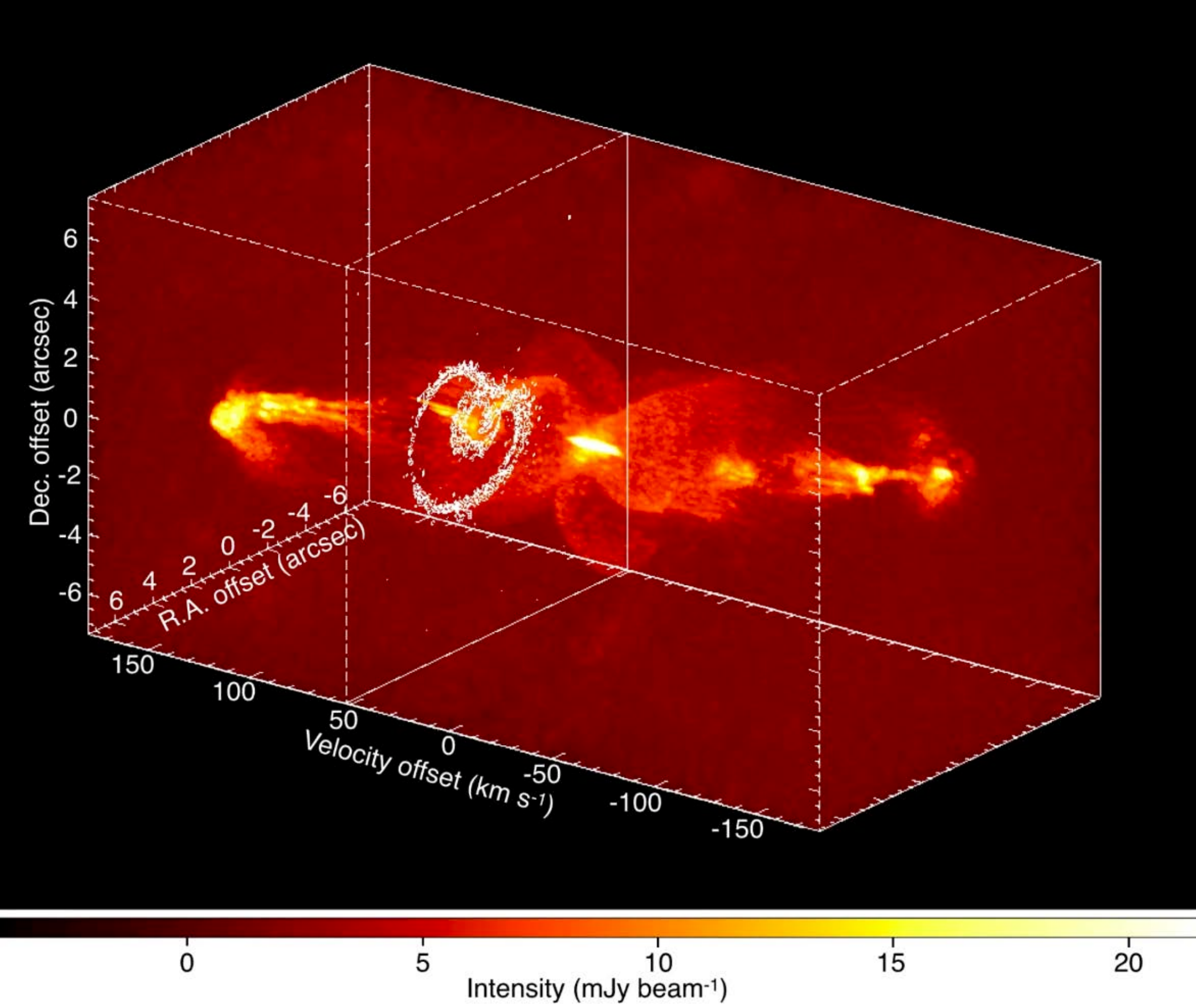}
      \caption{Left: The Boomerang Nebula \citep[reproduced by permission of the AAS]{Sahai2017}. Right: 3D reconstruction of HD~101584 \citep{Olofsson2019}. White contours indicate the CO channel map at that velocity (50\kms). This is one frame of the published movie.
              }
         \label{HD101584}
   \end{figure*}

\citet{Sahai2017} observed the Boomerang Nebula and obtained a detailed view of its highly-collimated inner bipolar nebula (Figure~\ref{HD101584}a). Towards post-RGB HD~101584, \citet{Olofsson2019} used ALMA to perform a 3D reconstruction of the almost pole-on object (right panel in Figure~\ref{HD101584}). Additionally, both oxygen-rich and carbon-rich chemistry was found towards this target. The observations of these complex systems benefit from as much sensitivity as possible for characterization of their spatially-differentiated chemistries. The increased sensitivity and spectral scan speed of the WSU will therefore help to provide a better understanding of the post-AGB and RGB evolutionary phases.

\subsubsection{Stellar Surfaces and Rotation}

\begin{figure*}[tbh]
   \centering
    \raisebox{0.3\height}{\includegraphics[width=4.7cm]{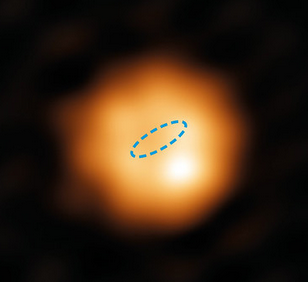}}
     \includegraphics[width=5.25cm]{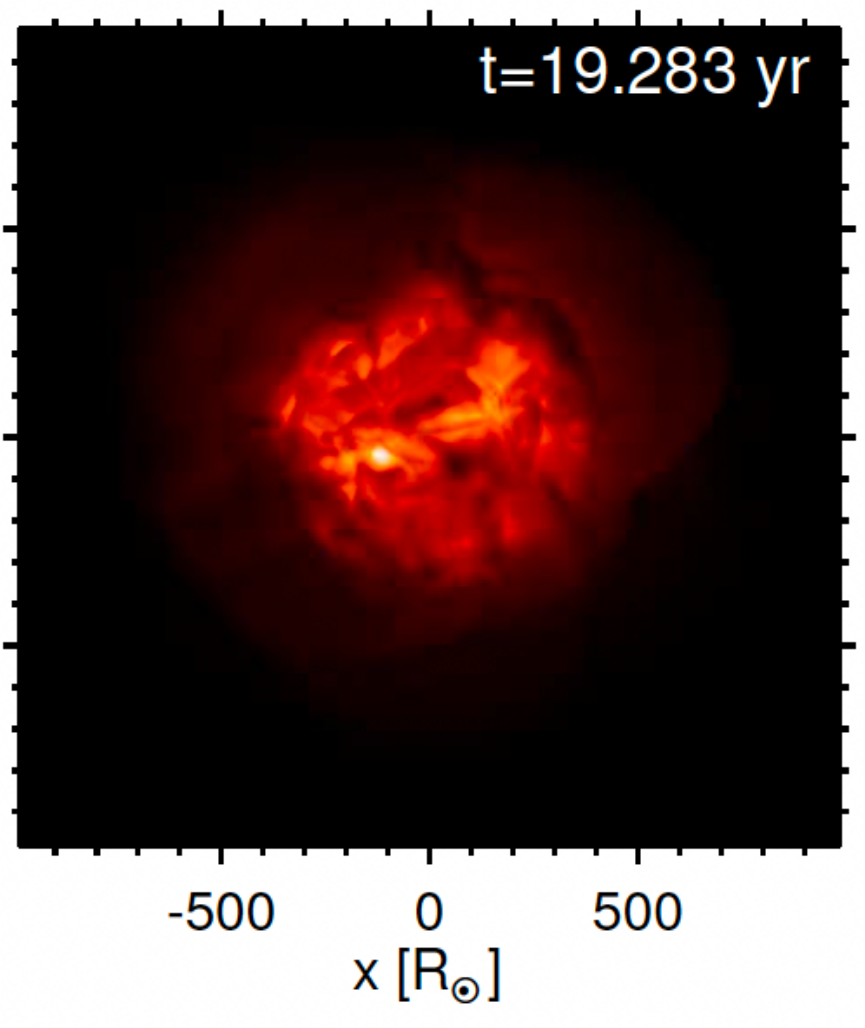}
      \caption{Left: Brightness temperature of the surface of W Hya derived from ALMA observations \citep{Vlemmings2017}. The dotted ring shows the size of the Earth's orbit. Credit: ALMA (ESO/NAOJ/NRAO)/W. Vlemmings. Right: Surface intensity distribution from a 3D-model \citep{Hoefner2019}.
              }
         \label{fig:hotspot}
   \end{figure*}

ALMA observations have detected hotspots on the surfaces of AGB stars; e.g., Mira \citep{Matthews2015,Vlemmings2015,Wong2016} and W Hya \citep[see Figure~\ref{fig:hotspot};][]{Vlemmings2017}. The hotpots are predicted by 3D stellar models and may be implicated in asymmetry of material in the inner CSE \citep{Hoefner2019}.
The increase in instantaneous bandwidth and continuum sensitivity provided by the WSU will enable faster imaging of stellar surfaces. This is important since a key aspect to be constrained observationally remains the lifetime of the features. Multi-epoch observations, while ALMA is in long baseline configurations needed to spatially resolve the central star, can therefore  be achieved. While the continuum from such observations can probe the stellar surface, the spectral lines probe the extended atmosphere and inner CSE. Indeed, rotation of material close ($<$ 2\,R$_*$) to the stellar surface has also been detected using ALMA \citep[e.g.,][]{Vlemmings2018rotation}.

%% file: origins_galaxies.tex
\section{Origins of Galaxies}
\label{sec:galaxies}

During its first 10 years of operation,
ALMA has unveiled a variety of physical phenomena in the early universe. With its superb sensitivity, ALMA has detected gas in normal and bright galaxies across nearly the full history of the universe, from the recent universe ($z< 1$), cosmic noon ($z=1\text{--}3$), phase of rapid growth ($z=3\text{--}6$), the cosmic dawn ($z > 6$), and as early as $z \approx 9$ when the universe was less than a billion years old. The baryonic matter found in these systems represents the cold and diffuse gas of the interstellar medium, possibly used as the fuel for Active Galactic Nuclei (AGNs) or the seed for future star formation across the galactic disk. The high spatial and spectral resolution capabilities of ALMA are now revealing the structure and kinematics of the gas, allowing astronomers to study galactic rotation, large-scale inflow and outflow, molecular/dynamical mass, and physical properties in exquisite detail.

\begin{figure}[h]
\centering
\includegraphics[width=15cm]{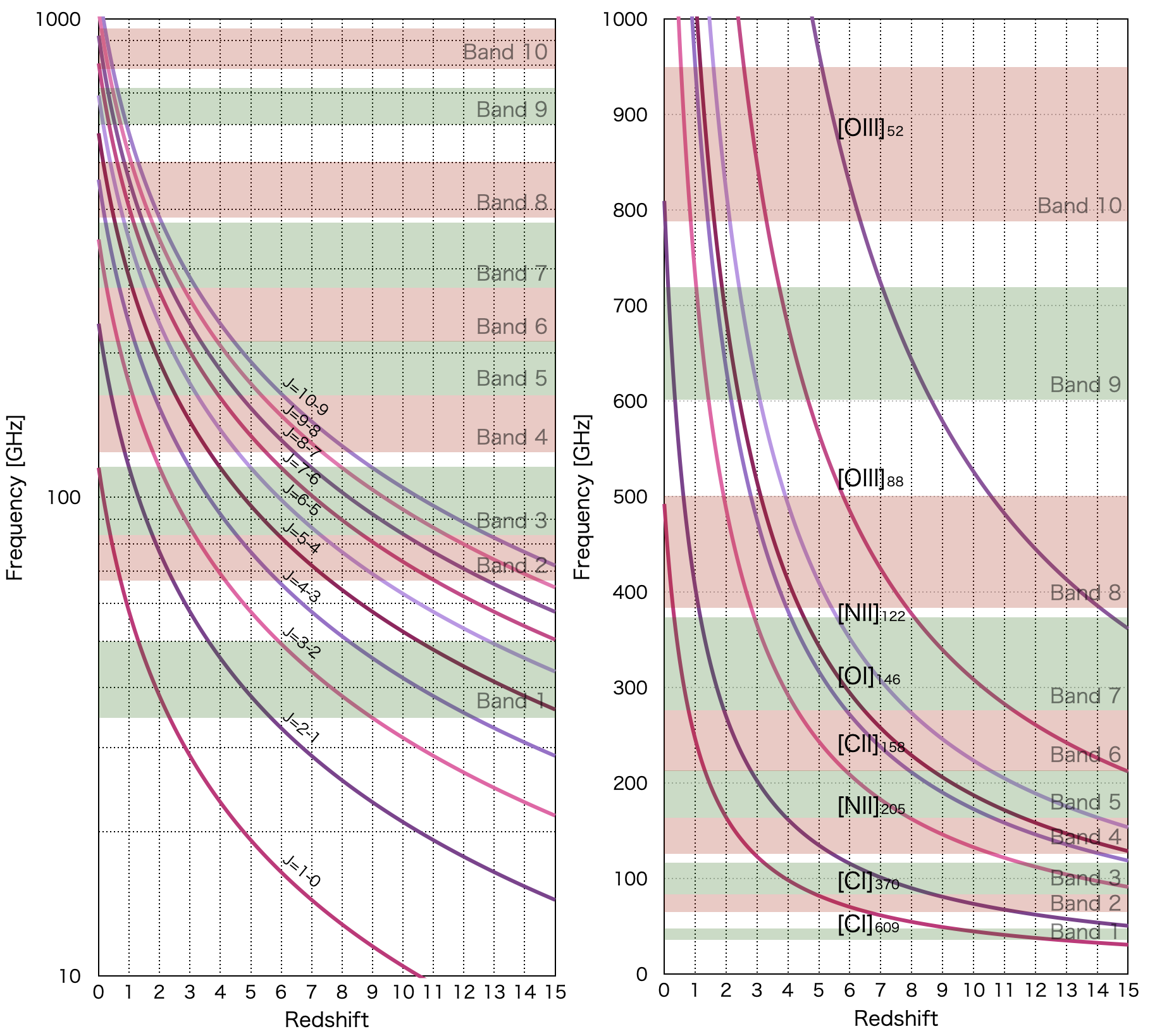}
\caption{(Left) The frequency vs.\ redshift relation for the CO lines, showing here the rotational transitions from $J=1-0$ to $J=10-9$. (Right) The same relation for notable atomic lines.}
\label{redshifted_lines}
\end{figure}

The frequency coverage of ALMA is ideal for detecting and imaging redshifted atomic lines such as carbon, oxygen and nitrogen, as well as molecular lines such as rotational transitions of carbon monoxide. Figure~\ref{redshifted_lines} illustrates the relation between the ALMA receiver bands and various redshifted spectral lines. The singly ionized carbon line (\cii), which occurs at a rest frame wavelength of 158\,\micron, is an important coolant that traces the cold interstellar medium in galaxies. It is known that the global \cii\ luminosity and star formation rate are correlated in local star forming galaxies \citep{Delooze11} and in distant sources to some degree \citep[e.g.,][]{Smit18,Harikane20}. The \oiii\ line at 88\,\micron\/ is often utilized to detect galaxies at $z > 6$ using ALMA Bands 7 and 8. In fact, the \oiii\ line appears brighter than \cii\ in lower mass galaxies (e.g., Lyman Break Galaxies, Lyman Alpha Emitters) by a factor of a few to 10, whereas the apparent flux ratio is comparable in more massive dusty star forming galaxies \citep{Bakx20}. The variation in the flux ratio may be related to the intrinsic properties of the host galaxy \citep{Hashimoto19}. Finally, the combination of the \oiii\ line at 52\,\micron, which is observable at ALMA Bands 8, 9 and 10 for sources at $5 < z < 15$, and the \oiii\ 
5007\,\AA\/ line observable with the JWST, can be a valuable tool for robust measurements of the metallicity in the early universe \citep{Jones20}. Table~\ref{tbl:redshifted_lines} provides a summary of the atomic and CO lines observable by ALMA at each epoch.  

Continuum imaging, produced by using the line-free channels in an ALMA data set, provides a strong complement to the line emission studies noted above, and helps complete the astrophysical puzzle. At (sub)millimeter wavelengths, the continuum emission is often dominated by reprocessed emission from dust at temperatures $T \sim 25$\,K.  Measurements of the dust thermal emission allows us to investigate, for example, the spatial distribution of star formation \citep[e.g.,][]{2020ApJ...901...74T}, derive the dust and gas masses \citep[e.g.,][]{2017ApJ...837..150S}, compile number counts \citep[e.g.,][]{2016ApJ...833...68A,2016ApJS..222....1F,2018ApJ...869...71Z,2022A&A...658A..43G}, and conduct clustering analyses through images obtained via wide field surveys \citep[e.g.,][]{2021MNRAS.504..172S}.  Massive environments (groups and clusters of galaxies) often show signatures of the Sunyaev-Zeldovich (SZ) effect \citep{1972CoASP...4..173S}, a broadband, diffuse signal due to hot ($T \gtrsim 10^6$\,K) ionized gas, discussed further in Section~\ref{subsec:continuum}.  Notably, broad spectral coverage across the ALMA bands is often required to separate the broadband continuum spectral components such as radio synchrotron and thermal dust emission from the SZ effect, and in doing so an upgraded ALMA will simultaneously better probe the coldest and hottest gas in large scale and distant structures, in addition to probing the warm atomic and weakly-ionized gas through \cii\ and \oiii\ line emission.

As highlighted in Section~\ref{sec:SC}, the WSU will significantly improve ALMA's ability to detect distant gas and dust, expediting surveys that were perceived difficult due to the highly competitive telescope time.  Here we elaborate on a few particular high redshift science cases that will significantly benefit from the WSU.
In the era of JWST, the Vera Rubin Observatory, Nancy Grace Roman Space Telescope (NGRST), Simons Observatory \citep{Ade2019_Simons}, Fred Young Submm Telescope (FYST, \citealt{CCAT-prime2021}), CMB-S4 \citep{Abazajian2019_CMB-S4}, and SKA, as well as potentially AtLAST, DSA-2000, and ngVLA, the number of candidate galaxies at the redshift of formation requiring ALMA follow-up will grow substantially, motivating the need for ways to search efficiently for line and continuum emission \citep[e.g.,][observed for 4 hours total, using 4 tunings and provide a tentative line detection]{Bakx2022}.
While \cite{Reuter20} presented results for 81 bright and, in most cases, highly magnified dusty star forming galaxies (DSFGs) detected using South Pole Telescope (SPT) data, both SPT and the Atacama Cosmology Telescope (ACT) have seen substantial upgrades since this study, and currently the number of similarly bright, highly magnified sources may approach 1000 (private communication).  Further, as the number of sources detected is a strong function of the flux limit of a survey, the predictions for Simons Observatory and CMB-S4 range from 10,000 to 100,000, with the latter expecting 10,000 strongly lensed DSFGs in the mm-wave regime \citep{Abazajian2019_CMB-S4}.  At the same time, FYST is expected to find 630,000 DSFGs, with 1200 of them at $z=5\text{--}8$ (i.e., during the epoch of reionization; see \citealt{CCAT-prime2021}).
Neglecting additional gains expected for improved receiver noise temperatures, the expected factor of 3--6 improvement in continuum imaging speed (see Section~\ref{sec:sensitivity} and Table~\ref{tbl:ICMode}) will facilitate the follow-up necessary to understand their molecular and dust compositions, star formation rates, and dynamical properties (i.e., for total mass constraints). 

\input{table_redshifted_lines}

\subsection{Unbiased Spectroscopic Redshift Surveys}

\begin{figure}
\centering
\includegraphics[width=15cm]{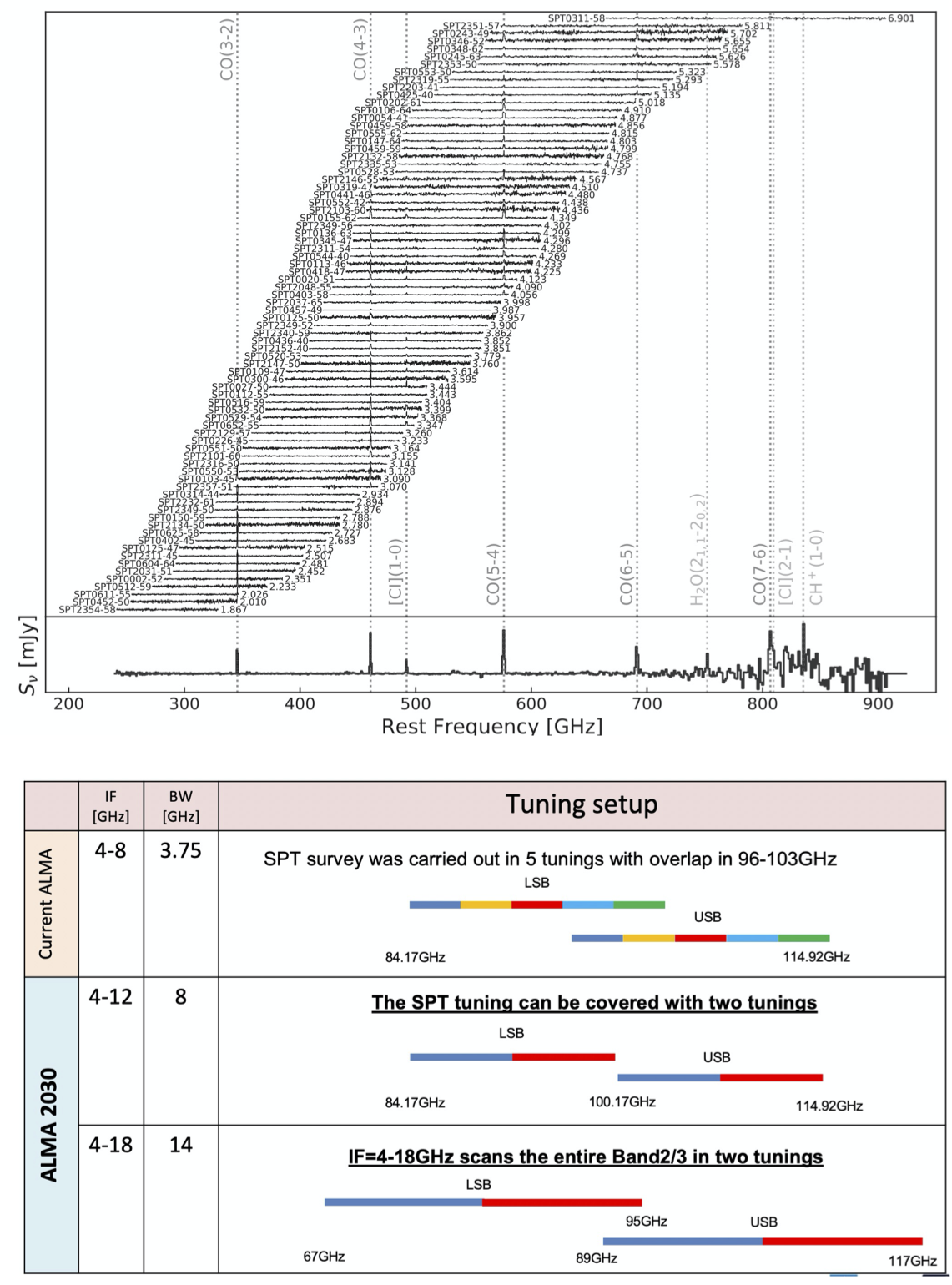}
\caption{Top: Redshift determination survey of 81 lensed sources discovered at the South Pole Telescope \citep[reproduced by permission of the AAS]{Reuter20}. Bottom: Illustration of the tuning setup adopted by the unbiased redshift survey of the SPT sources. 
The WSU will provide at least 3.6$\times$ faster survey speed for achieving the same sensitivity and bandwidth. }
\label{reuter_fig3}
\end{figure}

Precise redshift determination of distant dust obscured galaxies is often non-trivial as optical/UV light can be hampered by the large obscuration even in the deepest images. Interstellar dust is generally transparent at the millimeter/submillimeter frequencies where the important molecular rotational transitions and atomic emission lines are available. Even  without prior knowledge of the source photometric redshift, one can derive a precise redshift by detecting for example two or more consecutive rotational transitions of the CO line, and by measuring the frequency difference between the two lines; e.g., $z=115.27/\Delta \nu - 1$, where $\Delta \nu$ is the frequency separation in GHz.  Detecting even one CO transition or atomic line (e.g., \ci, \cii, \oiii) with other complementary information such as photometric redshift (discussed in Section \ref{sec:photoz}), optical spectroscopy, or another atomic lines in the mm/submm can yield a redshift with high confidence.  Unbiased surveys require a thorough spectroscopic scan over a wide frequency range and require a significant amount of telescope time. Up to the present, extremely bright sources or strongly lensed dusty submillimeter galaxies have been the main targets for redshift surveys using ALMA. As an example, a successful study was conducted toward lensed dusty star forming galaxies identified by the South Pole Telescope (SPT), yielding unambiguous redshifts in 81 distant galaxies by scanning the entire 85--115\,GHz range of Band 3 \citep[see Figure~\ref{reuter_fig3};][]{Reuter20}. The advantage of observing strongly magnified sources is that their CO lines are bright (10--20\,mJy at the peak), giving solid detection in a series of quick ($\sim30$\,min/source) integration time. With the higher observing efficiency realized by the WSU, the same SPT observations, which required five tunings in Band 3, can be covered in just two tuning setups, yielding observations that are at least 3.6$\times$ faster (including the factor 1.44 improvement in the digitizer and correlator efficiencies: see Section~\ref{sec:sensitivity} and Table~\ref{tbl:ICMode}). This is a significant speed improvement for redshift surveys, allowing larger surveys (and hence better statistics) of the fainter but more abundant star forming galaxy population in the early universe  \citep[e.g.,][]{Chen22}.

\subsection{Spectroscopic Confirmation of High-$z$ Galaxies with Photometric Redshift Estimates}\label{sec:photoz}

Analyzing multi-band photometry in conjunction with SED models allows the derivation of the photometric redshift of a distant galaxy. A wide range of photometry is usually available in optical/UV selected galaxies, providing a rough a priori guess of the relevant tuning frequency and the molecular/atomic emission line that falls into the corresponding ALMA band. However, photometric redshifts often contain relatively large uncertainties of a $z\sim$ few $\times$ 0.1, requiring ALMA to observe multiple tuning setups to ensure coverage of the full probable redshift range for a solid line detection, and thus securing the redshift. 
At $z > 6$, the high-J CO lines are available in Bands 3 and 4, but the brightness of high-J CO lines is often difficult to predict, especially for normal main sequence galaxies.
In contrast, atomic lines (notably \cii\/ and \oiii) are reliably bright and are good tracers of the gas in the ISM.
The \cii\/ line is redshifted to Band 6 (or lower) and the \oiii\/ line at $88\,\mu$m is redshifted to Band 8 (or lower). Recent ALMA observations have successfully demonstrated the utility of these lines for securing redshifts when the information is limited to photometric data.  \cite{Hashimoto18} employed a spectral setup that encompasses the frequency range of 314.4--340.5\,GHz which corresponds to $z=9.0$--9.8 for the redshifted \oiii\/ line, successfully detecting the \oiii\ line and a solid confirmation of $z = 9.1096 \pm 0.0006$ in MACS1149-JD. A spatially resolved [OIII] image of this source was subsequently obtained, finding tentative evidence of systematic rotation across the disk \citep{Tokuoka22}. \cite{Tamura19} adopted a similar strategy, securing the redshift of a Lyman break galaxy (LBG) MACS0416\_Y1 to be $z =8.13118 \pm 0.0003$. More recently, \citet{Harikane22} used four tuning setups to cover the 222--250\,GHz range of Band 6 to search for the \oiii\/ line from an \textit{H}-band dropout galaxy whose redshift is estimated to be $z\sim$12--13 based on their SED analysis.  They obtained a tentative $4\sigma$ spectral signature at 237.8\,GHz, which corresponds to $z=13.3$ if the feature is indeed tracing the \oiii\/ line. The broader IF bandwidth realised with the WSU will allow us to conduct the same spectral scan in Band 6 using only half the tuning setups, and 
reaching a comparable noise level at least 4.5$\times$ faster than currently possible (including the factor 2.25 from the improvement in the Band 6 spectral line speed (see Appendix~\ref{sec:Band6}).
Alternatively, by analyzing newly obtained Band 4 observations to search for \cii\ as well as reanalyzing the previous Band 6 data, \cite{Kaasinen2022} showed that it may be productive to pursue more statistically significant detections ($>$2$\times$) using similar integration times.
A wideband receiver for Band 4 will be critical to capture the \cii\/ line for high-$z$ galaxy candidates that could be located at $11 \lesssim z \lesssim 14$. 

\begin{figure}
\centering
\includegraphics[width=15cm]{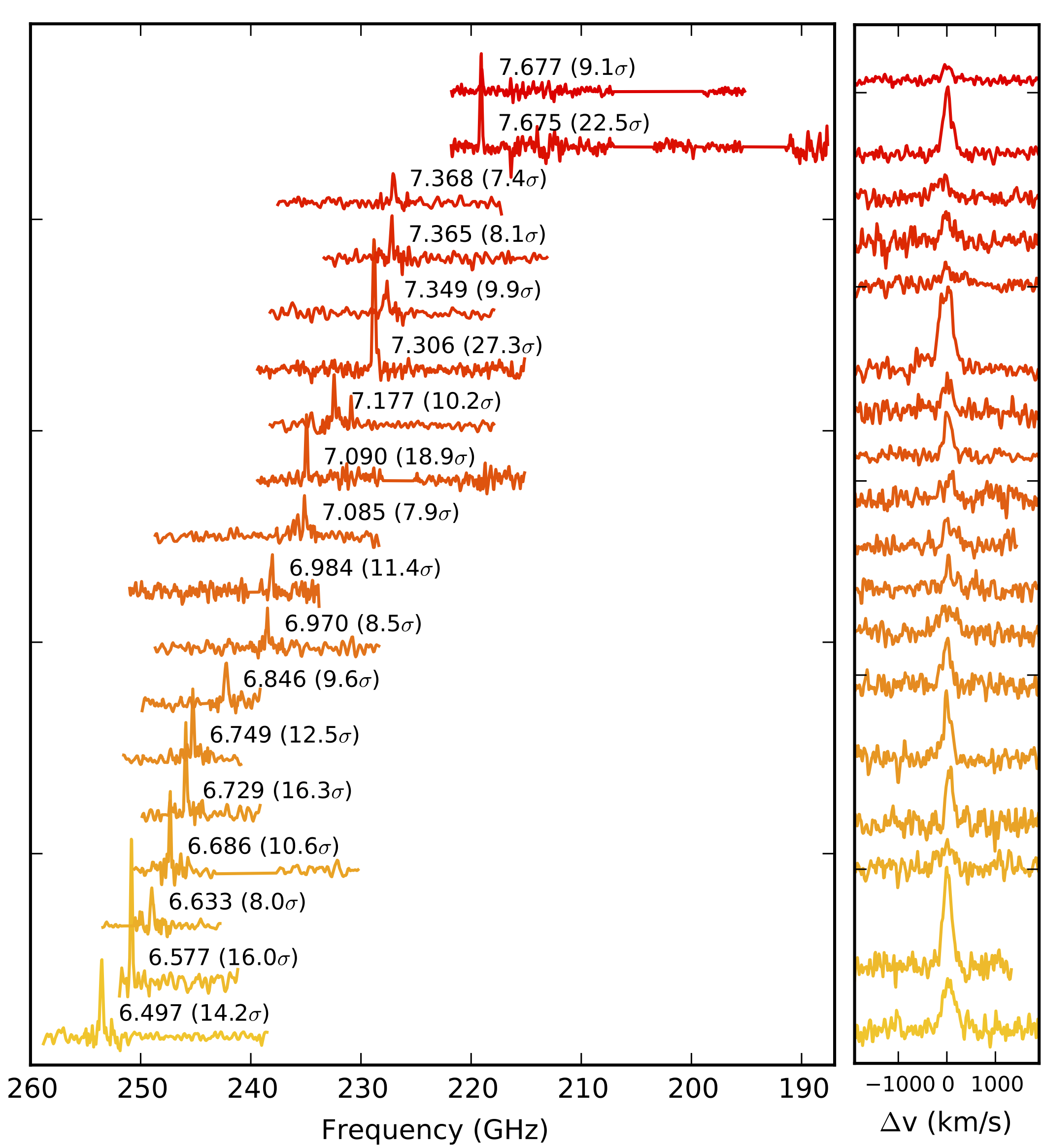}
\caption{\cii\ emission from $6.5 < z < 7.7$ obtained in the REBELS Large Program \citep{Bouwens22}.
}
\label{fig:bouwens_fig9}
\end{figure}

The Reionization Era Bright Emission Line Survey (REBELS) large program observed the \cii\/ and \oiii\/ lines in 40 UV-selected galaxies at $z > 6.5$, with the goal to generate a statistically significant sample of \cii/\oiii\/ emitting galaxies at the epoch of reionization (\citealt{Bouwens22}; see Figure~\ref{fig:bouwens_fig9}). With the current ALMA system, which allows an IF bandwidth of 4 to 5.5\,GHz (for Bands 3 to 8), a number of tunings are often necessary to thoroughly cover the probable frequency range estimated through photometric redshifts. For example, the REBELS program required one to eight tunings per source, depending on the redshift likelihood distributions derived based on the photometric redshift estimates. A total of 91 tunings for the \cii\ line and 22 for the \oiii\ lines were required for this study. A summary of the number of tunings and the corresponding number of REBELS targets can be found in Table~\ref{tbl:rebels_tuning}. 
The WSU will enable the same project with only $\sim60$\% of the tunings, which when combined with the improvement in spectral sensitivity from the Band 6 upgrade and digital improvements, would reduce the 70-hour program to 21 hours (i.e., only 30\% of the original time required).

\input{table_rebels}

\subsection{High Redshift QSO Outflow}
\label{sec:QSOs}

\begin{figure}
\centering
\includegraphics[width=11cm]{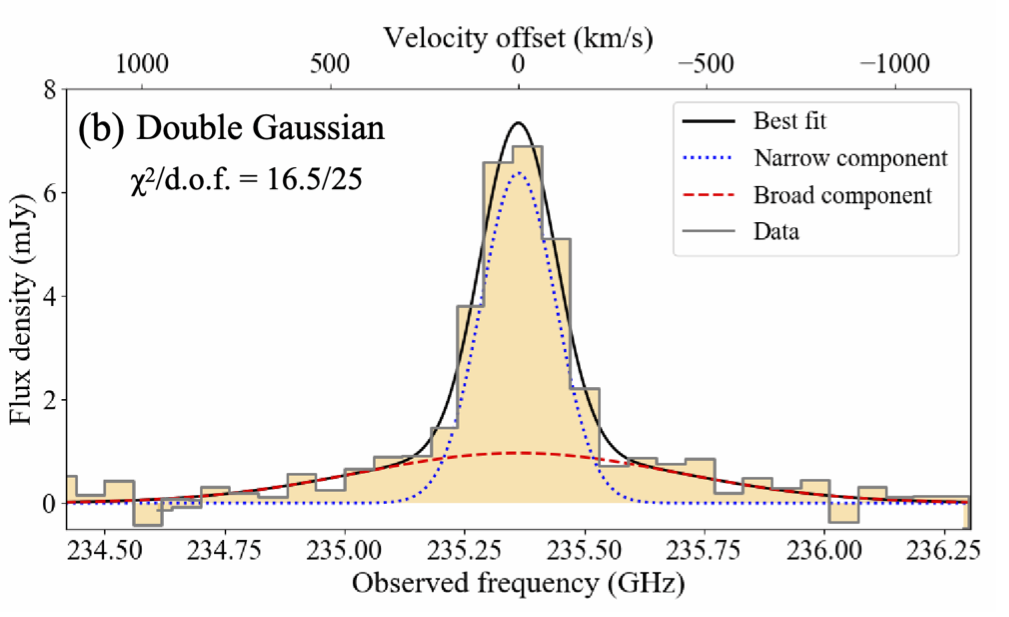}
\caption{Fast large-scale \cii\ gas outflow from a $z=7.07$ quasar \citep{Izumi21}. Reproduced by permission of the AAS.
}
\label{izumi_fig4}
\end{figure}

QSO outflows can significantly impact the intervening gas in the interstellar medium, and is an important mechanism to regulate star formation in early galaxies.
Large scale outflows may hold important clues to understand the known relation between the black hole mass and bulge mass \citep{Magorrian98,2013ARA&A..51..511K}, which may ultimately be related to solidifying our understanding of the galaxy-black hole co-evolution scenario.
Outflows can be very fast and broad, reaching the intergalactic medium with FWZI $\sim 2000$\kms\/ \citep[e.g.,][]{2022MNRAS.511.1462M}. Observing the redshifted molecular or atomic gas from the outflows is one of the essential ways to quantify and characterize the properties of the outflow; e.g., \citealt[see Figure~\ref{izumi_fig4}]{Izumi21, Bischetti21}.
One of the difficulties in capturing the broad outflow gas in a limited bandwidth is the associated risk of partially or completely missing the line in a single observational setup due to the uncertainty in the redshift. If the precise redshift is known a priori through spectroscopic techniques, this risk may not be so high. However, if the redshift is not known precisely, especially if based on photometric redshifts, it will be difficult to distinguish if a non-detection is purely due to lack of detectable gas in the host galaxy or simply due to an incorrect tuning. This problem is amplified at mid to high frequency bands. The WSU will significantly reduce this risk by simultaneous observations of a wide (8--16\,GHz) IF bandwidth which corresponds to $> 10,000$\kms\/ at Band 6 where the redshifted \cii\/ line from $z > 7$ sources can be detected. 

Similarly, the goal of 16\,GHz IF bandwidth will probe $> 5500$\kms\/ in velocity space for the crucial, bright \oiii\/ 88 and 52\,$\mu$m lines in Bands 7--10 (Figure~\ref{redshifted_lines}).
Perhaps more importantly, lines in Bands 8, 9, and 10 that are broadened by more than 2000\kms\/ have so far been difficult to measure with ALMA, as the current bandwidth limits the simultaneous measurement of the broadened lines and the continuum spectral baseline (i.e., for subtraction).  The ability to probe accurately both the line and continuum in a single sideband will open new possibilities for imaging higher frequency transitions and ionization states in such winds at the spatial resolution and sensitivity of ALMA.

Table~\ref{tbl:band_velocity} summarizes the corresponding velocity widths for different IF bandwidths. With the $4\times$ bandwidth WSU, the velocity coverage of a single sideband will vary between 5500\kms\/ (Band 10) to 112,900\kms\/ (Band 1). Thus the ability to trace large-velocity outflows will be improved significantly in all receiver bands. Appendix~\ref{sec:receivers} discusses the  upcoming and potential upgrades to the ALMA receivers.

\subsection{Cluster Membership}

Galaxy clusters are dynamically evolving entities in the early universe and important sites of active star formation and galaxy evolution.  Not only do galaxies impact their surroundings through AGN feedback, but their large scale environments impact their evolution.
Galaxies in proto-clusters, for instance, are expected to undergo multiple mergers and interactions with other members in the cluster as well as their nascent hot atmospheres, resulting in the dominance of quiescent early-type galaxies in mature clusters we see today \citep{2016A&ARv..24...14O}.
Characterizing the properties of the molecular gas in proto-clusters allows us to probe, for example, the gas mass, star formation efficiency, gas mass fraction, size of the cold ISM, and physical properties such as temperature and density, all of which are highly  critical physical information for better understanding the formation process of galaxies and the role of large scale environment \citep[e.g.,][]{2018ApJ...856...72O,2018Natur.556..469M,2019ApJ...870...56N,2022ApJ...933...11I}. The difference between the maximum and minimum redshift (recession velocity) of members in the proto-cluster can be $\Delta z \sim 0.03$ (or $\sim 9000$\kms; see e.g., \citealt{Hayashi18}). The current ALMA bandwidth is sufficiently broad if robust redshifts are determined a priori via other means such as optical spectroscopy. However, if exact redshifts are unknown, even with  photometric redshifts, the current bandwidth can be a problem especially at mid to high frequencies (see Table~\ref{tbl:band_velocity} for the velocity coverage of current ALMA), introducing a non-negligible risk that the emission line falls outside the band in a single scan. With the 2--4$\times$ expansion of the IF bandwidth, the chances of missing the line is significantly reduced, allowing a robust characterization of the molecular gas in all members associated with the cluster.

\input{table_origins_galaxies_2}

\subsection{Millimeter/submillimeter Continuum}\label{subsec:continuum}

The mm/submm continuum obtained by ALMA allows us to study important quantities such as massive star formation rate, dust opacity and mass, and the cold gas mass.  The cold gas mass is often derived by utilizing the conversion factor from Band 6 or 7 continuum flux density to molecular gas mass \citep{Scoville16}. Negative k-correction allows efficient observations of galaxies beyond $z \sim 1$, allowing large multi-field continuum surveys for statistical analysis such as number counts, clustering, and multi-wavelength analysis. Results from successful targeted and wide field continuum surveys using ALMA are now available in the literature \citep[e.g.,][]{Walter16,Hatsukade18,Franco18,Kohno19,Casey21,Shivaei22}. 
In \cite{Shivaei22} in particular, they found stacking of a handful of galaxies led to a robust detection, meaning a factor of 3--6$\times$ improvement could allow for direct, individual detections for the same time invested.
Continuum surveys after the WSU will be at least 3$\times$ faster (and more than 6$\times$ faster with the 4$\times$ bandwidth upgrade), enabling more comprehensive surveys of galaxies and deep field surveys than currently possible, possibly allowing for better control for the amount of cosmic variance affecting small area surveys. 

Another aspect of large scale structures such as galaxy groups, galaxy clusters, and proto-clusters that has been only been explored in a handful of ALMA 12-m and 7-m Array studies \citep[e.g.,][]{Kitayama2016, Basu2016, Ueda2018, DiMascolo2019, DiMascolo2019b, DiMascolo2020, DiMascolo2021} is the thermal SZ effect\footnote{See \citet{2019SSRv..215...17M} for a review of the theory and application in high-resolution, subarcminute scale observations.} \citep{1970Ap&SS...7....3S,1972CoASP...4..173S}.  The thermal SZ effect probes the pressure or thermal energy in hot ionized electrons in plasmas at temperatures $\sim 10^6$--$10^8$\,K, yielding an entirely complementary measurement of the hot gas (the dominate baryonic component) bound in the dark matter dominated potential of large scale structures, such as those listed above.  
The thermal SZ effect is a decrement at frequencies $\nu \lesssim 218$\,GHz, and becomes an increment at higher frequencies (i.e., $\nu \gtrsim 218$\,GHz). The exact frequency where the decrement crosses the null to become an increment depends weakly on the plasma temperature, while the ability to probe the overall spectrum will place constraints on the temperature.
A unique and salient feature of the SZ effect is that it has redshift-independent surface brightness, meaning that at the sensitivity of an upgraded ALMA, studies of the SZ effect can probe the hot gas in very high redshift structures that would be prohibitively long to observe in more traditional X-ray observations.

Low resolution SZ surveys in ALMA Bands 3, 4, and 6 have found thousands of massive galaxy clusters at $z<2$, which have SZ fluxes of order few to tens of mJy at 2--10\,mm wavelengths.  
Through stacking (often of tens of thousands of objects), SZ surveys have also probed massive galaxies ($\sim 10^{12}\,\rm{M}_\odot$) and low-mass ($\sim 10^{13}\,\rm{M}_\odot$) group scales.  However, as noted in \cite{2019SSRv..215...17M, Mroczkowski2019}, currently only ALMA has the sensitivity and resolution to image directly such lower mass structures, owing to their smaller signals (typically tens to hundreds of $\mu$Jy).  The wider IF bandwidth of Band 2 (see Figure~\ref{fig:Band2}) coupled with the WSU upgrades will improve the continuum imaging speed for SZ studies by at least a factor of 3$\times$ (for 2$\times$ bandwidth, including correlator and digitizer efficiency improvements).   For 4$\times$ bandwidth, the imaging speed improvement will be 6$\times$, while the improved receiver sensitivity will further help in both cases.

Finally, there is another flavor of SZ effect: the kinematic SZ effect \citep[kSZ;][]{1980MNRAS.190..413S}, which is a Doppler shift in the CMB due to the line-of-sight gas momentum.  A few studies have reported detections of the kSZ signal from, for example, quasar-driven winds  \citep{2019MNRAS.483L..22L, 2019MNRAS.490.5134B}, though it is clear that further multi- and wide-band studies are necessary for confirmation \citep{2021MNRAS.508.5259M}.
Wider Band 6 (see Appendix~\ref{sec:Band6}) measurements near the null in the tSZ effect will further aid in the component separation to allow cleaner measurements of both the kSZ and tSZ.

\subsection{Direct Imaging of Black Holes}
\label{sec:EHT}

The  Event Horizon Telescope (EHT)\footnote{\url{https://eventhorizontelescope.org/}}  has produced spectacular results in imaging the shadow of the supermassive black holes at the center of  M87 \citep{EHT2019} and the Milky Way \citep{EHT2022}. ALMA's participation in the EHT has been crucial, 
providing the largest collecting area and being situated at the best site in terms of both weather conditions and $uv$-coverage \citep{Goddi2019}. Over the next several years, the EHT will undergo extensive upgrades to provide wider bandwidths and dual-band operations, becoming the next generation Event Horizon Telescope (ngEHT).\footnote{\url{https://www.ngeht.org/technology}} 
The anticipated bandwidth of the ngEHT is 8\,GHz per sideband per polarization, well-matched to the initial WSU goal of 2$\times$ bandwidth, while the subarray capabilities of the next generation ALMA correlator will allow dual-band operations (see Appendix~\ref{sec:correlator}), allowing ALMA to continue to play a crucial, central role in the future. 

%% file: table_redshifted_lines.tex
\newcommand{\ispacestart}{0.1cm} 
\newcommand{\ispaceend}{0.1cm} 
\newcommand{\bspace}{-0.3cm} 
\newcommand{\miniwidth}{12.5cm} 
\newcommand{\leftmarginspace}{0.3cm} 

\begin{table}[ht]
\centering
\caption{Spectral line diagnostics at each cosmic epoch}
\begin{NiceTabular}{X[l]m{12.5cm}}[hvlines]
\CodeBefore
\columncolor{\headercolor}{1}
\Body
\Block{1-1}{\vspace*{\bspace}{\bf Cosmic dawn} \\ (Epoch of Reionization) \\ $\bm{z>6}$} & 
\begin{minipage}{\miniwidth}
\begin{itemize}[leftmargin=\leftmarginspace]
\vspace*{\ispacestart}
\item High-J CO ($\mathrm{J_{up}\geq6}$) in Bands 3--4, mid-J CO ($\mathrm{J_{up}=3\text{--}5}$) in Bands 1--2
\item \cii\/ in Band 6 for $6<z<8$, Band 5 for $8<z<10$, lower bands for $z>10$
\item \oiii\/ (88\,$\mu$m) in Bands 7--8, \oiii\/ (52\,$\mu$m) in Bands 9--10
\end{itemize}
\end{minipage}\vspace*{\ispaceend}\\
\Block{1-1}{\vspace*{\bspace}{\bf Phase of rapid growth} \\ $\bm{z=3\text{--}6}$} & 
\begin{minipage}{\miniwidth}
\begin{itemize}[leftmargin=\leftmarginspace]
\vspace*{\ispacestart}
\item Mid to high-J CO in Bands 3--5, $\mathrm{J_{up}=3\text{--}4}$ in Band 2, CO $J=2-1$ in Band 1
\item \cii\/ in Band 7--8, \ci\/ in Bands 2--4
\end{itemize}
\end{minipage}\vspace*{\ispaceend}\\
\Block{1-1}{\vspace*{\ispacestart}{\bf Cosmic noon} \\ (Peak of Cosmic Star Formation) \\ $\bm{z=1\text{--}3}$\vspace{\ispaceend}} & 
\begin{minipage}{\miniwidth}
\begin{itemize}[leftmargin=\leftmarginspace]
\vspace*{\ispacestart}
\item Mid to high-J CO $\mathrm{(J_{up}\geq3)}$ in Bands 4 or higher, low-J CO $\mathrm{(J_{up}\leq3)}$ in Bands 1--3
\item \cii\/ in Band 9, \ci\/ in Bands 4--7
\end{itemize}
\end{minipage}\vspace*{\ispaceend}\\
\Block{1-1}{\vspace*{\bspace}{\bf Recent universe} \\ $\bm{z<1}$} & 
\begin{minipage}{\miniwidth}
\begin{itemize}[leftmargin=\leftmarginspace]
\vspace*{\ispacestart}
\item A large range of CO $\mathrm{J_{up}}$ are available
\item \ci\/ (370 and 609\,$\mu$m) in Bands 7--10
\end{itemize}
\end{minipage}\vspace*{\ispaceend}\\
\end{NiceTabular}
\label{tbl:redshifted_lines}
\end{table}

%% file: table_rebels.tex
\begin{table}[h]
\centering
\caption{Number of tunings required for the REBELS survey}
\begin{NiceTabular}{lccccccc}[hvlines]
\CodeBefore
\rowcolor{\headercolor}{1}
\Body
Number of tunings & 1 & 2 & 3 &4 &5 & 6 & 8\\
Number of REBELS targets & 2 & 16 & 17 & 1 & 2 & 1 & 1\\
\end{NiceTabular}
\label{tbl:rebels_tuning}
\end{table}

%% file: table_origins_galaxies_2.tex
\begin{table}[ht]
\centering
\caption{Velocity bandwidths per sideband of the current and WSU correlators}
\begin{NiceTabular}{lccc}[tabularnote=Note: Velocity bandwidth is for the middle of the receiver band and one sideband.]
\hline
\hline
   \RowStyle[rowcolor=\headercolor]{\bfseries}
   & \Block{1-3}{Velocity width (\kmss)}\\ \cline{2-4}
  \RowStyle[rowcolor=\headercolor]{\bfseries}
   & Current & WSU & WSU\\
   \RowStyle[rowcolor=\headercolor]{\bfseries}
   & Bandwidth=3.75\,GHz & Bandwidth=8\,GHz & Bandwidth=16\,GHz\\
\hline
Band 1 & 26,500 & 56,500 & 112,900 \\
Band 2  & 14,300 & 30,600 & 61,100\\
Band 3  & 11,300 & 24,000 & 48,000\\
Band 4  & 7800 & 16,700 & 33,300\\
Band 5  & 6000 & 12,800 & 25,700\\
Band 6  & 4600 & 9900 & 19,800\\
Band 7  & 3500 & 7400 & 14,800\\
Band 8  & 2500 & 5400 & 10,800\\
Band 9  & 1700 & 3600 & 7300\\
Band 10 & 1300 & 2800 & 5500\\
\hline
\end{NiceTabular}
\label{tbl:band_velocity}
\end{table}

%% file: summary.tex
\section{Summary}
\label{sec:summary}

The WSU is the top priority of the ALMA Development Program for the upcoming decade. The broad goal is to initially double, and eventually quadruple, the bandwidth of the ALMA receivers and all associated electronics. The upgrade will require major changes to the ALMA signal chain, from the front-end receivers, to the digitizers and digital transmission system, the correlator, and the associated data processing and archive that delivers the data to the community. To accomplish these goals, the ALMA partnership has began a global, coordinated effort to ensure the success of the WSU. The first steps toward the WSU have already started. The first wideband receivers, Band 2 and Band 6 \citep{Yagoubov20,Navarrini21,Kerr2021}, are under development. A development project for the preliminary design of the digitizers and first Fourier transform for the correlator system has been approved by the ALMA Board. A project to build the new correlator has been approved by the National Science Foundation and will be submitted to the ALMA Board later this year. 

ALMA users will begin to reap the benefits of the WSU as soon as the correlator, digitizers, and data transmission system are installed later this decade and paired with the current suite of receivers. As the  receivers are upgraded and the final expansion of the correlator is completed, the full capabilities of the 2--4$\times$ bandwidth expansion will become available. The upgrade will increase the bandwidth of most receivers by a factor of 2--4, lower the receiver noise temperatures, replace the digitizers and data transmission system to handle the increased bandwidth, and replace the correlator with a far more powerful system. The WSU will benefit all ALMA observations by increasing digitization and correlation efficiencies, improving the spectral line and continuum sensitivity, greatly expanding the spectral grasp, and providing higher spectral resolution. The gains will be particularly extraordinary for high-spectral resolution observations ($\sim0.1$--0.2\kms) of molecular clouds in the Milky Way and nearby galaxies and for circumstellar disks, where the WSU will increase the bandwidth that can be correlated at such resolutions by 1 to 2 orders of magnitude compared to the current system. 

The WSU will expand the scope of the science programs that can be envisioned by the community. With the upgrade, singular spectral line observations will be $\sim2$--3$\times$ faster than the current system, continuum observations 3--6$\times$ faster (for the 2$\times$ and 4$\times$ bandwidth upgrade, plus any receiver speed gains from receiver improvements), and spectral scans 1--2 orders of magnitude faster in most receiver bands at high spectral resolution. ALMA users will be able to devise observing programs with vastly increased sample sizes, conduct extensive spectral line surveys, and obtain deep observations of individual sources more efficiently that will advance the three main themes of the ALMA2030 Roadmap: the Origins of Planets (Section~\ref{sec:planets}), the Origins of Chemical Complexity (Section~\ref{sec:chemistry}), and the Origins of Galaxies (Section~\ref{sec:galaxies}). The increased scope of ALMA observations will enhance the considerable synergies between ALMA and many of the existing and future facilities that share many of the same science themes with ALMA, including JWST, NGRST, ELT, TMT/GMT, and the ngVLA. ALMA remains the premier telescope for submillimeter astronomy, and the WSU will ensure it remains at the forefront of scientific discovery for years to come.

\vspace{1cm}
We thank Elizabeth Humphreys and Luciano Cerrigone for contributing the science case on evolved stars. We would also like to thank Todd Hunter for detailed analysis of optimal spectral scan tuning setups for the BLC and WSU and technical advice, as well as, Ryan Loomis for the continuum simulations of the IM~Lup-like disk. We appreciate the contributions from Ciska Kemper in an early draft of this paper. We also thank Enrique Mac\'ias, Sergio Mart\'in, James Miley, Miguel Vioque,  Alejandro Santamaria, Mar\'ia D\'iaz Trigo, Neil Phillips, and the ALMA Science Advisory Committee (ASAC) for their many insightful comments and suggestions.

%% file: AppendixTech.tex
\section{Technical Overview of the Wideband Sensitivity Upgrade}
\label{sec:tech}

This Appendix presents technical details of the WSU upgrade. The plan and status for upgrading the ALMA receivers is described in Appendix~\ref{sec:receivers}, the Back-End and DTS improvements are described in Appendix~\ref{sec:DigitizerDTS},  the capabilities of the next generation correlator (the ALMA TALON Central Signal Processor, or AT.CSP) are described in Appendix~\ref{sec:correlator}, and the upgraded ACA Total Power spectrometer (ACA TPS) is described in Appendix~~\ref{sec:TPspec}.
The information reported here necessarily represents our best understanding of the technical landscape at this point in time. Where possible, information provided by the development teams has been provided. For other less advanced aspects, we refer to the expectations codified by the three ALMA2030 WSU working groups, which have drafted detailed recommendations for the WSU requirements that are expected to be formally reviewed, adopted, and published later this year.

\subsection{Front-end Receiver Upgrades}
\label{sec:receivers}

\begin{figure}[ht]
\centering
\includegraphics[width=\textwidth]{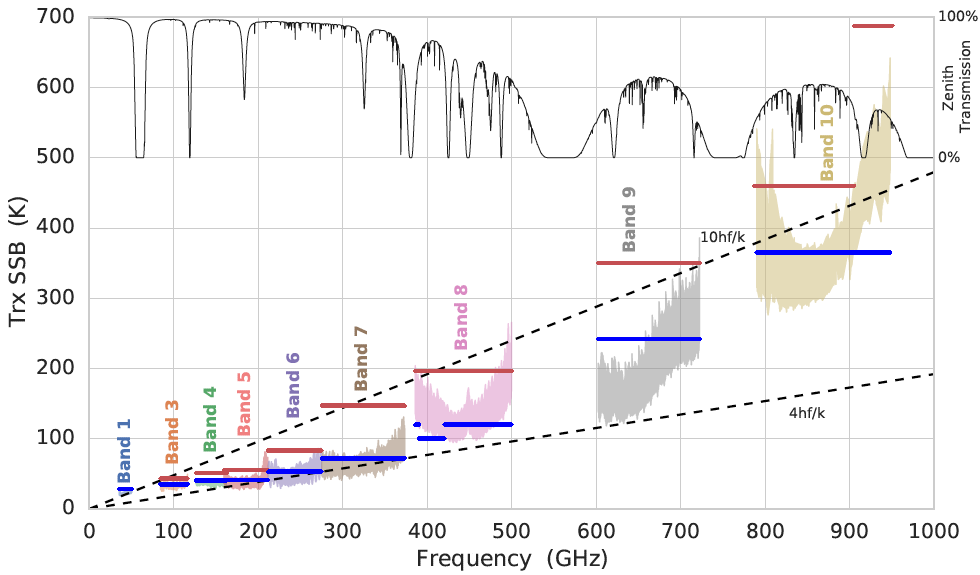}
\caption{Single sideband (SSB) equivalent receiver temperatures ($T_{\rm rx,\textsc{ssb}}$) for the current ALMA receivers as a function of frequency. The zenith transmission curve with Precipitable Water Vapor (PWV) of 0.472\,mm is also shown for reference. Red and blue horizontal line show the current and future $T_{\rm rx,\textsc{ssb}}$, respectively. Results for Band 2 will be available soon as the receiver is now in  pre-production.}
\label{fig:Trx}
\end{figure}

A key goal of the WSU is to increase both the sensitivity and instantaneous bandwidth (also known as the intermediate frequency, or IF, range) of the ALMA front-end receivers.  Figure~\ref{fig:Trx} shows the single sideband (SSB) equivalent receiver temperatures ($T_{\rm rx,\textsc{ssb}}$) for the current ALMA receivers as a function of frequency. Band 1 (35--50\,GHz) is also shown, and is expected to be available starting in Cycle 10. The shaded regions demonstrate the range of current $T_{\rm rx,\textsc{ssb}}$ performance for each receiver band. The zenith transmission curve at the ALMA site for the best quartile of  weather conditions is also shown for reference ($\mathrm{PWV}=0.472$\,mm). When Band 2 (not shown) is completed early in the second half of this decade, ALMA will have achieved the full frequency coverage accessible from the ground from 35 to 950\,GHz. Included on this figure are red horizontal lines (per band) that indicate the current system requirements for the $T_{\rm rx,\textsc{ssb}}$ applicable to 80\% of the band's frequency range. The dashed lines show the 4$\times$ and 10$\times$ Planck photon noise levels (a few times this limit is the best that can be achieved), demonstrating that the current suite of ALMA receivers already have excellent performance; in all cases the current performance already significantly exceeds the requirements over most of their tuning ranges. 

Also shown on Figure~\ref{fig:Trx} are blue horizontal lines for each band that indicate the $T_{\rm rx,\textsc{ssb}}$ goals (applicable to 80\% of the band's frequency range) for future receiver upgrades \citep{Asayama20}. Note that for the two current Double Sideband (DSB) receivers (Bands 9 and 10), the equivalent SSB performance and $T_{rx}$ are shown for ease of direct comparison. We expect that all future ALMA receivers (apart from Band 1) will use 2 Sideband (2SB) designs because of their superior noise performance, particularly in rejecting atmospheric noise from the opposite sideband, which is most impactful in the higher frequency bands \citep[see][]{Asayama20}. 
{\bf We expect that the sensitivity improvements for Bands 3--10 will be $\sim$20\%, when compared to the best IF/RF performance ranges of the current receivers, and growing to as much as $\sim$50\% improvement toward the edges of the current RF ranges for several receivers (see Figure~\ref{fig:Trx}).  At the same time, comparably significant imaging speed improvements will come from the 2--4$\times$ increase in instantaneous bandwidth.}

\input{table_frontend}

Table~\ref{tbl:FEstatus} summarizes the current radio frequency (RF) ranges and sideband properties for the ALMA receivers, as well as the future post-WSU properties and the current status. Additional information for the two projects that are already underway, to build Band 2 and upgrade Band 6, are described in Appendix~\ref{sec:Band2} and Appendix~\ref{sec:Band6}, respectively. Although it is likely to be well into the next decade before all of the ALMA receiver bands can be upgraded, after the digital aspects of the WSU are implemented, observations with the ``legacy'' ALMA receivers will enjoy a number of immediate improvements:
\begin{enumerate}
    \item The 1SB (Band 1) and 2SB (Bands 3--8) receivers will be able to access their full potential correlated bandwidth, rather than the present limitation of 7.5\,GHz of correlated bandwidth. This will be 8\,GHz of correlated bandwidth, apart from Band 6 observations, which will be able to use up to 11\,GHz of bandwidth (though as now, the lower IF portion of this expanded range will have poorer sensitivity until Band 6v2 is completed; see Appendix~\ref{sec:Band6}). Using the current DSB receivers (Bands 9 and 10), observations will be able to access 16\,GHz, instead of 15\,GHz of correlated bandwidth.   
    \item Improved sensitivity from digital efficiency improvements (see Appendix~\ref{sec:sensitivity}).
    \item Access to high spectral resolution without sacrificing bandwidth (see Section~\ref{sec:specres}).
\end{enumerate}

\subsubsection{Band 2: First Wideband ALMA Receiver}
\label{sec:Band2}

After Band 1 (led by East Asia) is offered to ALMA users starting in ALMA Cycle 10, Band 2 will complete the frequency coverage originally envisioned for ALMA --- resulting in the widest frequency coverage of any interferometric (sub)millimeter facility by more than a factor of four. As originally conceived, ALMA Band 2 would cover the frequency range of 67--90\,GHz. However, a large international collaboration led by ESO is near completion of a suite of pre-production receiver cartridges for Band 2 that will cover the 67--116\,GHz frequency range; i.e., also encompassing the Band 3 frequency range. Band 2 is expected to be deployed for science use early in the second half of this decade. The minimum and stretch goals for the IF bandwidth for this receiver are 4--16\,GHz and 4--18\,GHz, respectively, in each of two sidebands (2SB).

\begin{figure}[h!]
\centering
\begin{minipage}{0.65\linewidth}
\includegraphics[width=\textwidth]{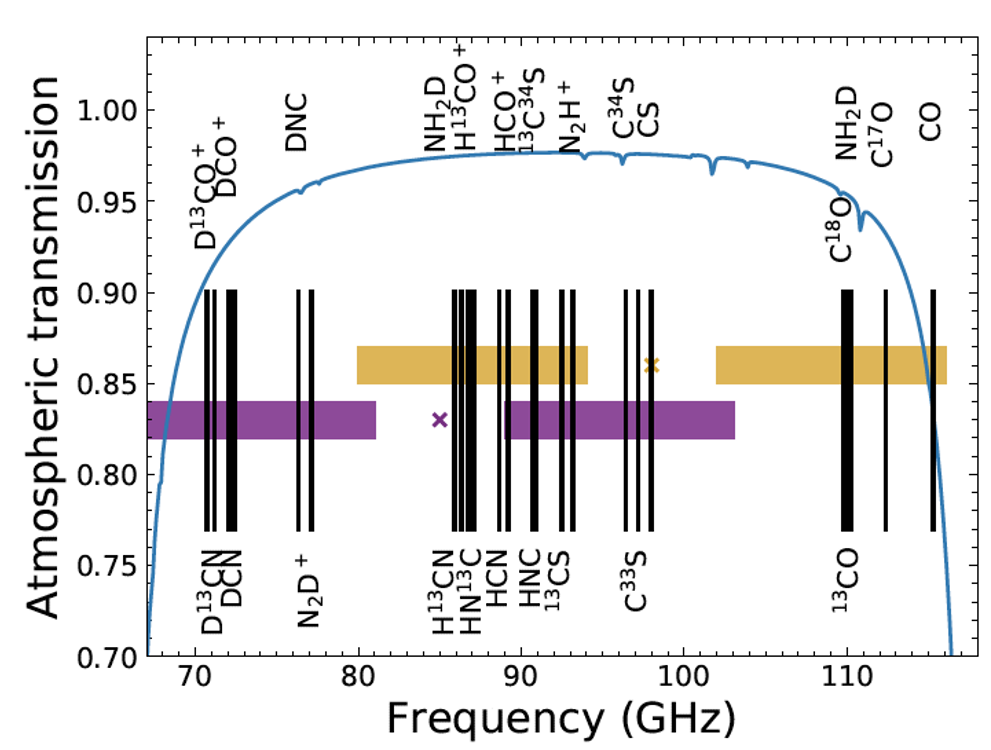}
\end{minipage}
\begin{minipage}{0.33\linewidth}
\caption{Overview of the frequency range for the new Band 2 receiver. A number of particularly diagnostic transitions in this frequency range are labeled for reference. The purple and gold line segments demonstrate two promising tuning options, assuming the receiver is able to meet its 4--18\,GHz IF stretch goal. The atmospheric transmission curve for this frequency range is also plotted for reference. Figure from \citet{Yagoubov20}.}
\label{fig:Band2}
\end{minipage}
\end{figure}

Band 2 will be ALMA's first wide-IF 2SB receiver, and will be critical in commissioning the WSU system. As demonstrated in Figure~\ref{fig:Band2},  Band 2 will enable observations of a wide-range of important diagnostic transitions from the cold local Universe. It will also allow observations of mid-J CO and \ci\ emission from the high redshift Universe (see Section~\ref{sec:galaxies}), as well as sensitive continuum observations in the critical transition zone between dominance of free-free versus dust continuum emission. Additional Band 2 science cases are presented in \citet{Mroczkowski2019} and a technical description of the receiver is provided by \citet{Yagoubov20}.

\FloatBarrier

\subsubsection{Band 6v2: First Upgraded (Wideband) ALMA Receiver}
\label{sec:Band6}

Band 6 has been chosen as the first receiver to be upgraded as part of the WSU, in part because it is presently ALMA's most popular receiver band, in terms of both the number of hours requested in observing proposals across all five proposal categories, as well as the appearing in more ALMA publications than any other receiver band. In November 2021, the ALMA Board approved a development project led by North America to build a ``Band 6v2'' prototype receiver that will provide the following enhancements: 
\begin{itemize}
    \item Extend the Band 6 frequency range from 211--275\,GHz to 209--281\,GHz; allowing observations of $^{13}$CS $J=6-5$, N$_2$H$^+$ $J=3-2$, $^{13}$C$^{18}$O $J=2-1$, and H31$\alpha$ with other Band 6 transitions for the first time.
    \item Expand the IF bandwidth from 4.5--10\,GHz to at least 4--16\,GHz, with a stretch goal of 4--20\,GHz, which will significantly expand the continuum range and number of lines that can be observed simultaneously. 
    \item Significantly improve the noise performance, especially in the 4.5--6\,GHz part of the IF that currently has quite poor noise. This improvement will be particularly important because the 4.5--5.0\,GHz  range is required to observe simultaneously the CO and $^{13}$CO $J=2-1$ transitions, which are crucial tracers of the molecular gas content of star forming regions and nearby galaxies. When the full WSU is considered (Band 6v2, digital efficiency improvements, and upgraded correlator):
    \begin{itemize}
        \item Spectral lines currently tuned in the 6--10\,GHz IF range will be 1.5$\times$ more sensitive for the same observing time; or 2.25$\times$ faster to achieve the same noise as today;
        \item Spectral lines  tuned in the currently poor 4.5--6\,GHz IF range will be 2.2$\times$ more sensitive for the same observing time; or 4.7$\times$ faster to achieve the same noise as today;
        \item Continuum observations will have 2.2$\times$ better sensitivity for the same observing time, or be 4.8$\times$ faster to achieve the same noise as today.
    \end{itemize}
\end{itemize}

\begin{figure}[h]
\centering
\includegraphics[width=0.95\textwidth]{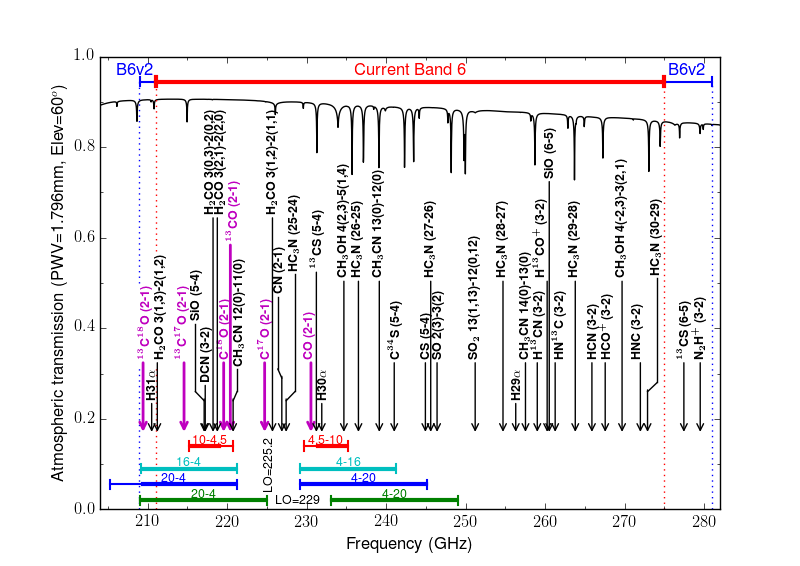}
\caption{Overview of the Band 6 frequency range and some of the diagnostic line transitions that fall in this band. The proposed extensions to the Band 6 RF range (red) are shown along the top of the plot in blue. Along the bottom, the current Band 6 IF range is shown in red, with the thicker part of the line indicating the better performance region --- presently it is just barely enables simultaneous CO and $^{13}$CO $J=2-1$ with the current Band 6, albeit with poorer than normal noise performance. Cyan and blue lines show the minimum and stretch goal IF ranges for Band 6v2, with positions chosen to demonstrate coverage of the isotopologues of CO $J=2-1$; the green line also uses the stretch IF goal and shows an alternative tuning option. These Band 6v2 IF range options demonstrate the dramatic increase in spectral transitions that will be possible with just one tuning after the upgrades.}
\label{fig:Band6}
\end{figure}

Figure~\ref{fig:Band6} shows the Band 6v2 receiver tuning capabilities
compared to the current Band 6. After the Band 6v2 upgrade, many more diagnostic spectral lines will be observable simultaneously, increasing the science that can be achieved with a single observation; at the same time both spectral line and continuum Band 6 observations will be significantly more sensitive or efficient as the science case warrants. In particular, it will be possible to sensitively observe almost all of the CO $J=2-1$ isotopologues with a single observation as demonstrated in Figure~\ref{fig:Band6}. Upon successful completion of the prototype, a project to manufacture the full complement of upgraded Band 6v2 receivers is expected to start mid-decade and be completed in $\sim 2030$. Upgraded receivers will be gradually deployed for science observing as they are commissioned. Additional technical details of the Band 6v2 receiver are provided by \citet{Navarrini21} and \citet{Kerr2021}.

\subsubsection{Other Receiver Developments}
\label{sec:rxother}

In addition to the ongoing Band 2 and Band 6 development projects, the WSU has the potential to improve dramatically the observing efficiencies across all ALMA bands thanks to the active receiver development programs across the ALMA partnership and the broader radio astronomical community.
As discussed in \cite{Asayama20} and summarized in Table~\ref{tbl:FEstatus}, several studies are underway to explore potential upgrade paths for the ALMA receivers, largely guided by the band upgrade prioritization set out in ALMA2030 Development Roadmap  \citep{Carpenter19}. 
These studies aim to meet the new receiver specifications defined in \citet{Asayama20}, including lower receiver noise temperatures, increased IF bandwidth, and upgrading Bands 9 and 10 to 2SB systems. In addition, several studies are exploring the feasibility of extending the RF bandwidth, which will enable new observational setups and access to simultaneous observations of spectral lines not accessible to ALMA currently. The upgrade to a 2SB design for Bands 9 and 10 will also enable spectral line observations using the Total Power Array, a key capability for observing extended sources since the field-of-view (and the maximum recovery scale) is small at high frequencies. 

As discussed in \cite{Asayama20}, many of the bands based on SIS mixers (Bands 3--10) would see immediate improvements simply through upgrades to lower noise, wider bandwidth IF amplifiers. 
Studies are underway in multiple ALMA partner regions that could deliver upgrades to Band 7 with expanded IF and RF frequency ranges.  For instance, \cite{Lee2021} recently presented a design that would cover 275--500\,GHz on sky, and could deliver $T_{\rm rx}$ well below the new requirements \citep[see][]{Asayama20} of 90\,K and 144\,K in Bands 7 and 8 respectively (see Figure~\ref{fig:Trx}). 

The Band 8 frequency range (385--500\,GHz) is currently covered by a 2SB receiver with an IF of 4--8\,GHz. The Band 8 receiver is expected to be upgraded in the coming years, expanding the IF bandwidth by 4$\times$ (IF=2--18\,GHz or 4--20\,GHz). Recently, a new 2SB receiver cartridge with an IF of 4--18\,GHz was installed in the ASTE telescope, and tests to improve the system performance are currently underway \citep{kojima20}.
With the WSU upgrade, the system temperature is expected 
to improve by 10--20\% over the current system (see Figure~\ref{fig:Trx}).
The expanded bandwidth will allow 80\% coverage of the atmospheric window from 385 to 425\,GHz (i.e., the portion of the Band 8 spectral window which is less susceptible to atmospheric absorption) in just one tuning, realizing efficient continuum and spectral line surveys in the highest angular resolution of ALMA available with high atmospheric transmission most of the time. 

For Band 9, the updated specifications given in \cite{Asayama20} require sideband separation and a $T_{\rm rx} \leq 242$\,K over 80\% of the bandwidth (see Figure~\ref{fig:Trx}). 
The expected $T_{\rm sys}$ improvement compared to the current Band 9 receiver is $20\text{--}50\%$ across the Band 9 RF range in good weather conditions. While still under study within the ALMA partner regions, \cite{Hesper2021} recently presented results for a sideband separating Band 9 cartridge that would expand the IF bandwidth to at least 12\,GHz of bandwidth, and possibly larger (e.g., 4--18\,GHz). This would also expand the RF bandwidth beyond the current 602--720\,GHz, improve the polarization performance, and deliver noise performance well within the new requirements defined in \cite{Asayama20}. This work builds on the work presented in \cite{Belitsky2019}, which presented a low noise 2SB Band 9 with 8\,GHz of IF bandwidth.

The current Band 10 receiver is a DSB system that covers the RF range of 787--950\,GHz with an IF bandwidth of 4--12\,GHz.
The Band 10 receiver is expected to be upgraded to a 2SB configuration, suppressing the noise from the image sideband and providing robust line detections with improved receiver temperature. The expected receiver temperature over $80\%$ of the RF band after the Band 10 upgrade is 365\,K, which will bring a $\sim20$--60\% improvement in the system temperature and hence sensitivity, assuming relatively good weather conditions at the ALMA site.
A receiver cartridge which was refurbished from a prototype ALMA Band 10 receiver with DSB mixers employing high critical current density SIS junctions was  installed at the ASTE telescope \citep{Ito20} and used for science verification observations \citep{Asayama22}. Technology developments are underway that will allow the receiver to be upgraded to a 2SB system  \citep{Fujii20} with expanded IF bandwidth \citep{Gonzalez20}.

\subsection{Back-end Processing and Data Transport}
\label{sec:DigitizerDTS}

Inside the  ALMA antennas, the ``Back-End'' subsystem takes the analog signal from the ``Front-End'' receivers and converts it to digital form. The  sensitivity of the conversion process is dependent upon the effective number of bits used to convert and store the digital data. The current system employs an analog down conversion to convert the receiver signal into $4\times2$\,GHz (per polarization) basebands, which are then fed to eight digitizers that employ 3-bits to convert the signal to digital format. The resulting data rate produced by each of the eight digitizers is 4\,GSamples/s (per polarization per baseband). 

The WSU will eliminate the need to perform a second analog down-conversion, simplifying the signal chain and thus removing some modes of failure and sources of spurious, locally generated signals commonly called ``birdies''.\footnote{See, for example, \emph{The Radio Communication Handbook} \citep{biddulph1994radio} which defines ``birdies'' as ``carriers heard within the tuning range of the receiver but stemming not from external signals but from the receiver's own oscillators.''} Each digitizer will handle an entire sideband per polarization (for 2SB receivers) and will do the digital conversion using 6-bits to ensure an effective number of bits $\gtrsim5$; see \citet{Asayama20} and Section~\ref{sec:sensitivity}. To achieve this level of performance, each of the four WSU digitizers must output 40\,GSamples/s per sideband per polarization, resulting in a Nyquist limit of 20\,GHz per sideband per polarization. After digitization, the data will likely be channelized via an over-sampled poly-phase filter bank into $10\times2$\,GHz subbands per sideband per polarization, with the inner 1.6\,GHz of each subband being of science quality.\footnote{Alternatively, $8\times 2.5$\,GHz subbands per sideband per polarization, with the inner 2\,GHz usable, are also under consideration.} The time-series nature of the data is preserved; i.e., there is no Fourier transform at this stage. {\bf Thus, the usable (science quality) portion of the digitized and subbanded data stream will be 16\,GHz per sideband per polarization wide, which is 4$\times$ the current digitized bandwidth}. The over-sampling means that there will be no gaps in spectral sensitivity across the full 16\,GHz per sideband per polarization. Including the higher bit-depth, the increase in the total digital data rate will be 10$\times$ higher than is presently transported to the correlator.

From the Back-End subsystem, the data then need to be transported down fiber optic cables to the Array Operations Site (AOS), where the current correlators operate at the high site at 5000\,m; the existing cables are sufficient to handle the 10$\times$ higher raw data rate.  A key aspect of the WSU is to locate the 2nd Generation ALMA Correlator at the Operations Support Facility (OSF) at 3000\,m rather than the AOS at 5000\,m in order to mitigate the cooling cost of a much more powerful correlator and to improve its overall maintainability. The current fiber optic cable between the AOS and OSF is not sufficient to handle its essential communications roles, as well as the increased data rates from the WSU, necessitating a new fiber run between these two locations, as well as new telecommunications equipment at the antennas, AOS, and OSF to place the WSU data onto the fibers, and deliver it seamlessly to the 2nd Generation Correlator. 

\subsubsection{Digital Efficiency}
\label{sec:digeff}

Improvements to the overall digital efficiency of the ALMA system translate directly to improved sensitivity for every ALMA observation (for a fixed observing time and spectral resolution), or shorter observing time (for a fixed sensitivity). The overall digital efficiency is largely determined by the number of bits used for each digital signal processing (DSP) stage. The dominant contributors are digitization, ``subbanding'' or channelization, and correlation. 
The WSU is timely because it is now affordable to employ DSP with much higher bit-depths than were feasible in the early 2000s when the baseline ALMA design was formulated. At minimum, the WSU hopes to deliver a final system efficiency of at least 0.96 (and potentially even somewhat better), which is a substantial improvement compared to the current efficiency (BLC FDM) of $\sim 0.811$. Accounting for the dominant contributions, the ALMA system efficiency for the WSU, when the correlator is in Imaging Correlation (IC) Mode, may be as high as 0.976, assuming a digitization efficiency of 0.978 \citep{Quertier2021}, 6 to 8-bits for subbanding stages, and 6-bit correlation. Thus, improvement in sensitivity due to the increased digitization and correlation efficiencies for standard Imaging Correlation Mode will be improved by a factor of up to $\sim1.2$ (compared to BLC observations), which amounts to a speed increase of a factor of 1.44 to achieve the same sensitivity as now (for a given bandwidth).

When the AT.CSP is in VLBI/Beamforming (VB) Mode, the beamformed data is expected to have a digital efficiency of $\sim 0.96$, and the visibilities that are created in VB Mode, will have an efficiency of $\sim 0.954$, due to the use of 4-bit rather than 6-bit correlation in this mode. The beamformed system sensitivity will also be affected by additional factors such as weather-related decorrelation and residual delay errors.

\subsection{2nd Generation ALMA Correlator: ALMA TALON Central Signal Processor (AT.CSP)}
\label{sec:correlator}

The proposed 2nd Generation ALMA Correlator is called the ALMA TALON Central Signal Processor (AT.CSP).\footnote{The AT.CSP information provided here is reproduced by permission from the currently under review NA ALMA Development Project Proposal ``Proposal for the 2nd Generation ALMA Correlator/Beamformer – the ALMA TALON Central Signal Processor''.} The AT.CSP comprises three main elements: the AT.CBF (ALMA TALON Correlator Beam Former), the AT.CDPs (ALMA TALON Correlator Data Processors), and the AT.HILS (ALMA TALON Hardware In the Loop Simulator). The AT.CBF (i.e., the correlator hardware and firmware) is based on the National Research Council of Canada's Square Kilometer Array - Mid telescope/beamformer \citep[Mid.CBF; see][]{Pleasance2017,Gunaratne2020,Pleasance2021} FFX architecture which passed critical design review in 2018. Using a common correlator architecture for two of the world’s largest interferometric radio telescopes, ALMA and SKA, is a unique opportunity to leverage existing extensive research and development work, and to potentially share in future technology, firmware, and software upgrades in the long term \citep[see also][]{Carlson2020}. Some of the details related to the adaptation of the Mid.CBF correlator design to suit the specific needs of ALMA2030 are described in \citet{Carlson2020} and \citet{Gunaratne2022}.

Below is a summary of some of the most important properties and capabilities of the AT.CSP:
\begin{enumerate}
    \item Doubles the maximum correlated bandwidth to 16\,GHz per polarization, and is readily expandable to 4$\times$ correlated bandwidth (32\,GHz per polarization) as additional funding becomes available in the future.  This increased bandwidth translates directly to improved continuum sensitivity and spectral grasp (correlated bandwidth at a given spectral resolution), which are two of the core goals of the ALMA2030 Development Roadmap.
    \item Correlates up to 70 antennas (the current total number of ALMA antennas is 66). 
    \item Provides flexible subarrays enabling independent observations with ALMA’s three main components: the Main 12-m array, the ACA 7-m Array, and the ACA Total Power Array.\footnote{The ALMA Compact Array Total Power Spectrometer (ACAS), an ongoing ALMA Development project led by East Asia, will also be upgraded as part of the WSU in order to observe  Total Power data efficiently.}
    \item Delivers improved digital sensitivity with an AT.CSP correlator efficiency of at least 99.88\% in interferometric observing mode (all processing stages are 6-bit or higher).
    \item Allows placement of up to 80 independent spectral windows in 200\,MHz bandwidth Frequency Slices (FS)\footnote{Up to 200.88\,MHz of each oversampled frequency slice (FS) is expected to be of science quality (i.e., limited sensitivity roll-off) which will allow ``stitching'' of frequency contiguous FSs, with each center separated by 200\,MHz, to create wider bandwidth science spectral windows with constant spectral sensitivity. The additional usable bandwidth of each FS will also allow seamless stitching of spectral scan tunings, since all upgraded receivers are expected to have a lower IF frequency of $\leq 4$\,GHz.}, each with easy sub-selection of native spectral resolution, spectral Zoom resolution (higher spectral resolution by successive factors of 2), or spectral averaging of channels to control data rates.
    \begin{itemize}
        \item Placement of FSs is flexible with respect to sideband. For example, if a given receiver has an instantaneous (IF) bandwidth of 14\,GHz per sideband (per polarization), then up to $60\times 200$\,MHz FS can be placed in one sideband and the remaining 20 FS can be placed in the other sideband. 
        \item If desired, frequency-contiguous FSs with the same channelization can be stitched together in the AT.CDPs to create wider spectral windows, for science or calibration purposes.
    \end{itemize}
    \item Enables high spectral resolution at maximum correlated bandwidth, with a native spectral resolution of 13.5\,kHz ($\sim 0.12$\kms\ at 35\,GHz), producing up to $80\times14880$  ($\sim 1.2$ million) spectral channels per polarization, 77$\times$ more channels than the BLC. All four polarization products are always produced, but the cross-products can be pruned if desired.
    \item Produces high spectral fidelity: 10$^{-6}$ (i.e., $-60$\,dB) spectral channel-to-channel isolation, removing the need to degrade spectral resolution with online windowing functions like the currently employed Hanning smoothing in the BLC.
    \item Applies all-digital, virtually perfect delay and phase tracking with no delay-dependent anomalies in visibilities and no post-correlation corrections needed.
    \item Provides full compatibility with, and significant improvements to, current VLBI and pulsar observing modes, including the ability to observe with a single antenna in VLBI mode (A1-VLBI) without disturbance to interferometric observing. Additionally, it will be possible to beamform up to two phase centers at a time, which for example, can be used to observe at two different frequencies in different phased arrays.
\end{enumerate}

\input{table_correlator_compare}

Table~\ref{tbl:ICMode} compares the key parameters of the BLC in FDM Mode and the AT.CSP in Imaging Correlation Mode. In addition to doubling the correlated bandwidth, the two most extraordinary improvements are in the number of available channels with full polarization (77$\times$ more) and the increased correlator efficiency ($\sim13.4$\%). The dramatic increase in the number of available channels will afford access to the full correlated bandwidth, even at high spectral resolution (see Section~\ref{sec:specres}) for the lower frequency bands of ALMA, and the improved correlator efficiency, together with other WSU digital improvements will translate directly into greater sensitivity for all ALMA observations (see Section~\ref{sec:sensitivity}). Table~\ref{tbl:SpecRes} presents a detailed comparison of the spectral resolution as a function of correlated bandwidth that can be achieved with the AT.CSP and the current ALMA correlator.

\input{table_atcsp}

\subsection{2nd Generation ACA Total Power Spectrometer}
\label{sec:TPspec}

The new Graphics Processing Unit (GPU) spectrometer for the TP Array of the ACA was installed at the AOS in early 2022, and it is expected to be used for science observations starting in Cycle 10. It is based on GPU technology, widely used in graphics processing, video gaming, machine learning, or other applications which require intense data analysis. The ACA spectrometer can process the data from the TP Array using 32-bit floating point arithmetic, allowing a higher level of linearity and dynamic range than the current ACA correlator, which calculates the FFT of the data using 16-bit fixed point arithmetic but decimates the result down to 4-bits before the calculation of the auto-correlation.  The total power spectrometer will allow more precise imaging of faint spectral lines surrounding a very bright source. It was developed and installed by the Korea Astronomy and Space Science Institute (KASI) and the National Astronomical Observatory of Japan (NAOJ), as a collaboration within the East Asian ALMA partnership.

The ACA spectrometer is designed to calculate full-polarization products of the science target and cross-antenna correlation of the calibration target. It is composed of 4+1 (spare) GPU servers and 1+1 (spare) computer server(s) for monitoring and control (M\&C) purposes, two-way optical splitters, and optical signal amplifiers. Each server has four GPU cards and two data acquisition cards. Each optical splitter bifurcates an optical signal path into two; one for the ACA Correlator and the other for the GPU spectrometer. The functionality of the spectrometer is provided by the code executed on the GPU cards, and it is therefore very flexible, allowing for future updates and improved performance.

The East Asian ALMA partnership plans to upgrade the ACA spectrometer toward the end of the decade. It will be upgraded to increase the processed IF bandwidth by a factor of four, accommodating the expanded IF bandwidth enabled at the front-end receivers, and allowing seamless data combination with the data products from the AT.CSP. The WSU will allow simultaneous imaging of multiple emission lines at high spectral resolution. The proposed  ACA spectrometer upgrade is critical for imaging extended sources in the Milky Way and nearby galaxies.

%% file: table_frontend.tex
\begin{table}[ht]
\centering
\caption{Properties of the current and future ALMA receivers}
\begin{NiceTabular}{YY[2]Y[2]YY[4,c]Y[2]Y[2]Y}[hvlines, code-before=\rectanglecolor[HTML]{\pastelgray}{1-1}{2-1} \rectanglecolor{\headercolor}{1-5}{2-8} \rectanglecolor[HTML]{\pastelorange}{1-2}{2-4}][tabularnote=Notes: Description of the current ALMA receiver bands on the left (shaded orange) and their future properties and upgrade status as of early 2022 on the right (shaded blue). 
For Band 2 and Band 6v2{,} the top future IF range is the minimum requirement and the bottom range is the goal. The ``Future'' RF Range with a ? indicates that upgraded receiver bands may expand their RF coverage{,} but they will not be narrower than the current receivers.]
\hline
\hline
  \RowStyle{\bfseries}
  \Block{2-1}{Band} & \Block{1-3}{Current} & & & \Block{1-4}{Future} \\ \cline{2-4} \cline{5-8}
    \RowStyle{\bfseries}
   & RF Range (GHz) & IF Range (GHz) & Type & ALMA2030 Status & RF Range (GHz) & IF Range (GHz)  & Type\\
\hline
1 & 35--50 & 4--12 & SSB & New capability in Cycle 10 & 35--50 & 4--12 & SSB\\
2 & 67--90 & -- & -- & Project underway & 67--116 & \Block{1-1}{\vspace*{0.1cm}4--16\\(4--18)\vspace*{0.1cm}} & \Block{9-1}{2SB}\\
3 & 84--116 & 4--8 & \Block{6-1}{2SB} & Pending & ? & $\geq4$--12 & \\
4 & 125--163 & 4--8 & & Pending & ? & $\geq 4$--12 & \\
5 & 163--211 & 4--8 & & Pending & ? & $\geq 4$--12 & \\
6 & 211--275 & 4.5--10 & & Project underway & 209--281 & \Block{1-1}{\vspace*{0.1cm}4--16\\(4--20)\vspace*{0.1cm}} & \\
7 & 275--373 & 4--8 & & Upgrade study underway & ? & $\geq 4$--12 & \\
8 & 385--500 & 4--8 & & Upgrade study underway & ? & $\geq 4$--12 & \\
9 & 602--720 & 4--12 & \Block{2-1}{DSB} & Upgrade study underway & ? & $\geq$ 4--12 & \\
10 & 787--950 & 4--12 & & Upgrade study underway & ? & $\geq 4$--12 & \\
\hline
\end{NiceTabular}
\label{tbl:FEstatus}
\end{table}

%% file: table_correlator_compare.tex
\begin{table}[ht]
\centering
\caption{Comparison of the BLC and AT.CSP in Imaging Correlation Mode}
\begin{NiceTabular}{lcc}[hvlines]
\hline
\hline
  \RowStyle[rowcolor=\headercolor]{\bfseries}
  Parameter & BLC ($\mathbf{2\times2}$ bit\tabularnote{The BLC is capable of $4\times4$ bit correlation in FDM mode but at the expense of bandwidth and channels (both halved){,} however it has not been fully commissioned because it would have limited scientific utility given its constraints.} FDM) & AT.CSP ($\mathbf{6\times6}$ bit) \\
  Number of antenna inputs & 64 & 70\\
  Maximum correlator bandwidth (Max CBW) & 7.5\,GHz per pol & 16\,GHz per pol\\
  Number channels per polarization: dual pol & $4\times3840$ & \Block{2-1}{$80\times14880$}\\
  Number channels per polarization: full pol & $4\times1920$ & \\
  Channel width\tabularnote{To suppress ringing it is necessary to apply Hanning smoothing online to BLC data{,} resulting in an effective channel width that is double the native channel width (i.e.{,} 2$\times$ lower spectral resolution). The AT.CSP{,} which has virtually independent channels{,} will not have this deficiency.} at Max CBW: dual pol & 488\,kHz & \Block{2-1}{13.5\,kHz}\\
  Channel width$^b$ at Max CBW: full pol & 976\,kHz & \\
  Finest channel width:$^b$ dual pol & 15.25\,kHz & \Block{2-1}{13.5\,kHz (2, 4, 8, .... 64)}\\
  Finest channel width:$^b$ full pol & 30.5\,kHz & \\
  Zoom windows\tabularnote{The BLC is capable of 16 zoom windows per baseband{,} but  only 4 were implemented because there are not enough channels available to populate them from a science point of view; AT.CSP's 80 FS can be shared flexibly between basebands.} & $4\times4$ & 80 \\
  Internal channel averaging\tabularnote{This channel averaging is done in the correlator hardware{,} which facilitates high-time resolution observations. Additional averaging will also be possible in the AT.CDPs (software).} & None & 2, 3, 4, 6, 8, 12, 16, 24, 48\\
  Correlator efficiency & 0.8810 & 0.99887\\
  Assumed digitization/DSP efficiency\tabularnote{We adopt the values presented in \citet{Quertier2021}{,} namely that the current system is limited to a digitization/DSP efficiency of 0.92{,} while the ALMA2030 WSU digitization/DSP efficiency with an ``effective number of bits'' $ENOB\gtrsim5$ is expected to yield an efficiency of 0.978 followed by a digital signal processing stage requiring a 6-bit requantization.} & 0.92 & 0.97698\\
  Resulting system efficiency & 0.811 & 0.976\\
  Sensitivity improvement factor & ... & 1.20\\
  Speed improvement factor &...  & 1.44\\
\hline
\end{NiceTabular}
\label{tbl:ICMode}
\end{table}

%% file: table_atcsp.tex
\begin{table}[ht]
\centering
\caption{Available spectral resolution of the current correlator and the WSU correlator}
\footnotesize
\begin{NiceTabular}{Y[2]cY[2]cY[3]cccccccccc}[hvlines, cell-space-limits=4pt, code-before=\rectanglecolor{\pastelred}{4-13}{4-15} \rectanglecolor{\pastelred}{5-11}{5-15} \rectanglecolor[HTML]{\pastelgreen}{4-1}{7-5} \rectanglecolor{\pastelred}{6-9}{6-15} \rectanglecolor{\pastelred}{7-7}{7-15} \rectanglecolor[HTML]{\pastelpurple}{9-1}{10-5}][tabularnote=\small Notes: Spectral resolution as a function of correlated bandwidth for the WSU AT.CSP correlator (top) and the current ALMA correlator (bottom). For the AT.CSP{,} up to 80 tunable zoom windows (TZWs) can be configured independently within the 16\,GHz per polarization of correlated bandwidth. Each TZW has a bandwidth of (200/Zoom Factor)\,MHz{,} and spectral channel width of 13.5\,kHz/Zoom Factor. The displayed examples use the indicated Zoom Factor for all spectral windows. Band/spectral resolution combinations that exceed the recommended ALMA2030 finest possible resolution requirement of 10\ms\ are highlighted in red. Unlike the BLC{,} the AT.CSP will have nearly independent channels and will not need Hanning smoothing to damp ringing{,} whereas for the BLC{,} the spectral resolution is twice the channel width; this factor has been applied to the BLC rows.]
\hline
\hline
\Block{1-15}{\bf a) Achievable velocity resolution with AT.CSP in imaging correlation mode}\\
\RowStyle[rowcolor=\headercolor]{\bfseries}
\Block{1-5}{Band} & & & & & 1 & 2 & 3 & 4 & 5 & 6 & 7 & 8 & 9 & 10\\
\RowStyle[rowcolor=\headercolor]{\bfseries}
\Block{1-5}{\bf Frequency (GHz)} & & & & & 35 & 75 & 100 & 150 & 185 & 230 & 345 & 460 & 650 & 870\\
\Block{4-1}{\bf Zoom\\Factor\\all spw} & \bf 1 & \Block{4-1}{\bf Max\\CBW\\per\\pol} & \bf 16\,GHz & \Block{4-1}{\bf Velocity\\width\\(\mss)} & 115.6 & 54.0 & 40.5 & 27.0 & 21.9 & 17.6 & 11.7 & 8.8 & 6.2 & 4.7\\
& \bf 2 & & \bf 8\,GHz & & 57.8 & 27.0 & 20.2 & 13.5 & 10.9 & 8.8 & 5.9 & 4.4 & 3.1 & 2.3\\
& \bf 4 & & \bf 4\,GHz & & 28.9 & 13.5 & 10.1 & 6.7 & 5.5 & 4.4 & 2.9 & 2.2 & 1.6 & 1.2\\
& \bf 8 & & \bf 2\,GHz & & 14.5 & 6.7 & 5.1 & 3.4 & 2.7 & 2.2 & 1.5 & 1.1 & 0.8 & 0.6\\
\Block{1-15}{\bf b) Achievable velocity resolution with the BLC at maximum and minimum FDM correlated bandwidth}\\
\Block{2-2}{\bf BLC} & & \Block{2-1}{\bf Max\\CBW\\dual\\pol}& \bf 7.5\,GHz & \Block{2-1}{\bf Velocity\\width\\(\mss)} & 8364.9 & 3903.6 & 2927.7 & 1951.8 & 1582.6 & 1272.9 & 848.6 & 636.5 & 450.4 & 336.5\\
& & & \bf 0.234\,GHz & & 261.4 & 122.0 & 91.5 & 61.0 & 49.5 & 39.8 & 26.5 & 19.9 & 14.1 & 10.5\\
\hline
\end{NiceTabular}
\label{tbl:SpecRes}
\end{table}